\documentclass[%
 reprint,
 amsmath,amssymb,
 aps,
]{revtex4-2}

\usepackage{graphicx}
\usepackage{dcolumn}
\usepackage{bm}
\usepackage{siunitx}
\usepackage{mhchem}
\usepackage{multirow}
\newcommand\Tstrut{\rule{0pt}{3.0ex}}         
\usepackage{placeins}
\usepackage{ragged2e}

\begin{document}

\preprint{APS/123-QED}

\title{A comprehensive screening of plasma-facing materials for nuclear fusion}

\author{Andrea Fedrigucci$^{1,2}$}
 \email{andrea.fedrigucci@epfl.ch}
\author{Nicola Marzari$^{1,3}$}%
\author{Paolo Ricci$^{2}$}%
\affiliation{%
 $^{1}$Theory and Simulation of Materials (THEOS), and National Centre for Computational Design and Discovery of Novel Materials (MARVEL), École Polytechnique Fédérale de Lausanne (EPFL), Lausanne CH-1015, Switzerland\\
 $^{2}$Swiss Plasma Center (SPC), École Polytechnique Fédérale de Lausanne (EPFL), Lausanne CH-1015, Switzerland\\
 $^{3}$Laboratory for Materials Simulations, Paul Scherrer Institut, Villigen PSI 5232, Switzerland
}%


\begin{abstract}
Plasma-facing materials (PFMs) represent one of the most significant challenges for the design of future nuclear fusion reactors. Inside the reactor, the divertor will experience the harshest material environment: intense bombardment of neutrons and plasma particles coupled with extremely large and intermittent heat fluxes. The material designated to cover this role in ITER is tungsten (W). While no other materials have shown the potential to match the properties of W, many drawbacks associated with its application remain, including: cracking and erosion induced by a low recrystallization temperature combined with a high ductile–brittle transition temperature and neutron-initiated embrittlement; surface morphology changes (He bubbles and fuzz layer) due to plasma-W interaction with subsequent risk of spontaneous material melting and delamination; low oxidation resistance in case of air contamination.
Exploring alternatives to W requires the design of a multivariable optimization problem. This work aims to produce a structured and comprehensive materials screening of PFMs candidates based on known inorganic materials. 
The methodology applied in this study to identify the most promising PFM candidates combines peer-reviewed data present in the Pauling File database and first-principle DFT calculations of two key PFMs defects, namely the surface binding energy and the formation energy of a hydrogen interstitial. The crystal structures and their related properties, extracted from the Pauling File, are ranked according to the heat-balance equation of a PFM subject to the heat loads in the divertor region of an ITER-like tokamak. The materials satisfying the requirements are critically compared with the state-of-the-art literature, defining an optimal subset where to perform the first-principles electronic structure calculations.
The majority of previously known and extensively studied PFMs, such as W, Mo, and carbon-based materials, are captured by this screening process, confirming its reliability. Additionally, less familiar refractory materials suggest performance that calls for further investigations. 
\end{abstract}

\maketitle

\section{Introduction}
One of the major challenges on the way to fusion energy - one of the few options we have to provide mankind with a large-scale, sustainable and carbon-free energy source \cite{iter1,banacloche2020socioeconomic} - is linked to the availability of in-vessel plasma-facing materials (PFMs) capable to sustain the severe operational condition of a fusion power plant \cite{fasoli2023essay} for an assumed lifetime of 1-2 full power years (fpy) \cite{knaster2016materials,arbeiter2018planned}. In particular, in modern tokamaks, such as ITER \cite{iter0}, the most external magnetic field lines are deviated toward the divertor, and the thermal loads that the divertor must withstand are the highest of all the components facing the plasma, making it one of the most challenging parts to build \cite{linke2019challenges}.
The divertor is expected to exhaust the heat and control the level of helium ash and other impurities, mainly resulting from erosion of the plasma-exposed surface, to minimize plasma contamination. In addition, the divertor protects the most delicate surrounding components from neutronic loads. The heat load on the divertor primarily results from a combination of five types of loads: ionic, neutral, radiation power, electrons, and surface recombination energy of incoming ion-electron pairs \cite{hoshino2012simulation}. The power deposited on the divertor surface is dominated by radiative and particle fluxes, with the latter mainly constituted by helium (He), hydrogen isotopes (deuterium and tritium), and impurities. Their flux at the divertor is the highest of the vacuum vessel, \num{e24} particles \si{\per\square\meter\per\second} with an average kinetic energy of 10 \si{\electronvolt} (during steady-state) \cite{federici2001plasma,behrisch2003material}. Last, 14.1 \si{\mega\electronvolt} neutrons bombard the PFM with a flux of \num{e18} particles \si{\per\square\meter\per\second} \cite{federici2001plasma}.

In ITER the thermal load at the divertor in quasi-stationary conditions is estimated to oscillate between 10 \si{\mega\watt\per\square\meter}, in steady-state conditions lasting around 400 \si{\second}, and 20 \si{\mega\watt\per\square\meter}, during the slow-transient thermal loads, where a transition from detached divertor operation to fully attached operation is observed ($<$ 10 \si{\second}) \cite{federici1997raclette,federici2001plasma}.
Moreover, intense and short transient loads are predicted to occur due to edge-localized modes (ELMs), that are quasi-periodic fluctuations  of the plasma pressure at the plasma edge \cite{federici2001plasma}. The thermal energy released on the divertor target during unmitigated ELMs can be as high as 10-15 \si{\mega\joule\per\square\meter} with a frequency of 1-2 \si{\hertz} and a duration of 0.1-1 \si{\milli\second} \cite{evans2013elm,loarte2003characteristics,igitkhanov2014modeling,ueda2017baseline,blanchard2018effect}. In fact, ELM mitigation is required in ITER, and it will be based on increasing ELM frequency to 25-50 \si{\hertz}, consequently lowering their heat flux to around 0.5 \si{\mega\joule\per\square\meter} and reducing the heat peak from tens to tenths of \si{\giga\watt\per\square\meter} \cite{evans2013elm,de2018influence,van2021impact}. 
Finally, off-normal events, which may not be excluded during ITER operations, include vertical displacement events (VDEs), which result in energy pulses up to 30 \si{\mega\joule\per\square\meter} in 1-5 \si{\milli\second}, plasma disruption, which can reach around 60 \si{\mega\joule\per\square\meter} and last 100-300 \si{\milli\second}, and runaway electrons, which lead to 50 \si{\mega\joule\per\square\meter} pulses for less than 50 \si{\milli\second} \cite{federici2001plasma,ueda2017baseline,linke2004performance}.
The simultaneous thermal, plasma, and neutron wall loads lead to complex dynamics and degradation process of the divertor armor. 
The initial ITER divertor design included both carbon fiber composite (CFC) and tungsten (W) as PFMs, but given the high-level of tritium retention and surface erosion, the ITER Organization (IO) decided to switch to a fully \ce{W} divertor since 2013  \cite{pitts2019physics,iter2}. While \ce{W} has the highest melting point of all the metals and outstanding thermal properties, it suffers of (i) low recrystallization temperature combined with high ductile–brittle transition temperature (DBTT); (ii) neutron-induced embrittlement that leads to cracking and erosion of the \ce{W} surface at high thermal loads; (iii) surface morphology changes (He bubbles and fuzz layer) due to plasma-\ce{W} interaction with subsequent risk of spontaneous material melting and delamination; (iv) neutron-induced damages and transmutation affecting thermal and mechanical properties; and (v) low oxidation resistance in case of air contamination \cite{linke2019challenges,coenen2020fusion}.
While the PFM community is working to propose solutions for some of these issues, it is still unclear if a \ce{W}-based divertor will be able to sustain operating fusion conditions in a fusion power plant. For instance, the design of the divertor is still open for DEMO \cite{you2016conceptual}, the power plant that will demonstrate the production of electric power from nuclear fusion. The primary PFM of DEMO divertor will withstand the same thermal loads as the one in ITER, but higher neutron loads (from 0.37–0.47 \si{\mega\watt\per\square\meter} to 1.8–2.4 \si{\mega\watt\per\square\meter}) \cite{asakura2018plasma,reiser2012optimization}. The same risks previously mentioned related to erosion, melting, and cracking of the \ce{W} surface will be faced. 
The more intense neutron bombardment will also lead to stronger displacement damage and transmutation effects, resulting in swelling, irradiation hardening, and radioactivity after exposure. These challenges pose even greater difficulties in the material choice for DEMO and future commercial fusion reactors.

Given the high cost and time required for both testing or simulating PFMs in such a complex environment \cite{brooks2015plasma,pintsuk2020tungsten,gilbert2021perspectives}, only few classes of materials have been evaluated as potential alternatives to \ce{W} \cite{rieth2019behavior} and a high-throughput protocol for PFMs screening is missing in the literature. This multivariable optimization problem requires to define a set of design criteria that a material need to satisfy in order to sustain divertor conditions. 
In this study we define the first methodology to assess the potential of a large class of materials as PFM. 
The heat loads that PFMs must endure are modeled using a heat balance equation, which takes into account both steady-state and transient loads. The thermal properties involved in this equation, such as thermal conductivity, melting temperature, heat capacity, and other related proxies are extracted from the inorganic crystal database MPDS \cite{mpds}, an online platform providing access to the PAULING FILE \cite{paulingfile}, which contains the peer-reviewed properties of, approximately, half a million experimentally-confirmed structures. The best performing materials are ranked and then selected to calculate, in the framework of density-functional theory calculations (DFT) \cite{marzari2021electronic}, surface binding energies (SBE), which are a proxy of the required energy for physical sputtering and therefore the erosion rate, and the hydrogen interstitial formation energies (H-IFE), as an indication of the hydrogen solubility into the crystal \cite{lee2015understanding} and therefore its tritium retention. 

The impact of neutron bombardment is not explored here, due to the significant computational burden associated with modeling its effects on candidate materials. Studies in the literature, such as those in Ref.~\cite{gilbert2017automated,gilbert2019waste}, have calculated the radioactivity of pure chemical elements following neutron bombardment in DEMO-like reactors. This information could serve as an additional design criterion, as threshold values have been proposed \cite{federici2017european}. 

The paper is organized as follows. In Sec.~\ref{methods}, we discuss the design criteria, data processing of the data extracted from MPDS, and the ranking procedures used to select the PMFs.
In Sec.~\ref{results}, we show the best candidates emerging from the ranking criteria and in Sec.~\ref{literature} we present a comprehensive literature review on their applicability as PFMs. Only the materials that pass both selection procedures undergo ab-initio calculations of SBE and H-IFE, and their results are presented in Sec.~\ref{dft}. In Sec.~\ref{comprehensive} we offer a conclusive evaluation of the top PFM candidates.
Finally, Sec.~\ref{conclusions} provides a perspective on the development of the field of PFMs.

\section{Methods} \label{methods}
The set of design criteria needed for finding an exhaustive list of PFMs candidates should include the effects produced by the three loads on the exposed material: heat, neutrons, and plasma particles. In this work, a first material screening, we consider known or easy-to-compute properties of experimentally observed crystal, therefore excluding the analysis of complex dynamics emerging at the atomic-scale level from degradation processes. The design criteria include the constraint dictated by thermal loads (Sec.~\ref{thermal}), and the plasma-wall interactions (Sec.~\ref{workflows}).
For the heat load, the heat balance equation of a solid material exposed to a base load of steady-state heat flux and short-transient loads (ELMs) is being investigated as the primary design criterion. In Sec.~\ref{thickness}, the maximum thickness that a given material possesses to avoid surface's melting is used as first ranking procedure of the material retrieved from MPDS. If this cannot be estimated due to a lack of available properties, the concepts of Pareto fronts and win-fraction are defined in Sec.~\ref{Pareto} to determine how materials performs in terms of thermal properties. To compare the consistency of the two ranking procedures the idea of a comparative ranking is presented in Sec.~\ref{comparative}.
Plasma-wall interactions are then explored by calculating the formation energy of two common PFM defects, the surface binding energy (SBE) and the hydrogen interstitial formation energy (H-IFE). The relations between these crystal defects, the sputtering phenomena, and the tritium retention process are explained in Sec.~\ref{SBE} and \ref{H-IFE}, together with the Python workflows built to automate the first-principle calculation of these properties starting from a crystal structure. 

\subsection{Thermal loads} \label{thermal}

All materials-related data of this section are retrieved from the inorganic crystal database MPDS \cite{mpds}. In MPDS multiple values can be associated to a single material property. Indeed, the database is composed of entries, and each entry corresponds to the value of a peer-reviewed measured (or, occasionally, calculated) property of a specified crystal. Multiple entries for the same property and material can differ because of their literature reference and/or experimental or simulated conditions, which are not always reported. For this reason, in this work, the value of a property of a material is defined through the use of the kernel density estimation method (KDE) \cite{terrell1992variable}. This choice is based on the empirical observation that most of the reported values of a given property are around standard conditions ($T$ = 298 \si{\kelvin}, $P$ = 1 atm). The KDE creates a probability density function estimation, $f$, from the finite data sample of the available entries, $x_i$, that is:
\begin{equation}\label{eq3}
f(h,x)=\sum_{i=1}^n K\left(x-x_i;h\right)
\end{equation}
where $h$ is the bandwidth parameters setting the level of smoothness of $f(h,x)$ ($h$ = 0.05 in this work) and $K$ is a kernel function (here a Gaussian $K = exp [-9(x-x_i)^2/(2h^2)]$). The value of a property of a crystal used in this study is thus obtained as the value that maximises $f(h,x)$, unless otherwise specified.
We note that some entries are calculated under high-pressure conditions; sometimes, MPDS identifies such entries with the "hp" flag in the crystal structure chemical formula. All entries with the "hp" flag are excluded from the screening process. 
In addition, 17 more materials, selected based on ranking criteria, are excluded from the dataset because their reported properties are measured under high-pressure conditions, as indicated in the literature references provided within the database.

The underlying assumption of the design criteria defined in this section is that a candidate material cannot melt during the steady-state (SS) tokamak operating conditions and ELM events, in order to keep its shielding functions and to avoid plasma contamination.
In SS conditions the heat flux $\Phi_q$ to the PFM equals the heat flux that is removed on the cooling side, by assuming one dimensional thermal conduction through a material of thickness $d$ \cite{behrisch1993heat}:

\begin{equation}\label{eq1}
    \Phi_q^{SS}=\frac{1}{d} \int_{T_{bott}}^{T^{SS}_{surf}} k(T) \mathrm{d}T
\end{equation}
where $T^{SS}_{surf}$ and $T_{bott}$ are the temperature on the plasma-facing and the cooling sides. Assuming constant thermal conductivity $k(T)$ = $\langle k(T)\rangle$, Eq.~(\ref{eq1}) leads to:

\begin{equation}\label{eq1.1}
    T^{SS}_{surf}=T_{bott}+\Phi_q^{SS}\frac{d}{\langle k(T)\rangle}
\end{equation}
In contrast, during short thermal transients, such as ELMs, the heat is deposited on the surface as a power pulse. The temperature at the end of the ELM event ($T^{ELMs}_{surf}$) can be estimated as:

\begin{equation}\label{eq2}
    T^{ELMs}_{surf}=T^{SS}_{surf} + \Phi_q^{ELMs} \sqrt{t} \frac{2}{\sqrt{\pi \rho \langle C_p(T) \rangle \langle k(T) \rangle}} 
\end{equation}
if the penetration depth $d^{\ast}$ of the heat in the material surface is smaller than the material thickness $d$ \cite{behrisch1980evaporation}:

\begin{equation}\label{eq2.1}
    d \geq d^{\ast} = 2 \sqrt{t} \sqrt{\frac{\langle k(T)\rangle}{\rho\langle C_p(T)\rangle}}
\end{equation}
where $t$ is the duration of an ELM event, $\Phi_q^{ELMs}$ is the power flux density during ELMs, $\langle C_p(T)\rangle$ is average heat capacity of the material and $\rho$ its density. Eq.~(\ref{eq2}) can be used to define a filter to assess  the potential of a material to withstand the divertor thermal loads without melting ($T_m > T^{ELMs}_{surf}$), having excluded off-normal events. The material-dependent properties appearing in Eq.~(\ref{eq2}) are retrieved from the MPDS database.
The thickness $d$ is a target parameter, that is partially constrained by the erosion rate at the divertor, its assumed lifetime (e.g.~twice as much CFC is needed compared to \ce{W} metal \cite{federici2001assessment} for a comparable lifetime of the armor material), and the material response to thermal stress.

Two ranking procedures are then pursued to select the most promising PFM candidates. The first is the direct application of the heat balance equations and is called thickness ranking; the second builds on the various material-related properties extrapolated from the same equations and selects materials based on a Pareto optimization procedure refined by a win-fraction ranking equation, and is called Pareto ranking.
A third ranking procedure is applied with the main goal of comparing the accuracy of the two ranking methods and establish a unique ranking for all the materials retrieved from MPDS, which we denote as comparative ranking.

\subsubsection{Thickness ranking}\label{thickness}
The first ranking method comes directly from the application of the heat balance equation in Sec.~\ref{thermal}. The list of materials for which is possible to collect all the variables in Eqs.~(\ref{eq2}) and (\ref{eq2.1}) are ranked accordingly to the maximum thickness allowed to avoid melting during type-I ELMs ($d^{ELMs}$).
If  $C_p(T)$ 
is not available, then the maximum thickness allowed to avoid melting during steady-state ($d^{SS}$) is used as ranking criterion.
The input values assumed to compute the material thickness are reported in Tab.~\ref{tab_par}.

\begin{table}
\caption{Input parameters used to compute the heat balance equations: Eq.~(\ref{eq1.1}), (\ref{eq2}) and (\ref{eq2.1}).}
\label{tab_par}
\centering
\begin{tabular}{llcc}
\hline
\hline
\textbf{Parameter}                            & \textbf{Unit}                     & \textbf{Value}            & \textbf{Ref.}            \\
\hline
$T_{bott}$                                      & \si{\kelvin}                      & 415                       & \cite{raffray1999critical}                         \\
$Q^{SS}$                                        & \si{\mega\watt\per\square\meter}  & 20                        & \cite{federici2001plasma,de2018influence}                          \\
$t$                                             & \si{\milli\second}                & 0.5                       & \cite{evans2013elm,de2018influence}                          \\
$Q^{ELMs}$                     & \si{\giga\watt\per\square\meter}                   & 1                        & \cite{evans2013elm,de2018influence}      \\
\hline
\hline
\end{tabular}
\end{table}
Given the large uncertainty around the real value of each property extrapolated from MPDS, the thicknesses calculated from Eqs.~(\ref{eq1.1}) and (\ref{eq2}) are reported with a standard deviation (SD) that is obtained by the propagation of the standard deviation of every single property.
Materials with $d^{ELMs} >$ 0 \si{\milli\meter} and $d^{ELMs} - d^{\ast} >$  0 \si{\milli\meter} and materials with $d^{SS} \geq 3$ \si{\milli\meter} are considered PFM candidates; materials not passing this test are discarded. The threshold value of 3 \si{\milli\meter} for steady-state conditions is chosen since no material with a positive $d^{ELMs}$ has a lower $d^{SS}$ thickness.

It is important to remember that these values are calculated using three assumptions: (i) the value at the maximum of the KDE probability density function captures well the real value of the material property, (ii) these properties remain constant over large temperature ranges, (iii) degradation processes and materials defects are neglected. Therefore, the calculated $d^{ELMs}$ and $d^{SS}$ remain indicative of a trend and should not be  considered for the engineering design.

\subsubsection{Pareto ranking} \label{Pareto}
For materials where neither $d^{ELMs}$ nor $d^{SS}$ can be calculated a ranking process that does not require the knowledge of an objective function is used. This is performed in order to extract materials from MPDS that are showing promising performance in a subset of the properties identified through thermal balance analysis. In order to expand the number of available PFM candidates, the properties retrievable from MPDS that showed a high Pearson correlation coefficient ($\geq$ 0.7) with $T_m$, $k$, $C_{p}$ are added to the list of properties. Of all the properties available on MPDS, $T_m$ shows a Pearson correlation coefficient greater or equal than 0.7 along with 6 other properties: shear modulus ($G$), Young's modulus ($E$), adiabatic and isothermal bulk modulus ($B_S$ and $B_T$), Mohs hardness ($MH$), Knoop hardness ($KH$); $C_{p}$ with the heat capacity at constant volume ($C_{v}$). 
At this stage, the total number of properties is 10 and they are categorized as follows: (i) melting temperature + proxies, (ii) thermal conductivity, (iii) heat capacity (at constant pressure and constant volume). In the following, for the sake of clarity, these 3 macro-categories are identified with the term categories ($c$), whereas the 10 features are called interchangeably features or properties ($p$) (see Tab.~\ref{tab_category}).
\begin{table}
\caption{The 10 properties ($p$) retrieved from MPDS are classified into 3 categories ($c$). This classification is applied inside the Pareto and comparison ranking systems.}
\label{tab_category}
\centering
\begin{tabular}{lc}

\hline
\hline
\textbf{Category ($c$)}                            & \textbf{Property ($p$)}                   \\
\hline
\multirow{7}{*}{Melting temperature + proxies}           & $T_m$                           \\
                                                         & $G$                                  \\
                                                         & $E$                                  \\
                                                         & $B_S$                                \\
                                                         & $B_T$                                \\
                                                         & $MH$                                 \\
                                                         & $KH$                                 \\
\hline
Thermal conductivity                                     & $k$  \\
\hline
\multirow{2}{*}{Heat capacity}                           & $C_p$                                \\
                                                         & $C_v$                                \\
\hline
\hline
\end{tabular}
\end{table}
Note that the density is always available for every crystal structure reported in MPDS, therefore instead of using the heat capacity in molar units, it is converted in unit of volume as Eq.~(\ref{eq2}) suggests.

A rigorous way to treat an optimization problem for which the objective function is unknown is through the so-called Pareto set (or Pareto front), a collection of solutions (i.e.~materials) that cannot be improved in one objective (i.e.~property) without worsening at least one of the others \cite{lejaeghere2013ranking}. 
Given a set of materials $m$ of size $M$ that share the same set of properties $p$ of size $P$, the Pareto set $\mathcal{P}$ of this system is defined as follow:
\begin{gather}\label{eq4}
\begin{aligned}
&m_i \in \mathcal{P} \Leftrightarrow \nexists m_j:\left\{\begin{array}{l}
p_{j,k} \geq p_{i,k} \quad \forall k=1, \ldots, P \\
\exists k_0: p_{j,k_0}>p_{i,k_0}
\end{array}\right. \\
&\forall i,j=1, \ldots, M
\end{aligned}
\end{gather}
First, upstream of the Pareto optimization process, the data are standardized by subtracting its mean, so that the mean of the standardized values vanishes, and divided by the standard deviation, ensuring that the standard deviation of the standardized values is one.
Then, different sets of materials that posses the same set of properties to apply the Pareto optimization are defined. Since the whole set of 10 $p$ is not available for all the materials, it is crucial to define their subsets for which a Pareto set optimization is pursued.
Here, we apply Pareto for every material possessing at least 1 property out of the 10 previously defined. Therefore, three subsets of materials are defined as follows: $3p$, which includes materials possessing at least one property for each of the three category; $2p$, which encompasses materials having at least one property in, at least, two different category; and $1p$, which includes materials with only one property per category. Then, for every subset of $3p$/$2p$/$1p$, all possible combinations of $n$ properties that belong to $n$ different categories are used to define the feature space. For instance, in the $3p$ subset an allowed combination of features can be either $T_m$-$C_p$-$k$ or $T_m$-$C_v$-$k$ but not $T_m$-$B_S$-$k$ or $T_m$-$C_v$-$C_p$ since $T_m$-$B_S$ and $C_v$-$C_p$ are part of the same category (respectively melting temperature + proxies and heat capacity).

If the number of features is exhaustive for the description of a given problem, any solution that is not a Pareto front can be discarded from the list of interesting candidates, because it is worse in terms of performance. In the context presented here, the list of properties used as design criteria for an armor divertor material is far from being comprehensive; therefore, multiple Pareto set are considered as optimal solutions. Once a Pareto set is found, it is discarded from the list of candidates and those are re-evaluated. Each Pareto set can be conceptualized as a Pareto layer ($l$), which, once identified, is peeled off, and the subsequent layer is selected. This process is repeated for a specified number of cycles.

To distinguish between any of the candidates thus selected, the concept of win fraction, as proposed in Ref.~\cite{lejaeghere2013ranking}, is revisited and applied. 
\begin{equation}\label{eq5}
WF_{i}=\frac{\sum_{j}\sum_{k}\left|p_{i,k}-p_{j,k}\right| \times \mathcal{H}\left(p_{i,k}-p_{j,k}\right)}{\sum_{j}\sum_{k}\left|p_{i,k}-p_{j,k}\right|}
\end{equation}
where $WF_{i}$ is the win fraction of material $i$, $p_{i,k}$ is the value of the property $k$ of material $i$ and $\mathcal{H}(x)$ is the Heaviside step function that is equal to 0 for $x < 0$ and to 1 for $x > 0$.
Contrary to the original reference, the ranking equation is not defined as the smallest win fraction that each materials $i$ posses when compared with the other candidates $j$; instead, it corresponds to a summation extended over all candidates of the selected Pareto fronts. In addition, it is implied that an unitary weight is assigned to every property of the summation.
Once the ranking is pursued, the best candidates are selected. 
The number of Pareto sets $l$ and the threshold value of $WF_i$ used to identify the best candidates for each category of Pareto solutions is reported in Tab.~\ref{tab_pareto}.
\begin{table}
\caption{A Pareto ranking is utilized to identify the most promising candidates that do not meet the criteria for thickness ranking due to insufficient retrievable properties. The materials were categorized into three subsets based on the number of available properties. For each subset, the first $l$ Pareto fronts were selected, and only those with a $WF_i$ \ref{eq5} value exceeding a specified threshold were considered as PFM candidates.}
\label{tab_pareto}
\centering
\begin{tabular}{ccc}

\hline
\hline
\textbf{Material class}                            & \textbf{$\boldsymbol{l}$}      & \textbf{$\boldsymbol{WF_i}$}             \\
\hline
3p                                     & $\leq$ 5            & $\geq$ 50\%              \\
2p                                     & $\leq$ 5            & $\geq$ 70\%              \\
1p                                     & $\leq$ 100          & $\geq$ 90\%              \\
\hline
\hline
\end{tabular}
\end{table}

\begin{figure*}
  \includegraphics[width=\textwidth]{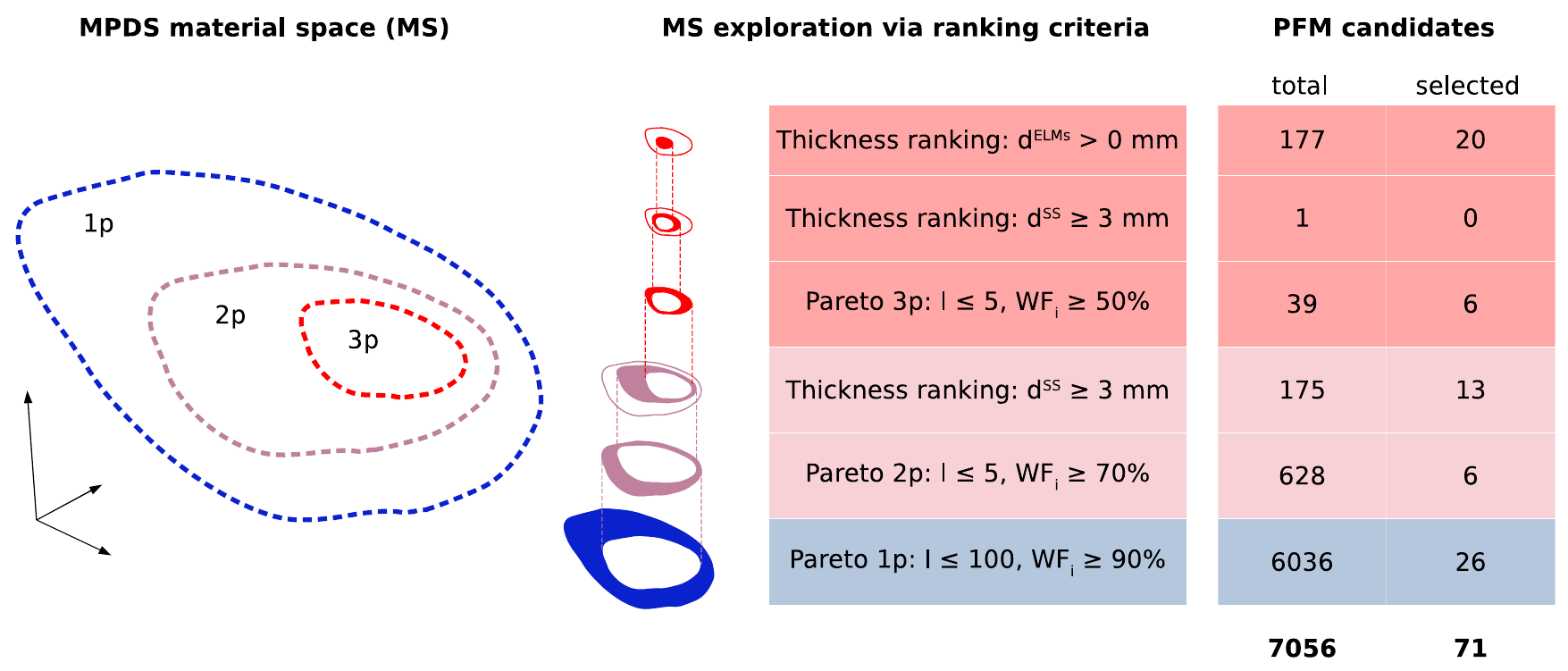}
  \caption{Schematic representation of the material space in MPDS and its exploration based on the number of available properties and the corresponding ranking procedure, allowing for the identification of promising candidates through the application of multiple ranking criteria. 
  }
  \label{fig:mat_space}
\end{figure*}

In order to avoid false positive, materials discarded from more stringent criteria of selection should not be captured by weaker filters. Hence, materials that are excluded by the thickness ranking scheme or by a Pareto optimization with a higher number of properties are removed from the list of materials subject to a less stringent procedure. Fig.~\ref{fig:mat_space} introduces a visual representation that illustrates the materials space and the process of exploring it.

\subsubsection{Comparative ranking} \label{comparative}
The third ranking procedure is then created as a visualization tool to collect all available materials in a single table, to rank them within a unique framework, and as a benchmark to compare the results of the previous two ranking scheme. For this procedure, the main issue is connected to the fact that materials with different available properties must produce a comparable ranking value. The win fraction concept might be re-used to define a ranking value for every available material, after applying some changes to the formulation presented in the previous section.
Eq.~(\ref{eq5}) is intended to be applied to the set of properties that a material possesses, and the absence of certain features does not impact the final value of $WF_{i}$. If $WF_{i}$ is used directly, materials with one single outperforming feature might rank at the top of the list, obscuring the rest of the candidates. 
On the other hand, if the win fraction is utilized for every material across the 10 retrieved properties, even in the presence of missing data, it becomes essential to specify how the missing values affect the win fraction. This depends on two factors: the weight assigned to each property and the value to be assigned to the missing ones.
A unitary weight assigned to all the properties would reward materials with a lot of available data at the expense of others that have a smaller set of essential and high-performing features.
A good compromise was found by subdividing the properties in three different categories ($c$), as proposed in Tab.~\ref{tab_category}. Then, the weights are built by multiplying two coefficients:
\begin{itemize}
    \item The first term is a constant value that applies to all materials and is defined as 1/$C$, where $C$ is the number of categories. In this work, we always consider three categories, so the first term is equal to 1/3.
    \item The second term is material-dependent and it is defined as 1/$P_{i,z}$, where $P_{i,z}$ is the number of available properties of the $z$-th category of the $i$-th material.
\end{itemize}
In this way, materials lacking in the key categories - melting temperature + proxies, thermal conductivity, and heat capacity - are penalized, but the number of properties retrievable within a category does not affect the weight.
Then, the weight of a given property is multiplied by the corresponding single-property win fraction $WF_{i,k}$, and the summation over the all features builds the weighted average win function $\overline{WF_i}$ as presented in the following equation:

\begin{gather}\label{eq6}
\begin{aligned}
\overline{WF_i} &= \sum_{z,k}w_{i,z} WF_{i,k} \\
&=\sum_{z=1}^{3}\frac{1}{3}\sum_{k=1}^{P_{i,z}}\frac{1}{P_{i,z}}WF_{i,k} \\
\end{aligned}
\end{gather}
where $k$ represent the $k$-th property of the $z$-th category of the $i$-th material. If a category is not present for a given material, fictitious $WF_{i,k}$ = 0 and $P_{i,z}$ = 1 are assigned.
This methodology can be generalized for any number of categories and any of their subsets. For instance, the melting temperature and proxies category might be split between actual $T_m$ and proxies, generating a subset with two sub-categories. The results remain very similar whether or not this additional subcategory is included. Therefore, for the sake of simplicity, this approach is not discussed further.

\subsection{Plasma-wall interactions}\label{workflows}
The interactions between hydrogen isotopes, alpha particles and the PFM have a complex dynamics, which requires large-scale (10$^{5}$-10$^{6}$ atoms) and long-timescale (10$^{2}$-10$^{3}$ \si{\nano\second}) molecular dynamics (MD) simulations \cite{hammond2018large,hatton2022importance}. For instance, the entire formation process of a \ce{W} fuzz layer under plasma bombardment is out of the timescale explorable even with classical MD \cite{dasgupta2019origin}. With the goal of finding indicators that reflect some of the undesired phenomena PFMs may experience when interacting with a plasma particle flow, while being computationally efficient, we use two key defects, as proxies, surface binding energies (SBE) and hydrogen interstitial formation energies (H-IFE), whose formation energies can be calculated within the DFT framework.

\subsubsection{Surface binding energies} \label{SBE}
As mentioned above, the minimum thickness of a PFM is highly constrained by the rate of erosion of the materials surface under the plasma impingement, which in turn is measured through the sputtering flux of the surface atoms.
For physical sputtering, the energy distribution of the sputtered particle flux, $\Phi (E_e)$, can be described within the binary collision approximation (BCA) by the Thompson energy spectrum \cite{thompson1981physical}. The Thompson energy spectrum highlights an important proportionality between $\Phi (E_e)$ and the surface binding energy (SBE or $E_{sb}$) of a material:

\begin{equation}\label{eq7}
    \Phi (E_e) \propto \frac{E_e}{(E_e+E_{sb})^3}
\end{equation}
This relation has a distribution that peaks at $E_e$=$E_{sb}$/2, where $E_e$ is energy of the sputtered (emitted) particles leaving the surface. It is notable that the tail of the Thompson energy spectrum deviates from experimental values when the projectiles bombarding the surface have low-energy ($<$ 500 \si{\electronvolt}) and it is corrected by the Falcone energy spectrum \cite{goehlich2000anisotropy}. In both cases, SBEs highly affect the sputtered particle flux, therefore the erosion rate and the minimum thickness required from a given PFM. The SBE is defined as the energy that is necessary to remove an atom from the top surface layer (in vacuum) during the physical sputtering process.
We calculate SBEs from first principles, and we use these as a proxy of the material's resistance to physical sputtering. SBEs are often estimated using elemental enthalpies of sublimation in sputtering calculations. However, recent literature emphasizes that first-principles SBEs are essential to achieve more accurate outcomes \cite{yang2014atomic,morrissey2021simulating,gyoeroek2016surface}. 
In the context of density-functional theory (DFT), these are calculated through the following relation:

\begin{equation}\label{eq8}
E_{sb}=E_{a}+E_{s+v}-E_{s}
\end{equation}
where $E_{a}$ is the DFT total energy of the single isolated (sputtered) atom $a$, $E_{s}$ is the total energy of the crystal slab with a perfect surface, and $E_{s+v}$ is the energy of the unrelaxed (for efficiency) slab with one surface vacancy at the position that was atom $a$. 
$E_{sb}$ is obtained by developing the Surface Binding Energy (SBE) workflow, a semi-automated Python script created to make standardized SBE calculations (in-house, forthcoming). Once the candidate is selected and its crystal structure is defined, 6 main steps follow:
\begin{enumerate}
    \item Relaxation of atomic positions and cell shape of the primitive crystal structure as obtained by MPDS, and calculation of its total energy.
    \item From the relaxed lattice constants, all the inequivalent slabs that can be generated with Miller indices $hkl$ up to 1 are created and relaxed and their surface energy is obtained through the following relation:
    \begin{equation}\label{eq9}
    E_{surf}^{hkl,\alpha}=\frac{E_{s}^{hkl,\alpha}-\frac{N_{s}^{hkl,\alpha}}{N_{b}}E_{b}}{2A_{s}^{hkl}}
    \end{equation}

    where $E$, $N$, $A$ are the total energy, the number of atoms, and the area of the crystal slab, $s$ and $b$ are the subscripts indicating slab and bulk structures, and $hkl$ and $\alpha$ indicate the Miller index and the termination of a specific facet of the crystal structure. A minimum distance of 10 \si{\angstrom} of vacuum between repeated slabs is imposed as discussed in Refs.~\cite{tran2016surface,singh2009surface}.
    \item The slab with the lowest $E_{surf}^{hkl,\alpha}$ is selected together with all possible terminations of the same Miller index, supercells are created and their total energy ($E_{s}$) is obtained. A minimum distance of 9 \si{\angstrom} between periodic images is imposed to reduce vacancy-vacancy interaction.
    \item One atom is removed from the surface of each supercell (per species and inequivalent positions) and the total energy ($E_{s+v}$) is calculated. One atom is considered a surface atom if it lies within 1 \si{\angstrom} from one of the two surfaces. Atomic relaxation is neglected within this approximation.
    \item The total energy of a single atom ($E_{a}$) in a cubic cell of side $\sqrt[3]{4000}$ \si{\angstrom} is calculated for each element of the crystal.
    \item $E_{sb}$ is calculated following Eq.~(\ref{eq8}); if inequivalent sites of the same chemical element are found on the surface of given terminations $\alpha$ and/or multiple termination $\alpha$ are available, $E_{sb}$ is calculated for each element as the arithmetic mean between all inequivalent sites of each termination.
\end{enumerate}
All DFT simulations are performed by using the open-source computer code for electronic-structure calculations Quantum ESPRESSO (QE) \cite{QE,giannozzi2009quantum} and all parameters required to run the calculation are retrieved by Quantum Espresso Input Generator website \cite{MatCloud,prandini2018precision,garrity2014pseudopotentials} set for spin-unpolarized calculations, with maximum k-points distance of 0.2 \si{\per\angstrom}, a smearing width for metals of 0.2 \si{\electronvolt} 
and using the pseudopotentials library SSSP Efficiency PBEsol (version 1.2.0), where exchange and correlation effects are approximated using the Perdew-Burke-Ernzerhof revised for solids (PBEsol) functional. The kinetic energy cutoff for both charge density and wavefunctions is consistent across all calculations for a specific crystal and is set as the highest value suggested by the Quantum Espresso Input Generator among the elements present in the crystal.

Once $E_{sb}$ is known, the energy that a projectile must posses to sputter one atom from a perfect-crystal surface in a head-on elastic collision has to be greater or equal to the kinetic energy transferred to the recoil atom $E_T$:
\begin{equation}\label{eq10}
E_p \geq E_T = E_{sb} \frac{\left(m_{p}+m_{a}\right)^{2}}{4 m_{p} m_{a}}.
\end{equation}
where in the case of equality $E_{p}$ is the minimum energy that the projectile (a plasma atom) of mass $m_{p}$ must posses to produced physical sputtering on a target surface atom of mass $m_{a}$ that is bonded to the surface with energy $E_{sb}$.

\subsubsection{Formation energy for interstitial hydrogen and its isotopes}  \label{H-IFE}
A similar design is used to provide an estimate of the interaction between a hydrogen isotope (hydrogen, deuterium or tritium), denoted as $H$ for the sake of simplicity, and the host crystal structure ($M$). The solubility of hydrogen isotopes in a perfect crystal structure besides its porosity, is of particular relevance inside the reactor to estimate hazardous tritium retention. The formation energy of an interstitial hydrogen isotope $\Delta E^{f}(H_i)$ in a perfect crystal at 0 \si{\kelvin} can be calculated as follows:

\begin{equation*}
  M_x + H \rightarrow M_xH_i
\end{equation*}
and
\begin{equation}\label{eq11.0}
\begin{aligned}
    \Delta E^{f}(H_i) &=  E^{DFT}(M_xH_i) + E^{ZP}(M_xH_i)  \\
    &- E^{DFT}(M_x) - E^{ZP}(M_x) \\
    &- E^{DFT}(H) - E^{ZP}(H)
\end{aligned}
\end{equation}
where $E^{DFT}$ is the ground-state total energy calculated with a DFT code, $E^{ZP}$ is the zero-point energy (ZPE).
This expression can be approximated by considering $E^{ZP}(M_xH_i) \approx E^{ZP}(M_x) + E^{ZP}(H_i)$. Therefore, Eq.~(\ref{eq11.0}) can be re-written as reported in Ref.~\cite{van2000hydrogen}:

\begin{equation}\label{eq11}
\begin{aligned}
    \Delta E^{f}(H_i)
    &\approx E^{DFT}(M_xH_i) +  E^{ZP}(H_i) \\
    &- E^{DFT}(M_x) - E^{DFT}(H) - E^{ZP}(H)
\end{aligned}
\end{equation}
As demonstrated in Refs.~\cite{reuter2003composition,shao2017solubility,kirchheim2014hydrogen}, the solubility of a gas within a crystal under low pressure conditions ($<$ 100 atm) can be derived for a system in thermodynamic equilibrium between an ideal gas and the bulk of the perfect crystal. It is expressed as a function $\Delta E^{f}(H_i)$, along with additional entropic corrections ($s_{corr}$). Specifically:

\begin{equation}\label{eq12}
\begin{aligned}
    \frac{N_{H_i}}{N_i} \approx exp -\left( \frac{\Delta E^{f}(H_i) - s_{corr}}{K_bT} \right) \\
    s_{corr} = Ts^{vib}(H_i) + \Delta\mu (H)(T,p_0) + K_bT\ln\frac{p_H}{p_0}
\end{aligned}
\end{equation}

where $N_{H_i}$ and $N_i$ are the number H atoms and interstitial sites respectively, $Ts^{vib}(H_i)$ is the vibrational entropy of a single $H$ atom in an interstitial site, $\Delta\mu (H)(T,p_0)$ and the last factor account for the entropic contribution of $H$ at temperature $T$ and pressure $p_H$.
Eq.~(\ref{eq12}) is derived by the separating vibrational and conformational entropy of the crystal with an interstitial hydrogen isotope ($WH_i$), as the sum of the entropies of the perfect crystal ($W$) and the hydrogen interstitial ($H_i$), while assuming $N_i \gg N_{H_i}$.
To maintain computational affordability for screening purposes, we calculate $\Delta E^{f}(H_i)$ neglecting zero-point energy (ZPE) and entropic contributions, as outlined in Eq.~(\ref{eq13}). Consequently, under this formulation, the computed interstitial formation energy remains equivalent for any hydrogen isotope.
However, it is worth noting that within the context of PFMs, a more accurate approach would require to redefine the hydrogen reservoir as the hydrogen stored in the crystal surface at equilibrium, resulting from the impingement of high-energy plasma particles.

A second semi-automatized Python workflow, called the Hydrogen Interstitial Formation Energy (H-IFE) workflow, is built to calculate $\Delta E^{f}(H_i)$.
The main challenge of this calculation lies on the identification of the interstitial hydrogen ($H$) sites with the lowest formation energy. Below, are the steps used to accomplish this task:
\begin{enumerate}
    \item A supercell of the primitive crystal structure of a given material is created with a minimum size of 9$\times$9$\times9$ \si{\angstrom}$^3$. The supercell is relaxed and its $E^{DFT}(M_x)$ is calculated.
    \item $N$ supercells of the same size of step 1 but containing one interstitial H atom are generated.
    The interstitial sites are randomly chosen from within the primitive cell of the supercell, with the only constraint being that the interstitial site of the $n+1$ supercell is at a minimum distance of 1 \si{\angstrom} from all crystal atoms and the interstitial sites of all previous supercell. The number of interstitial sites, $N$, is determined based on the size of the primitive cell, and the random search for interstitial sites is terminated if more than 5,000 attempts are required. This approach ensures that the primitive cell of each crystal structure is uniformly explored in terms of interstitial site density. The interstitial supercell are then relaxed and their total energy is calculated.
    \item $\Delta E^{f}(H_i)$ is calculated from the total energy of the interstitial supercell with lowest $E^{DFT}(M_xH_i)$, using the following equation:
    \begin{equation}\label{eq13}
    \begin{aligned}
    \Delta E^{f}(H_i) &= E^{DFT}(M_xH_i) \\
    &- E^{DFT}(M_x) - E^{DFT}(H)
    \end{aligned}
    \end{equation}
\end{enumerate}
The same DFT code and set up of the input parameters used for the SBE workflow are used for the H-IFE. For comparison with literature reference $\Delta E^{f}(H_i)$ is also computed for a hydrogen molecule ($H_2$) reservoir ($\frac{1}{2}E^{DFT}(H_2)$).

\section{Ranking results} \label{results}

The MPDS database contains more than 500,000 crystal structures. When only the materials possessing at least one out of the ten properties previously selected are retained and radioactive and high pressure crystals are discarded, the total number of available candidates is reduced to 7,056.

The $d^{ELMs}$ thickness ranking procedure is applied to the 177 crystal structures having the required properties. These materials are ranked as a function of their maximum allowable thickness to avoid melting during type-I ELMs, and it is found that only 20 materials meet the set prerequisites. Out of the 6879 remaining materials, 176 had the properties to be ranked on $d^{SS}$. Of these, 13 candidates are retained due to their ability to withstand steady-state conditions without melting. The remaining available structures are, instead, classified using the Pareto front procedure. 
From the latter, we identify 38 best-performing candidates, bringing the total number of selected materials from MPDS to 71.
Fig.~\ref{fig:mat_space} illustrates a schematic representation of the material space in MPDS and its exploration, taking into account the number of available properties and the corresponding ranking procedure. All 7,056 candidates are then ranked according to the comparative ranking scheme, to show consistency across different ranking procedures and identify any candidates that exhibit discrepancies among the various rankings.

\begin{figure*}
  \includegraphics[width=\textwidth]{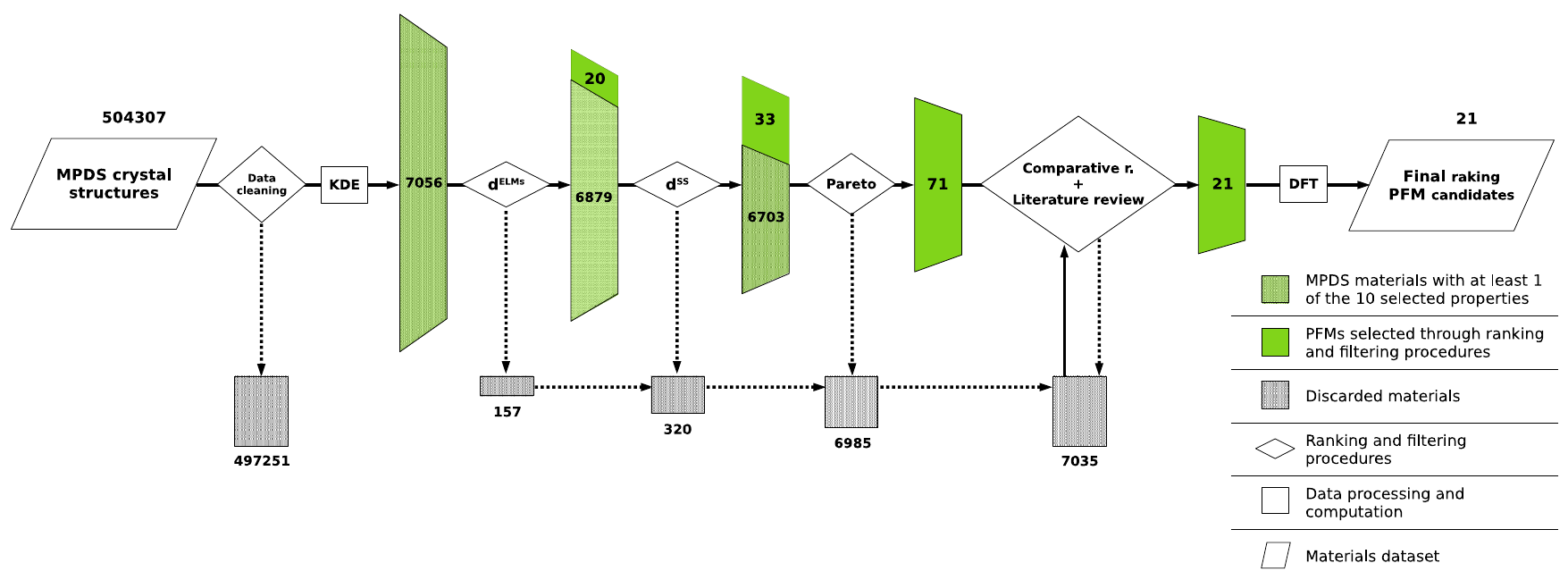}
  \caption{Flowchart illustrating the PFM selection process, which involves all the steps described in Section \ref{methods} and the literature review. It starts from the MPDS materials and narrows down to a set of selected PFM candidates. The material selection is visually represented using a funnel chart.}
  \label{fig:funnel}
\end{figure*}

We then assess the potential of the top candidates thanks to the analysis of literature on divertor materials.
This serves as an additional benchmark to validate our findings and identify the materials that have already been explored and the corresponding advantages and disadvantages of their application. New $d^{SS}$ and $d^{ELMs}$ values are calculated for the most promising PFM candidates identified through the literature review. These calculations are based on the literature values of the properties required from heat balance equation (Eq.~\ref{eq2}). Then, SBE and H-IFE workflows are applied to highlight possible drawbacks of the selected structures. Ultimately, a comprehensive ranking is provided to identify the most promising PFMs.
Fig.~\ref{fig:funnel} depicts the flowchart encompassing the entire process, beginning from the retrieval of MPDS data to the generation of a list comprising the top candidates for PFM.
It should be noted that the chemical formulas of the materials presented in the tables and figures of this chapter correspond exactly to those documented in the MPDS database.

\subsection{Thickness ranking during ELMs} \label{trELMs}
In Fig.~\ref{fig:d_rank} and Tab.~\ref{tab:d_rank} in the Appendix, we present 20 PFM candidates able to sustain the divertor conditions during type-I ELMs without melting and satisfying Eq.~(\ref{eq2.1}).
These selected materials can be summarized into 3 main categories.
\begin{enumerate}
    \item Carbon-based materials: diamond (\ce{C dia}).
    \item Pure transition metals: \ce{W}, \ce{Mo}, \ce{Ir}, \ce{Cu}, \ce{Os}, \ce{Ru}, \ce{Rh}, \ce{Re}, \ce{Ta}, \ce{Cr}.
    \item Ceramics: (carbides) \ce{TaC}, \ce{HfC}; (nitrides) \ce{AlN}, \ce{GaN}, \ce{TiN}; (borides) \ce{HfB2}, \ce{TiB2}, \ce{ZrB2}.
\end{enumerate}
This list includes the most extensively studied PFMs proposed as divertor materials in numerous experimental projects.
In addition, other materials, such as \ce{TiB2} and \ce{TiN}, which have been employed as PFMs in tokamak limiters \cite{federici2001plasma} subject to comparable thermal loads as divertor materials, appear in the selection.

It might result surprising that a few of the most well-known PFMs are not included in this initial selection.
For example, a few complex alloys, such as stainless steel and Inconel \cite{federici2001plasma}, that have been utilized in past as PFMs do not appear just because they are not included in the MPDS database. However, these alloys do not withstand the high heat loads present in the divertor region. On the other hand, while \ce{Ti} and \ce{Be} are listed in the MPDS, not all the required properties are available to calculate $d^{ELMs}$. However, \ce{Be} exhibits a thermal conductivity that falls just short of satisfying divertor conditions \cite{wilson1990beryllium}, while the thermal conductivity of pure \ce{Ti} is significantly inadequate for the intended purpose \cite{haynes2016crc}.
Another group of materials previously employed as PFMs but excluded through thickness ranking are \ce{TiC}, pure \ce{B}, \ce{Al}, and boron carbide (\ce{B4C}) \cite{federici2001plasma}, as they are designed to withstand lower temperature regimes. Graphite and \ce{SiC}-3C are excluded due to the underestimate of their actual thermal property values in the MPDS database.

\begin{figure*}
  \includegraphics[width=\textwidth]{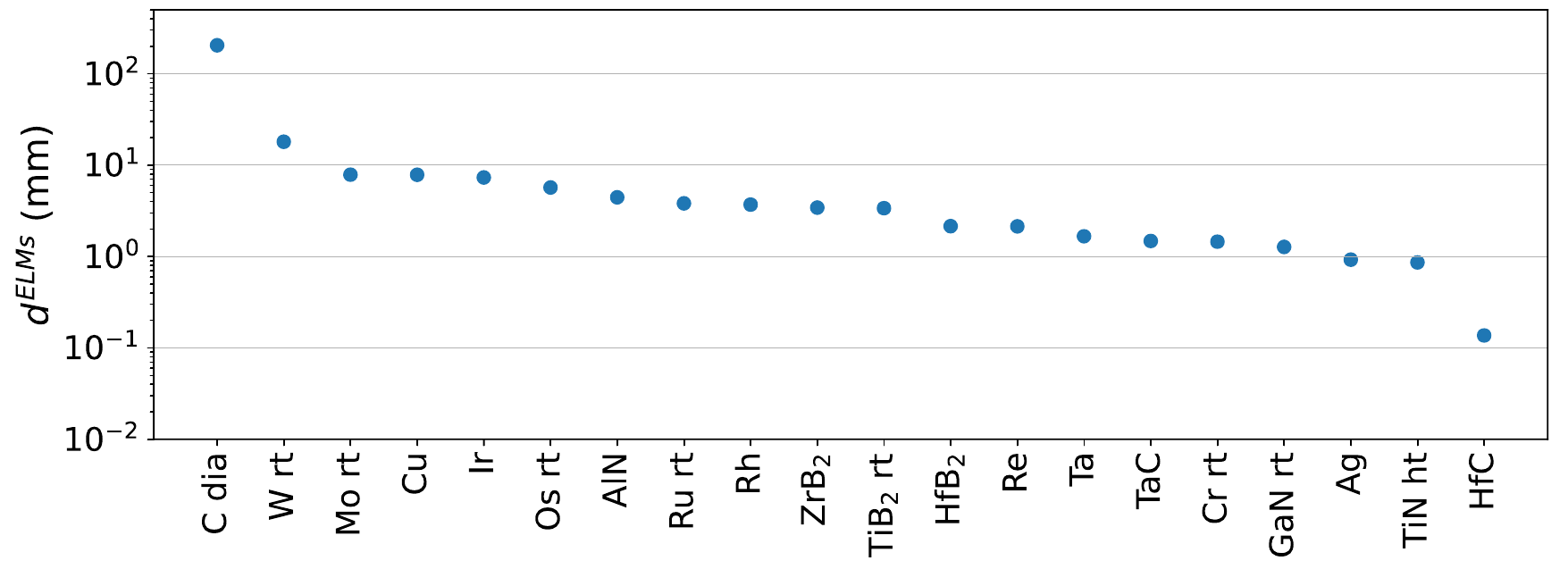}
  \caption{$d^{ELMs}$ logarithmic scale for the 20 selected PFM candidates extracted from MPDS. $d^{ELMs}$ defines the maximum thickness for a material to withstand type-I ELMs without melting. Only materials with $d^{ELMs} > 0$ \si{\milli\meter} are selected.}
  \label{fig:d_rank}
\end{figure*}

\subsection{Thickness ranking in steady-state conditions} \label{trSS}

\begin{figure*}
\centering  
  \includegraphics[width=\textwidth]{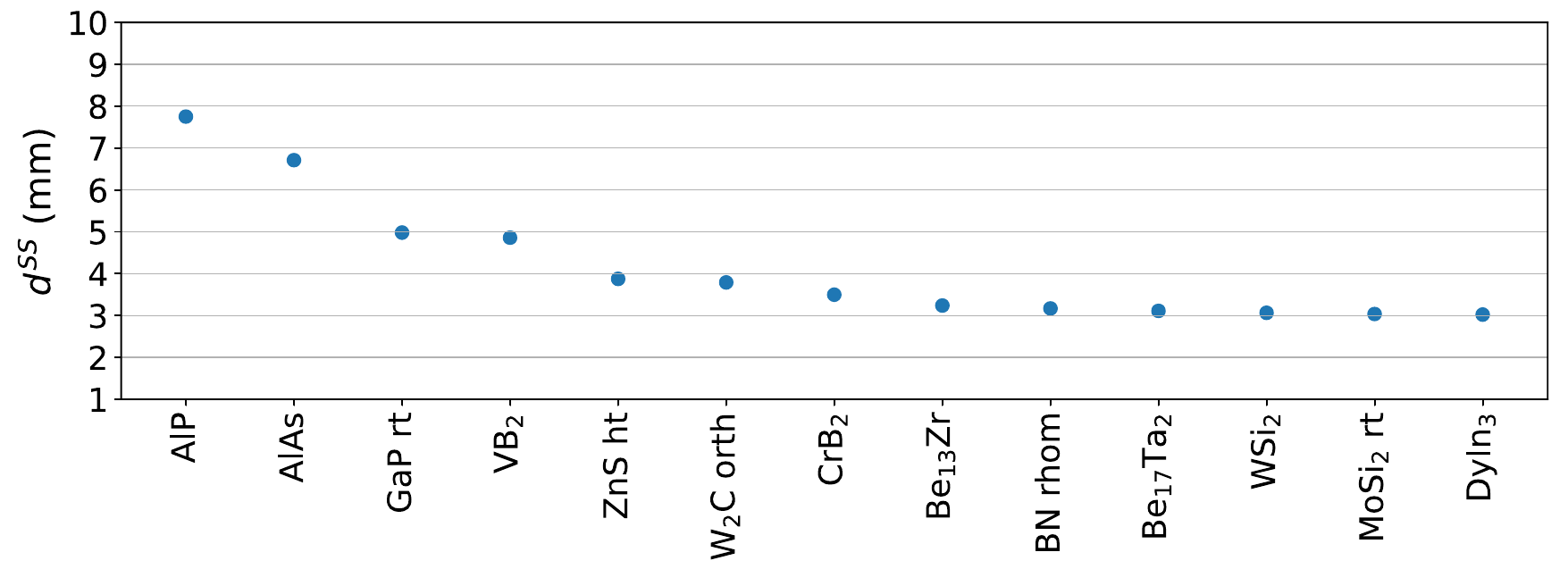}
  \caption{$d^{SS}$ defines the maximum thickness for a material to withstand steady-state condition without melting. The 13 materials showing $d^{SS} \geq$ 3 \si{\milli\meter} are selected from MPDS once materials ranked through $d^{ELMs}$ criteria are excluded.}
  \label{fig:d_rank2}
\end{figure*}

Once the materials ranked with $d^{SS}$ are excluded, we identify 13 materials with $d^{SS} \geq$ 3 \si{\milli\meter}; these are reported in Fig.~\ref{fig:d_rank2} (and Tab.~\ref{tab:d_rank2}).
A comparison of their characteristics with those of the $d^{ELMs}$ ranking candidates reveals that most of these materials rank at the bottom of the list in terms of performance, and they could potentially be excluded altogether unless unreasonably high heat capacities are considered. Additional literature investigations substantiate this hypothesis, except for three materials, \ce{W2C} and \ce{BN}, which possess adequate melting temperatures, and \ce{VB2}, which exhibits sufficiently high thermal conductivity, warranting further scrutiny.

\subsection{Pareto ranking}\label{PR}

Pareto ranking is performed with materials whose properties cannot be assessed by the two thickness rankings. Of the 6,703 crystal structures analyzed, 6 are in the $3p$ category (Tab.~\ref{tab2}), 6 in the $2p$ category (Tab.~\ref{tab3}), and 26 in the $1p$ category, including 5 polymorphs of \ce{BC2N} and \ce{C3N4} which are excluded from further analysis because they exhibit highly similar thermal properties (see Table \ref{tab4}). We increase the portfolio of properties for the selected materials with values found in the literature. Comparing these properties with those of materials selected through $d^{ELMs}$ and $d^{SS}$ rankings, we find that the number of remaining candidates qualifying as PFM is reduced to the following:
\begin{enumerate}
    \item Metal alloys: \ce{Ta_{0.5}W_{0.5}}.
    \item Ceramics: \ce{BC2N}, \ce{C3N4}, \ce{HfN}, \ce{SiC}-2H, \ce{Ta2C}, \ce{TaN}, \ce{VC}, \ce{VN}, \ce{WC}, \ce{W2C}.
    \item Carbon-based materials: graphite intercalated with \ce{CoCl2}, \ce{FeCl3}, \ce{SbCl5}.
    \item Boron-based materials: \ce{Mg2B25C4}, \ce{YB66}.
    \item Perovskite nitrides: \ce{TiFe3N}, \ce{VFe3N}.
\end{enumerate}
Among these materials, it is worth noting that the x-\ce{Fe3N} perovskite nitrides (\ce{TiFe3N} and \ce{VFe3N}) exhibit a high value for the isothermal bulk modulus. However, the literature lacks a comprehensive analysis of their thermo-mechanical properties, which is crucial for determining their applicability as PFMs. For these reasons, they are not further considered as potential PFM candidates.

\subsection{Comparative ranking}\label{CR}

\begin{figure*}
  \includegraphics[width=\textwidth,height=0.9\textheight,keepaspectratio]{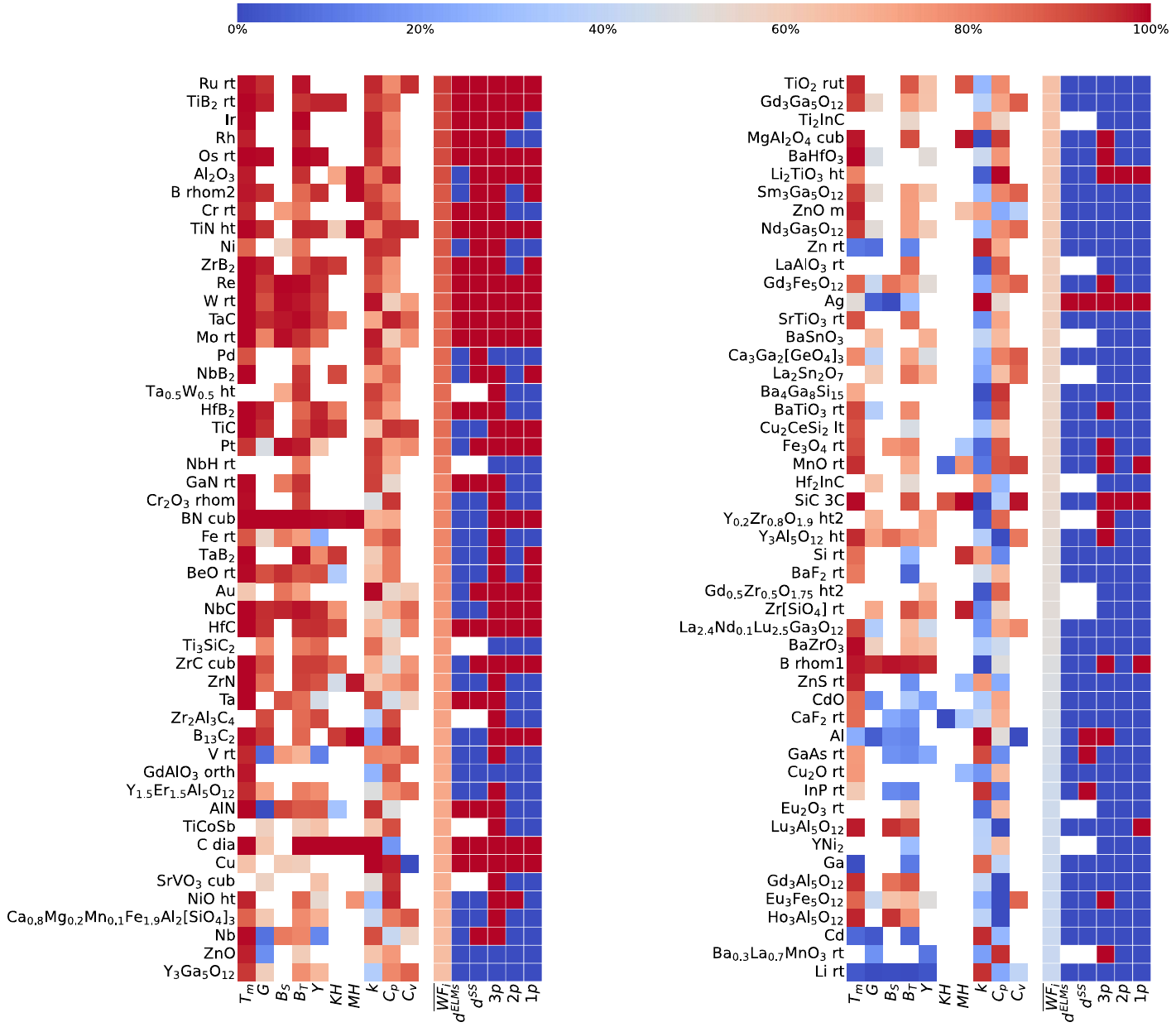}
  \caption{Heatmap representation of the materials ranked from 1 (top left) to 100 (bottom right) according to the comparative ranking procedure. The materials are classified into the $3p$ subset and arranged in order based on their weighted average win-fraction ($\overline{WF_{i}}$) scores, with a maximum potential score of 100\%. The figure also includes the individual win-fraction ($WF_{i,k}$) values for each property of the materials.}
  \label{fig:wf_3.1}
\end{figure*}

The comparative ranking procedure is used as benchmark for the previous ranking processes. 
From the weighted average win-fraction definition ($\overline{WF_{i}}$), materials belonging to the subsets $3p$, $2p$, and $1p$ naturally separate, with their scores limited to 100\%, 66.7\%, and 33.3\% respectively. The results of the comparative ranking procedure are presented in Fig.~\ref{fig:wf_3.1}, illustrating the materials belonging to category $3p$ ranked from the 1\textsuperscript{st} to the 100\textsuperscript{th} position. Additionally, the Appendix includes materials from the 101\textsuperscript{st} to the 200\textsuperscript{th} position of category $3p$, together with the top 200 materials from category $2p$ and $1p$ (Fig.~\ref{fig:wf_3.2} - \ref{fig:wf_1.2}).
$\overline{WF_{i}}$ is compared with the result coming from the thickness and Pareto ranking procedures. For the latter, a binary value is assigned to every material in order to show its positive or negative detection by the ranking process. 
In order to emphasize the comparison between different ranking procedures, each of these is applied to the entire set of available materials, regardless of whether they are excluded by more stringent rankings.

The first 63 materials (out of 218) outperforming the $3p$ $\overline{WF_{i}}$ ranking scale (Fig.~\ref{fig:wf_3.1}) contain the whole set of the PFM candidates able to sustain divertor ELMs conditions, as indicated by the heat balance equation. Of the top 15 materials, 12 of them are selected by the $d^{ELMs}$ ranking. All of these candidates are also selected by the $d^{SS}$ ranking and one or more Pareto rankings, indicating an excellent consistency between the three ranking procedures. 
The importance of a comparative scheme is further emphasized by its ability to avoid false negatives in screening studies of this nature. Inconsistencies among different selection criteria can bring to light previously dismissed candidates. Indeed, the two materials erroneously discarded according to the $d^{ELMs}$ ranking (as indicated in Sec.~\ref{trELMs}), graphite and \ce{SiC}-3C, are well-captured by the whole set of Pareto rankings when included.
Consistently, other known refractory materials exhibiting excellent thermal properties are selected by lower level ranking procedures when included: \ce{TiC}, pure \ce{B}, \ce{BeO}, \ce{ZrC}, \ce{B4C} (as \ce{B13C2}), and \ce{Li2TiO3}. 
In particular, the latter material has been proposed as a tritium breeding material for the test blanket module of ITER \cite{hoshino2011development}.
Furthermore, another class of materials recently proposed as \ce{W}-alternative plasma-facing materials is identified by examining the top 50 materials ranked by the 3p $\overline{WF_{i}}$ and in Fig.~\ref{fig:wf_3.1}; this is the class of MAX phase materials, represented by \ce{Ti3SiC2} \cite{coburn2019surface,zhang2017helium}. Despite being these materials, and especially \ce{Ti3SiC2}, refractory and possessing a good thermal conductivity, their decomposition temperature severely limits their suitability as divertor PFMs \cite{barsoum1999thermal}.
In the $2p$ and $1p$ $\overline{WF_{i}}$ ranking scales, which consist of 803 and 6,063 materials respectively, the top 200 candidates include the complete set of $d^{SS}$ and $2p$ Pareto top candidates from the $2p$ set, and 20 out of 26 $1p$ Pareto candidates from the $1p$ set. The clustering of the previously selected candidates near the top of the comparative ranking provides additional evidence of the reliability of the ranking procedure we have established.

\section{Literature review of the PFM candidates} \label{literature}

We analyse the four main macro-categories of materials, identified using the previous ranking criteria, based on previous literature results:
\begin{enumerate}
    \item carbon-based materials;
    \item boron-based materials;
    \item transition-metal-based materials;
    \item ceramics, which further divide into three main classes: borides, carbides, and nitrides. 
\end{enumerate}
Among the most promising materials, we select specific prototypical crystal structures as the foundation for calculating the formation energy of two commonly observed defects, the surface binding energy and the hydrogen interstitial formation energy, as described in Sec.~\ref{dft}.

\subsection{Carbon-based materials}\label{carbon-based}
The ranking procedure highlights two distinct allotropes of pure carbon: diamond and various types of intercalated graphite. Surprisingly, pure graphite is discarded by the $d^{ELMs}$ ranking because of its low assigned KDE thermal conductivity value (see Fig.~\ref{fig:wf_3.2}).
In fact, the majority of the thermal conductivity values found in the MPDS database for graphite pertain to the direction perpendicular to its hexagonal plane. However, in the case of polycrystalline samples, such as those used for this applications, the primary contribution to thermal conductivity arises from heat transport parallel to the basal plane \cite{taylor1968thermal}, which is thousands of times higher than the perpendicular one \cite{fugallo2014thermal,haynes2016crc}. The resultant thermal conductivity of these samples, accounting for the limiting factors inherent to the polycrystalline averages, remains of the order of hundreds of \si{\watt\per\meter\per\kelvin} \cite{taylor1968thermal}, considerably higher than the value extrapolated from the database (5.7 \si{\watt\per\meter\per\kelvin}) and sufficient to avoid melting during type-I ELMs. Consequently, graphite can be considered as a PFM candidate according to the $d^{ELMs}$ ranking.

Graphite and carbon fiber composites (CFCs) were identified as the primary choices for (PFMs) for the ITER divertor \cite{matthews2007overview}. Due to their remarkable thermal conductivity and high sublimation temperature  \cite{abrahamson1974graphite}, these materials have demonstrated the ability to withstand significant thermal loads with no structural damage or, at most, minimal degradation of the polyacrylonitrile fibers in the case of CFCs \cite{riccardi2013effect}.
Furthermore, graphite-based materials pose low concerns regarding plasma core contamination due to their low-Z number, which minimizes radiative energy losses and allows for higher tolerance levels within the plasma. For this reason, fusion experiments worldwide continue to adopt graphite as the preferred PFM \cite{linke2019challenges}. However, a material composed of low-Z elements is subject to surface erosion by physical sputtering processes, with the additional drawback of an enhanced chemical sputtering for graphite, due to the high reactivity between carbon atoms and the hydrogen isotopes contained in the plasma \cite{roth1991erosion}. This leads to the formation of co-deposited layers of tritium-containing hydrocarbon on all in-vessel components, which increases the inventory of tritium \cite{roth2009recent}. In addition, the high thermal conductivity of graphite deteriorates rapidly under the influence of energetic neutrons \cite{maruyama1992neutron,bonal1996neutron}, an  effect only partially mitigated by defects recombination that occurs at high-temperature.

Graphite-based materials have been tested for many years \cite{snead2020graphite} and, even recently, carbon-based candidates such as diamond or diamond-like carbon structures are re-considered and proposed as alternatives to \ce{W} \cite{de2011thermal,hughes2021degradation}.
However, effective suppression of the surface sputtering process, has not been demonstrated.
Despite many attempts on suppressing chemical sputtering \cite{kato1999laboratory,garcia1992chemical,de2009interactions} and radiation enhanced sublimation \cite{begrambekov1990development}, physical sputtering process plays a crucial role per se. 
Based on the literature data from the early 2000s \cite{federici2001assessment,federici2001plasma}, one can see that even when considering only physical sputtering, the tritium co-deposition rate constant in graphite drops to approximately one-fifth of its original value. This reduction is inadequate to offset the tritium inventory limit that exceeds the threshold by two orders of magnitude \cite{roth2008tritium,roth2009recent}.
For carbon too be the main constituent of a candidate material, this constraint cannot be circumvented unless: (i) significantly higher SBE values are assumed for the surface carbon atoms of the candidate material, (ii) a different preferential target element is considered for the sputtering process (e.g.~low-Z carbides), (iii) the kinetic energy of plasma particles is reduced below the threshold value of carbon sputtering, (iv) new tritium desorption techniques are proposed. 

Carbon-based materials are, therefore, not considered optimal as divertor PFMs in ITER-like devices. However, due to their impressive ability to withstand high thermal loads and their well-documented characterization in the literature, they are included as reference materials in the list of top PFM candidates. 
In particular, carbon diamond (\ce{C} dia) and graphite (\ce{C} gra) are chosen as representatives of the carbon-based materials category for the purpose of calculating defect formation energies.

\subsection{Boron-based materials} \label{boron-based}

The boron-based materials reported in this study belong to the family of compounds known as icosahedron-based borides. These materials consist of a main structure of \ce{B12} icosahedra, which may be interconnected through chains of carbon (or other nonmetallic) atoms. The voids within this framework can also accommodate metal atoms. 
Boron-based materials are selected through the $1p$ Pareto ranking procedure, due to their outstanding Knoop hardness, and fall into two classes.
The first class includes \ce{RB66}, with R being a metallic atom. \ce{RB66} materials combine the physical and chemical properties of refractory crystals with the typical electrical, optical, and thermal properties of amorphous semiconductors. Despite their interesting stability under high temperature and abrasive conditions, they posses a really low thermal conductivity value because of their complex crystal structure \cite{mori2006effect,novikov2014thermal}.
An example of the second class is \ce{Mg2B25C4}, where, beside the presence of metal in the boron structure, carbon acts as a bonding element between the various icosahedral \ce{B12} structures. The materials of this class have similar thermo-mechanical properties to the \ce{RB66} crystal class \cite{adasch2010synthesis}.
Boron carbide \ce{B4C} has one of the highest thermal conductivity within the second class, reaching a value of approximately 28 \si{\watt\per\kelvin\per\meter} at room temperature. Under neutron irradiation, this value is significantly reduced \cite{qu2022thermal}, but it can be largely restored through high-temperature annealing, especially in the case of $^{11}$B enriched samples \cite{morohashi2008neutron}. However, this value of thermal conductivity is insufficient to avoid melting during ELM event, as confrmed by $d^{ELMs}$ ranking procedure correctly discarding of this material from the list of potential candidates. It is worth noting that \ce{B4C} is identified by a compositional variation of boron carbide, \ce{B13C2}, in MPDS. Due to its limited thermal conductivity, this class of materials is deemed unattractive for withstanding high heat loads and frequent thermal shocks and, therefore, it is not included in the list of top PFMs candidates. It is worth noting that in combination with a highly conductive substrate, such as graphite, \ce{B4C} has been proposed as protective layer of the main divertor armor material \cite{buzhinskij1995performance}.

We finally remark that the presence of $^{10}$B (approximately 20\% of natural elemental boron) poses significant limitations on the use of boron-based PFMs. In fact, $^{10}$B is known to undergo transmutation through the (n, $\alpha$) reaction, resulting in the production of helium and lithium atoms and the formation of cavities or bubbles in the matrix and grain boundaries \cite{grinik1996fracturing,bhattacharya2019nano}. This process leads to void-swelling, high-temperature helium embrittlement, and compromises the structural integrity of the crystal due to progressive loss of $^{10}$B. The negative effects of $^{10}$B transmutation can be mitigated by using isotopically enriched $^{11}$B, which has a considerably lower transmutation cross-section to thermal neutrons \cite{bhattacharya2019nano,koyanagi2019response}

\subsection{Transition metal-based materials}\label{trans-metal}

\paragraph{Tungsten}
The future and current generations of PFM candidates for the divertor region are \ce{W}-based. Despite the fact that \ce{W} outperforms all other alternatives tested so far, many downsides in its application remain, and are reported in literature. 
\ce{W} can sustain the divertor thermal loads in a very narrow temperature window compared to the operating temperature interval of the PFM surface. In steady-state conditions the surface of the metal is close to its ductile–brittle transition temperature (DBTT), 200 to 600 \si{\degreeCelsius}, depending on the manufacturing processes, and below the DBTT the reduced thermal stress resistance leads to intense crack networks \cite{wirtz2017transient}. While \ce{W} does not encounter any surface changes at heat loads of 10 \si{\mega\watt\per\square\meter}, even during typical slow-transient of 20 \si{\mega\watt\per\square\meter}, it shows severe surface damage and deep crack formations \cite{pintsuk2013qualification}. In addition, when the low recrystallization temperature of \ce{W} (roughly 1,300 \si{\degreeCelsius}) is surpassed, grains growth and consequent increase of DBTT, brittleness and surface roughening, and decrease of thermal shock resistance and mechanical strength are observed \cite{li2015interpretation,suslova2014recrystallization}. Finally, power pulses produced during ELMs can result into cracking of the surface, partial melting of the tiles edges, and depletion of the PFM surface \cite{wirtz2017transient}; in particular, alarming levels of \ce{W} erosion are reached when ELMs are combined with slow-transient loads \cite{klimov2011experimental,makhlaj2013dust}.

The heat load is associated with a flux of plasma particles and their implantation in the \ce{W} armor, for a distance of a few nanometers \cite{kreter2019influence}. The energy of the plasma particles (approximately 10 \si{\electronvolt}) is not sufficient to physically sputter the surface atoms of the \ce{W} armor, as it is much lower than the \ce{W} sputtering threshold energy (209 \si{\electronvolt} for D, 102 \si{\electronvolt} for He, at 20 \si{\degreeCelsius} \cite{eckstein1993sputtering}). Furthermore, the weak chemical reactivity between \ce{W} bulk, hydrogen isotopes and He atoms does not lead to any additional chemical sputtering \cite{lasa2012md,de2013helium}. Nevertheless, the erosion of the \ce{W} surface is a reason of concern due the surface morphology changes produced by the interaction between \ce{He} atoms and \ce{W} crystal defects in the presence of hydrogen, which leads to the formation of helium bubbles and the so-called a fuzz layer, a process that is not yet completely understood \cite{wright2022review}. Despite the well-known decrease of the effective thermal conductivity of the nano-fiber layer and consequent ejection of melted droplets \cite{rieth2019behavior}, the effects on the net material erosion rate is still under debate \cite{de2013helium,sinclair2018erosion,hwangbo2017erosion}.
Plasma particles can be retained by the PFM, which is problematic as mentioned above for CFCs. ELMs and fuzz formations seem to play an opposite role in terms of tritium retention in the \ce{W} wall: the damaged surface produced by ELMs enhances its retention compared to steady-state conditions \cite{poskakalov2020influence}; in contrast, the open porosity of the fuzz layer serves as an additional release channel for the hydrogen isotopes \cite{kreter2019influence}. However, in the presence of \ce{He} defects in \ce{W} can affect its retention capacity; dislocation loops punched by helium bubbles are responsible for tritium trapping sites \cite{iwakiri2002effects} and the total retention amount can increase drastically.

In this scenario, high-energy neutrons also play a role and irradiated-\ce{W} samples showed more than double tritium retention compared with undamaged \ce{W} \cite{hodille2017estimation}. Vacancy clusters of the neutron-damaged materials can act as additional traps for hydrogen isotopes \cite{toyama2018deuterium}.
14.1 \si{\mega\electronvolt} neutrons are also responsible for the degradation of the wall armor under operational conditions via reduction of the thermal conductivity and embrittlement, due to defect accumulation and transmutation. For both properties, the worsening of performance dictated by the contribution of dislocations and voids is in large part recovered at higher temperature, thanks to the annihilation of the irradiation-induced defects \cite{hu2016irradiation,cui2018thermal}. \ce{W} transmutations leads to the production of rhenium (Re) and osmium (Os) in different percentages as a function of the neutron's flux \cite{cottrell2004sigma}.  In ITER, divertor replacement is planned after a displacement per atom (dpa) of 0.7 with 0.15\% of Re content and less than half of that for Os content \cite{bolt2002plasma}; in DEMO this could be up to 20 times higher \cite{stork2014developing}. Thermal diffusivity is negatively affected by the presence of Re in the \ce{W} bulk (around -25\% of unirradiated \ce{W} at 1,000 \si{\kelvin}) and temperature changes seem to be negligible at the relevant concentration \cite{fujitsuka2000effect,tanabe2003temperature}. In terms of ductility (and recrystallization temperature) Re addition to \ce{W} is beneficial \cite{rieth2019behavior}, but in presence of Os and dpa above 0.6-1 their concentration is enough to produce an intermetallic second phase precipitate, which leads towards strong hardening and embrittlement of the PFM, regardless of temperature variations \cite{hu2016irradiation}. Transmutation products dominate over the neutron-induced defects for the strengthening effect as long the irradiation dose is higher than 0.6 dpa, while the opposite is true below 0.3 dpa.

To summarize the aforementioned points, \ce{W} is presently considered the most viable option to withstand the expected heat loads at the divertor of a tokamak reactor, securing its position among the most promising PFM candidates. However, its performance under extreme thermal stress conditions and its long-term structural integrity when exposed to the combined effects of plasma and neutrons remain unresolved issues and crucial areas of ongoing research in the field.

\paragraph{Molybdenum} The most studied high-Z divertor PFM alternative to \ce{W} is molybdenum (\ce{Mo}). \ce{Mo} was used as divertor PFM in the Alcator C-Mod tokamak reactor and its behavior extensively analyzed \cite{brooks2011analysis}. While \ce{Mo} has a melting temperature and thermal conductivity lower than \ce{W}, it shows some advantages when compared with \ce{W}, such as a lower transmutation rate and therefore reduced formation of precipitates \cite{cottrell2004sigma}, and slightly lower H retention \cite{sharpe2009retention}. In addition, as in the case of \ce{W}, \ce{Mo} thermal conductivity is barely affected by high temperatures \cite{rasor1960thermal} and it can be restored to its high values thanks to annihilation of defects on neutron-irradiated samples \cite{yi2019defect}. However, neutron irradiation poses a clear drawback on the application of \ce{Mo} as plasma-facing material. Both the US and UK waste disposal rating classify a \ce{Mo} armor exposed to divertor neutron irradiation conditions as high-level-waste \cite{brooks2015plasma,gilbert2019waste,gilbert2017automated}.
Despite this, plasma-\ce{Mo} interactions under divertor conditions are still under investigation, and shows many similarities with plasma-\ce{W} interaction. The erosion in steady-state conditions is mainly due to impurities in the plasma that have mass sufficently large to sputter \ce{Mo} (i.e.~\ce{Mo} itself and oxygen). Even if this erosion is predicted to be slightly higher than the case of \ce{W}, it remains tolerable (1.5 \si{\milli\meter}/burn-year) \cite{brooks2015plasma}. As in the case of \ce{W}, the interaction with He ions results in the creation of fuzz nanostructures on the surface exposed to the plasma \cite{de2012nanostructuring}. The temperature window for their formation is narrower and lower than for the case of \ce{W} \cite{tripathi2015temperature}, but it seems to follow the same pattern \cite{de2012nanostructuring}. Also, under unmitigated  type-I ELMs-like pulses the difference in erosion between the two metal with a pristine surface has not been fully elucidated yet. A recent study reported that the volume of \ce{Mo} ablated from the surface by a single laser pulse is 10 times larger than \ce{W} \cite{montanari2020surface}, whereas another study reports that after 100 laser pulses (of similar energy) the two materials do not show appreciable difference in the amount of mass loss \cite{sinclair2017melt}. In both experiments the energy required from the laser pulse for melting the two metals was found to be above the energy of mitigated type-I ELMs, 1.0 \si{\mega\joule\per\square\meter} for \ce{Mo} and 1.4 \si{\mega\joule\per\square\meter} for \ce{W} \cite{sinclair2017melt}. Similar effects are produced by pulsed lasers, simulating unmitigated ELMs on fuzz form \ce{Mo} surfaces \cite{sinclair2016structural}. When covered with a fuzz layer, the mass loss appears to be a linear function of the laser pulse, instead of exponential as for the pristine surface \cite{sinclair2017melt}. Under mitigated type-I ELMs, the fuzz layer gradually disappears, without any melting, with a mass loss that increases as a function of the surface temperature \cite{de2013helium}. The extent to which this phenomenon, when combined with arcing, could pose a risk of core plasma contamination remains unclear. Additionally, it should be noted that a higher concentration tolerance is predicted for \ce{Mo} (\num{e-3}-\num{e-4} \cite{mazzitelli2019experiments,winter1991test}) compared to \ce{W} (\num{5e-5} \cite{kallenbach2005tokamak}) before plasma quenching. 
Overall, \ce{Mo} is still considered an attractive alternative to \ce{W} and is therefore included in the list of top PFM candidates for further investigations.

\paragraph{Copper} A high-strength copper alloy material is envisaged for the cooling tube system running through the central region of the \ce{W} monoblock of the ITER divertor target. Divertors designed for DEMO envisage the use of pure copper (Cu) as stress-relieving interlayer between \ce{W} and the \ce{Cu}-alloy \cite{li2015low,zinkle2015applicability}. In fact, despite its impressing thermal conductivity, \ce{Cu} cannot be exposed directly to the plasma flux, primarily because of its low hardness and softening temperature \cite{zhang2022microstructure} and the very intense erosion to which it is subject to under hydrogen plasma flux \cite{bondarenko2000behaviour}. In addition, even if \ce{Cu} and \ce{Cu} alloys are known to maintain high thermal conductivities at elevated temperatures \cite{nath1974thermal}, the exposure to neutron radiation can damage their crystal structure and cause transmutations, which can negatively affect their thermal diffusivity \cite{zinkle1994copper,fabritsiev2005effect}. Furthermore, the induced radioactivity caused by the high-energy fusion neutrons raises concerns about safety and waste disposal considerations \cite{gilbert2017automated,gilbert2019waste}. 
Due to these reasons, pure \ce{Cu} is not directly employed as PFMs in fusion tokamak devices and is not included in the list of promising PFMs. However, application of \ce{Cu} in the form of alloy as a thermal conductor material that is not exposed to plasma has been extensively investigated, and it is important to mention that, in a small number of studies, copper-alloys are also proposed as primary divertor and first-wall materials, in particular \ce{W}-\ce{Cu} \cite{bondarenko2000behaviour,lee2012thermal} and \ce{Cu}-\ce{Li} \cite{krauss1985self,krauss1987temperature,schorn1989suitability}.

\paragraph{Platinum-group and noble metals} A set of transition metals, belonging to the so-called platinoids and noble metals classes, are reported here for their valuable thermal properties. We consider in descending order of performance (see Fig.~\ref{fig:d_rank}): iridium (\ce{Ir}), osmium (\ce{Os}), ruthenium (\ce{Ru}), rhodium (\ce{Rh}), rhenium (\ce{Re}), and silver (\ce{Ag}). The main limit to their use as pure metals is dictated by their high price (they typically range in cost from \$1,000 to \$100,000 USD per kilogram) and scarcity \cite{ContributorstoWikimediaprojects2023Apr} and their thermo-mechanical properties, typically slightly inferior than more affordable alternatives (\ce{W} and \ce{Mo}). The main application of these elements is thus as dopants or alloying elements. \ce{Re}, being one of the cheapest of this list, second only to \ce{Ag}, has been proposed as alloying element for \ce{W} due to its enhanced thermo-mechanical performance: higher ductility, high-temperature strength, plasticity, and corrosion resistance, stabilization of the grain structure with increased recrystallization temperature and weldability, and reduced degree of recrystallization embrittlement \cite{rieth2019behavior}. However, the high cost and scarcity of \ce{Re} is already affecting other industrial sectors (platinum–rhenium catalysts and high-temperature superalloy for jet engines) and strategies to reduce its application, finding alternatives and recycling processes are under investigation \cite{fink2010rhenium,shen2021review}.
Despite its cost-related limitations, \ce{Re} is included in the list of leading PFM candidates as the most favorable alternative within this class of materials, in terms of cost and performance.
The least expensive \ce{Ag} is discarded due to inferior thermal properties when compared to the aforementioned transition metals, primarily dictated by a low melting temperature.

\paragraph{Early transition metals}
Together with the already discussed \ce{Mo}, \ce{W}, and \ce{Re}, there are a few more elements from the first 5 groups of the transition-metal class that exhibit notable thermal capabilities, in particular \ce{Ta} and \ce{Cr}. 
\ce{Ta} was suggested as a potential PFM in the past: however two significant drawbacks are present: (i) its inferior thermal properties when compared with \ce{W} \cite{hirai2003testing} (ii) the exothermic hydrogen absorption process, which yields larger tritium retention than other metals (such as \ce{W}) with positive heat of solution for hydrogen \cite{takagi2003trapping}. In contrast, \ce{Ta} demonstrates a higher energy threshold for the formation of a "fuzz" structure when exposed to divertor-like He ion flux, which suggests that surface thermal and mechanical properties may not degrade as quickly as for \ce{W} in extreme fusion environments \cite{novakowski2016effect}.
Chromium has been recently proposed as PFM for the DEMO-divertor structural body of the monoblock. \ce{Cr}-10\%\ce{W} displays superior mechanical properties to \ce{Cr} at low temperatures, where \ce{W} is brittle, and higher strength compared with pure \ce{Cr}. Further investigations of its behaviour under actual divertor conditions must thus be carried out \cite{terentyev2020development}.
For this reasons, both \ce{Ta} and \ce{Cr} are added to the list of promising PFM candidates.

\paragraph{Transition metal alloys}

Only one material stands out in this category, exhibiting features comparable to those of the positive thickness materials under type-I ELMs, this is the \ce{Ta_{0.5}W_{0.5}} solid solution, possessing high level thermal conductivity, heat capacity, and isothermal bulk modulus, with a melting temperature of 3,450 \si{\kelvin} is reported in the literature \cite{cahn1991binary}. The two constituents of this alloy exhibit complete solubility \cite{okamoto1990binary}, but given the low mass contrast between its components, the phonon contribution to its thermal conductivity is not significantly diminished, in contrast to many other alloy systems. Previous studies also investigate the thermo-mechanical properties of \ce{Ta}-\ce{W} composites, wherein \ce{Ta} fibers are dispersed within a nanostructured \ce{W} matrix to enhance fracture toughness \cite{dias2013synergistic}. However, the ability of these composites to withstand plasma irradiation conditions remains a topic of ongoing debate \cite{dias2013synergistic,dias2017helium,konuru2020deposition}. Our screening process did not identify any other promising alloys; nevertheless, one of the most promising research areas in terms of alloys is represented by the high-entropy alloys (HEAs). These materials are currently not included in MPDS but, in a recent study \cite{el2019outstanding}, a \ce{W}-\ce{Ta}-based HEA alloy (38\% W, 36\% Ta, 15\% Cr, 11\% V) was able to show outstanding neutron radiation resistance properties, high hardness and small irradiation-induced hardening \cite{el2019outstanding}. The material capacity to sustain high-dose neutron irradiation is attributed to the equal mobilities of vacancies and self-interstitials, which maximizes recombination and reduces defect concentration. This is likely due to the rough energy landscape induced by local lattice distortion and the disparity in chemistry. While lattice distortions can affect the thermal conductivity (a common drawback of this class of materials \cite{wang2021recent}), very recent promising results for thermal conductivity have been shown for the class of medium-entropy alloy \cite{cui2022studies}. As a prototype of this class, the ordered cubic structure (B2-superstructure \cite{turchi2001first}) of \ce{TaW} is added to the list of top PFM candidates for which defect formation energies are then calculated using the DFT workflows developed here.

\subsection{Ceramics}\label{ceramics}
Most of the ceramics identified by the three ranking criteria can be classified as ultra-high temperature ceramics (UHTCs). This family includes a variety of materials, such as borides, carbides, and nitrides of early transition metals, and are characterized by their extremely high melting temperatures, typically above 3,000 \si{\celsius}, and/or their ability to withstand continuous use at temperatures exceeding 1,600 \si{\celsius} \cite{fahrenholtz2012oxidation}. One of the major challenge for UHTCs in thermostructural applications is their poor resistance to thermal shocks. However, the manufacture of fibrous monolith ceramics limit the formation and propagation of cracks to the cell boundaries and can partially address this problem \cite{fahrenholtz2007refractory}.

\paragraph{Borides}
Refractory diborides, such as the here \ce{HfB2}, \ce{TiB2}, and \ce{ZrB2} reported here, exhibit an unusual combination of high strength, moduli, melting temperature, and thermal conductivity. The possibility of using diborides as PFMs is now again under discussion, since the production of isotopically separated $^{11}$B diborides is now demonstrated \cite{garrison2018response,koyanagi2019response}. In fact, neutron irradiation of naturally occurring \ce{B} is indicated as the primary causes of swelling and catastrophic cracking of borides, due to the helium and lithium accumulation deriving from the (n, $\alpha$) reaction of $^{10}$B, as showed in Sec.~\ref{boron-based}.
Neutron-irradiation experiments on \ce{Ti}$^{11}$\ce{B2} demonstrated reduced swelling (especially at high temperature), suppressed macroscopic damage and moderate micro-cracking formation \cite{koyanagi2019response}.
Concerning the thermal properties, \ce{ZrB2} shows the capability to maintain a level of thermal conductivity at high temperature that is comparable to \ce{W}. Furthermore, since the electronic contributions mostly determine the total thermal conductivity value, neutron-induced defects are expected to have a minimal effect \cite{garrison2018response}. Both \ce{TiB2} and \ce{ZrB2} show encouraging results in terms of erosion resistance under high-energy He ions: the \ce{TiB2} sputtering yield is reduced when compared with those of pure \ce{Ti} \cite{kaminsky1980sputtering,nenadovic1996sputtering}, while \ce{ZrB2} show comparable erosion depth with that of \ce{W} \cite{garrison2018response}. \ce{ZrB2} also shows an interesting oxidation mechanism that might lead to oxygen absorption at higher temperatures \cite{fahrenholtz2012oxidation,fahrenholtz2007refractory,garrison2018response}. However, experiments at divertor conditions are still lacking. While showing the same oxidation mechanism as \ce{ZrB2}, \ce{HfB2} has raised some concerns regarding the production of 20 \si{\mega\electronvolt} gamma rays when irradiated with 14 \si{\mega\electronvolt} neutrons \cite{nasseri2016behavior}. 

Inspired by the growing field of high-entropy alloys (HEAs) where five or more elements are mixed in similar proportions to form a new alloy, \ce{ZrB2}-based ceramics are also proposed as high- and medium-entropy ceramics (HECs) \cite{oses2020high}, and their thermo-mechanical properties are investigated \cite{wu2023zrb2}. (Zr\textsubscript{x}Ti\textsubscript{y}Nb\textsubscript{y}Ta\textsubscript{y})B2 systems containing \ce{Zr}, \ce{Ti}, \ce{Nb}, and \ce{Ta} at the highest mixing entropy posses the highest hardness and moduli, but also the highest phonon scattering with the lowest thermal conductivity. Similarly, carbon-boron-titanium compounds with varying \ce{B}/\ce{C}/\ce{Ti} ratios have been studied. It is observed that the highest thermal shock resistance is achieved for compositions where the constituents are mixed in similar proportions, which also exhibited lower thermal conductivity compared to pure borides and carbides \cite{tanabe1992thermal}.
\ce{ZrB2} is also proposed as alloying ceramic to produce \ce{W}-\ce{ZrB2} alloys, with improved bending strength and hardness thanks to the strengthening of the originally fragile \ce{W} grain boundaries \cite{wang2020microstructure}.
An additional diboride, \ce{VB2}, is also of interest, according to the $d^{ss}$ ranking, due to its remarkable values of high melting temperature and thermal conductivity. Considering the combination of these properties, with the noteworthy volumetric heat capacity reported in the literature \cite{krikorian1971estimation,cao2022theoretical} (exceeding 5 \si{\joule\per\kelvin\per\cubic\centi\meter} at high temperatures), \ce{VB2} has been previously suggested as a protective layer to mitigate the ingress of hydrogen isotopes into structural materials in fusion devices \cite{pontau1980vb2,doyle1981hydrogen}.
Consequently, the crystal structures of all the presented diborides, namely \ce{HfB2}, \ce{TiB2}, \ce{ZrB2}, and \ce{VB2}, are included in the compilation of the most favorable PFM candidates.

\paragraph{Carbides}
Carbides of the same transition metal have usually larger melting temperature and hardness than the corrisponding borides and nitrides \cite{wang1995electrical}, at the price of a
substantially reduced thermal conductivity when compared with borides \cite{wuchina2004designing}. Despite obvious concerns associated with high carbon content, linked the high level of tritium retention caused by carbon sputtering and co-deposition, carbides have mainly been investigated as PFMs for doping or alloying \ce{W}.
Three polymorphs of \ce{SiC} are reported in MPDS, but only the hexagonal polytype, \ce{SiC}-2H, is selected according to the Pareto ranking thanks to its excellent melting temperature, around 3,100 K. In fact, the cubic polytype \ce{SiC}-3C is discarded by the thickness (ELMs) ranking, due to a reported very low thermal conductivity. This is quite surprising, since \ce{SiC}-3C is known for its outstanding thermal conductivity (\ce{SiC}-3C possesses the highest thermal conductivity value \cite{cheng2022high}, along with a comparable melting temperature and slightly inferior mechanical properties \cite{yang2022effects} than the other polymorphs). Indeed, an erroneous value of thermal conductivity is reported in the MPDS database, 0.49 \si{\watt\per\meter\per\kelvin}, instead of the correct value, 490 \si{\watt\per\meter\per\kelvin}. Once this is accounted for, \ce{SiC}-3C appears at the top of the $d^{ELMs}$ ranking.
\ce{SiC} is largely investigated as refractory and plasma-facing material due to its excellent thermal properties and low activation under fusion or fission neutron bombardment \cite{rocco1991criteria}. It is also considered as divertor PFM \cite{abrams2021evaluation}, motivated by its favourable erosion properties when compared with graphite. In particular, \ce{SiC}-3C has been used as a coating PFM of graphite tiles during the 2017-19 DIII-D run campaigns \cite{abrams2021evaluation}, and the extrapolation of results to a DEMO-type device shows a 70\% decrease of the erosion yield of carbon is expected with \ce{SiC} as PFMs at the divertor with respect to graphite. The reduced erosion was also confirmed in a later study \cite{abe2022computational}. This effect, combined with the low solubility and diffusivity of tritium in \ce{SiC}, \cite{causey1995use} associated with a lower predicted tritium inventory in the co-deposited layer when compared with graphite (2/3 of the carbon material) \cite{causey2003sputtering}, could motivate further investigations of \ce{SiC} as divertor PFM. However, the reduced erosion has been extrapolated from very low concentration of C atoms in the plasma. In addition, the value of thermal conductivity of 490 \si{\watt\per\meter\per\kelvin} is not realistic under operational conditions where high temperature and neutron bombardment occur. The high temperature reduces significantly the thermal conductivity of the most well-known \ce{SiC} polytypes \cite{protik2017phonon,wang2017strong}. In addition, the presence of neutron-induced defects in \ce{SiC} results in detrimental swelling and a significant decrease of the thermal conductivity, as demonstrated in several studies \cite{price1973thermal,katoh2011stability,katoh2007current,snead2011silicon}. Despite the fact that these defects are partially recovered at high temperatures, the thermal conductivity values obtained for $\beta$-SiC at dpa levels comparable to ITER/DEMO devices decreases to around 1 to 10 \si{\watt\per\meter\per\kelvin} (depending on the irradiation temperature), making this material almost unsuitable for the expected thermal loads.  This issue is ameliorated in modern \ce{SiC} fiber composites that exhibit minimal degradation of their thermo-mechanical properties even when subject to high levels of neutron damage \cite{abrams2021evaluation,katoh2014continuous}. However, their thermal conductivity in the pristine state values are low, and comparable to those of neutron-damaged monolithic CVD \ce{SiC}.
As a consequence, based on the current state of the literature, \ce{SiC} does not meet the criteria to serve as a PFM in the divertor region.

With small percentages of \ce{TaC} that fill up the grain boundaries, sintered \ce{W} shows increased microhardness as a function of the \ce{TaC} content, no cracking formation under ELMs-like thermal shock, and even increased thermal conductivity at the low carbide percentage of 1\% \cite{tan2015development,feng2018effect}. Similar effects on the mechanical properties, thermal stability, and thermal conductivity are also reported for this \ce{W}-\ce{HfC} alloy
\cite{wang2017thermal}. However, the retention of hydrogen isotopes is generally increased under divertor-like deuterium bombardment \cite{zibrov2017high} and can be one order of magnitude larger when compared with pure \ce{W} in the case of \ce{W}-1.1\%\ce{TiC} and \ce{W}-3.3\%\ce{Tac} at 800 \si{\kelvin}, decreasing with increasing temperature. This might arise concerns on the tritium inventory.
The direct use of \ce{TaC} and \ce{HfC} as PFMs has not been extensively investigated. Despite concerns related to their high carbon content, their thermal properties are at the threshold of meeting the requirements for withstanding divertor conditions without undergoing melting.
However, the pure alloy system \ce{TaC}-\ce{HfC}, shows the highest melting point of the known materials at the right ratio of \ce{Ta} and \ce{Hf} (4:1) \cite{andrievskii1967melting,cedillos2016investigating} and decent values of thermal conductivity that, in contrast to pure metals or diborides, increase with rising temperature \cite{cedillos2016sintering}. Further studies, including tritium inventory and behaviour under neutron irradiation, should be pursued, since at the experimentally reported values of the melting temperature, heat capacity and thermal conductivity \cite{cedillos2016sintering}, the 4\ce{TaC}-1\ce{HfC} alloy system would withstand type-I ELMs heat loads at the divertor with an armor thickness of about 5 \si{\milli\meter}. 
Rather than for their individual potential, \ce{TaC} and \ce{HfC} are added to the list of PFMs to calculate defects formation energies as a proxy of their ultra-high temperature alloys.

Given the melting temperature, heat capacity, and the thermal conductivity reported in literature \cite{li2015spark}, vanadium carbide (\ce{VC}), can only withstand steady-state conditions, contrary to its boride form. Furthermore, its high-temperature stability is moderate, but its ability to act as a barrier for hydrogen permeation is noteworthy, which has led to the exploration of various type of vanadium carbides (\ce{V_xC_y}) as a coating layer for structural fusion materials \cite{liu2019vanadium,huang2021atomic}. The small hydrogen diffusion is attributed to its high hydrogen penetration barriers and large endothermic detrapping energy of H atoms in C vacancies. Previous test of the \ce{V_xC_y} behavior under neutron irradiation reports hardening and increased electrical resistivity \cite{takahashi1988effects,matsui1991neutron}. Even if more investigation is required to understand degradation of the thermal properties induced in the presence of a plasma flux, this material does not qualify to sustain divertor heat loads and is discarded from the list of top PFM candidates.

\ce{Ta2C}, along with the other carbides emerging from the previous ranking selection criteria - \ce{TaC}, \ce{HfC}, \ce{WC}, and \ce{W2C} - has also been proposed in the PMF field, but solely as a dispersion agent to enhance mechanical properties, suppress the embrittlement of \ce{W}, and improve the migration of grain boundaries \cite{yan2021microstructural}. \ce{Ta2C} has been explored also as a bonding material for joining the structural materials or as a cooling system to the \ce{W} armor \cite{matvejivcek2011development}. Although \ce{Ta2C} is not extensively investigated as an alternative to W, its promising melting temperature raises interest. However, the current lack of available literature on the heat capacity and thermal conductivity of this material hinders the determination of its capability to withstand divertor conditions and to compare it with the top PFMs. Therefore, we refrain from conducting additional follow-up studies on \ce{Ta2C} at this time.

The last class of ceramic materials that emerges from the Pareto set analysis is that of tungsten carbides, in particular \ce{WC} and \ce{W2C}. As previously mentioned, they might be applied as alloying compounds of a pure \ce{W} armor \cite{odette4,vsestan2022non}, but their thermo-mechanical properties and response under plasma and neutron irradiation are considered good enough to be considered and studied as an alternative PFMs at the divertor \cite{traskelin2007radiation,humphry2019shielding} and therefore these are retained in our ranking. At high temperatures, \ce{WC} has a lower value of thermal conductivity when compared with pure \ce{W} (reaching around 2/3 of its value). However, \ce{WC} has a thermal shock coefficient that is 3 times higher, indicating better resistance to thermal shocks, and a Vickers hardness that is higher by a factor of 8 at room temperature to 20 at operating temperature. \ce{WC} shows a good tolerance to neutron irradiation when compared to other carbides in terms of cracking resistance. Regarding the interaction with plasma, \ce{WC} shows slightly higher sputtering rate and hydrogen retention than \ce{W}. \ce{W2C} has been mainly applied as composites compound of both \ce{W} \cite{vsestan2022non} and \ce{WC} \cite{dash2014preparation}. When \ce{W2C} is subject to \ce{He} ion bombardment, contrary to \ce{W}, \ce{He} bubbles do not form inside its crystal grains \cite{vsestan2022non}. These bubbles are the early stages of the fuzz layer. The improved mechanical properties and surface layer stability of tungsten carbides are potentially offset by reduced thermal properties and increased sputtering and hydrogen retention, which has led to ongoing debates over their applicability as substitutes for \ce{W} or as composites with the latter metal.

\paragraph{Nitrides}

The rhombohedral structure of boron nitride (\ce{BN}) is selected according to the $d^{SS}$ ranking, contrary to the cubic phase which is discarded due to the $d^{ELMs}$ ranking. Other two polymorphs of boron nitride are well-known: the hexagonal polytype, discarded during the data cleaning procedure due to discrepancies in reported properties that matched another polymorph according to references, and the wurtzite polytype, not documented in MPDS.
The melting temperatures and heat capacities of the polymorphs are in agreement with the KDE value extracted from MPDS \cite{gavrichev1993low,haynes2016crc}, but significant discrepancies arise in the thermal conductivities. Computed values of isotopically pure crystals of \ce{BN} are around 1,900 \si{\watt\per\meter\per\kelvin} \cite{chakraborty2018lattice,chen2020ultrahigh}, while experimental measurements report values that fall within the range of 150-550 \si{\watt\per\meter\per\kelvin} \cite{yuan2019modulating,corrigan1979thermal,sichel1976heat,duclaux1992structure,buzhinskij1992rhombohedral}. Thermal conductivities of this order of magnitude would justify the potential application of boron nitride as PFM, even at the divertor region, contrary to what the KDE values suggest. Although the effect of neutron bombardment on the thermal conductivity of \ce{BN} is not extensively investigated, thermal annealing shows limited impact on the neutron-induced damage of an irradiated \ce{BN} structure \cite{cataldo2017neutron}, indicating a potential irreversible degradation of its thermal conductivity.
Nonetheless, the chemical and thermal stability of \ce{BN}, its limited physical sputtering, high thermal conductivity, and resistance to thermal shocks, has motivated its application as first-wall protective layer in fusion devices in the past \cite{buzhinskij1990application,buzhinskij1992rhombohedral}. 
The interaction of \ce{BN} with the plasma requires further investigation. On the one hand, cubic and hexagonal \ce{BN} are highly promising for hydrogen barrier applications within the refractory nitrides class, exhibiting a diffusion barrier to hydrogen migration of approximately 3.0 \si{\electronvolt} \cite{bull2022ab}. On the other hand, \ce{BN} shows high hydrogen permeability and solubility, which may lead to significant tritium inventories \cite{causey1992tritium}.
The ability of \ce{BN} to efficiently store hydrogen molecules is linked to specific structural characteristics, such as nanotubes and hexagonal structures with high porosity and structural defects \cite{lale2018boron}. While these features are intentionally pursued for certain applications, they should be avoided when considering \ce{BN} suitability as a PFM.
Given its intriguing thermal properties and the necessity to conduct further investigations under operational conditions, \ce{BN} is included in the list of leading PFM candidates, specifically in its cubic polymorph (c-\ce{BN}).
Recently, a material sharing the same crystal structure as c-\ce{BN}, boron arsenide (\ce{BAs}), has emerged as the second-highest thermally conducting material among all known isotropic crystals, after diamond \cite{kang2018experimental,garg2011high}. The highest experimentally measured thermal conductivity for \ce{BAs} single crystals is 1,300 \si{\watt\per\meter\per\kelvin}. However, the thermal stability of \ce{BAs} is compromised as it decomposes into \ce{B13As2} at 1,200 \si{\kelvin}, releasing arsenic gas under atmospheric pressure \cite{chu1974preparation}. This negates its application as PFM.

Another promising material, but quite unexplored for PFM application, is represented by the  hexagonal, cubic, rhombohedral, orthorhombic, and amorphous boron-carbon nitride \ce{BC2N}. This material posses thermal and mechanical properties that lie between those of carbon diamond and boron nitride (c- and h-BN). Cubic, hexagonal and rhombohedral \ce{BC2N} show impressive Vickers hardness value (greater than c-\ce{BN}) \cite{luo2007body,liu2018hexagonal,sadeghi2020ultra,wang2023superhard}. A thermal conductivity that is almost comparable with carbon diamond is calculated for h-\ce{BC2N} \cite{sadeghi2020ultra} and close to that of c-\ce{BN} is estimated for orthorhombic \ce{BC2N} \cite{ju2021exploring}. The Debye temperature of both hexagonal and orthorhombic \ce{BC2N} is estimated close or even higher than diamond, with values between 1,700-2,000 \si{\kelvin} \cite{chang2010first,li2019elastic}. The heat capacity of the orthorhombic phase is predicted to lie between that of diamond and boron nitride \cite{chang2010first,cheng2008ab}.
It can be assumed that the behavior of this class of compounds under divertor conditions is similar to their well-known precursors, as low-Z materials. However, differences are expected due to their chemical reactivity, and specific studies are necessary to determine their suitability as PFMs. For these reasons, the hexagonal phase h-\ce{BC2N} is added to the list of top PFM candidates.

The \ce{C3N4} compound has been proposed in the past as potential low-Z PFM alternative to graphite and CFCs, and a possible reduction of hydrogen diffusion and retention inside the material was hypothesized \cite{ross2003improvement}.  
Given the difficulties connected with its synthesis \cite{montigaud2000c3n4}, its thermo-mechanical properties are not well-known. Only estimates of its thermal conductivity are reported in literature, which range from 150 to 500 \si{\watt\per\meter\per\kelvin} as a function of the phase considered \cite{ju2021exploring,morelli2002thermal,morelli2006high,zhu2020charting}. Bulk moduli rivaling that of diamond are also predicted for this crystal \cite{morelli2006high,ruan2014elastic}. Given the interesting thermo-mechanical features, the potentially reduced hydrogen inventory (a bottleneck in the application of carbon-based materials), and that synthesis paths to produce microcrystalline and noncrystalline \ce{C3N4} have been proposed \cite{montigaud2000c3n4,uddin2013formation}, the hexagonal structure $\beta$-\ce{C3N4} is also added to the list of top PFM candidates.

Vanadium nitride (\ce{VN}) exhibits high heat capacity and moduli, but its inferior thermal properties, in terms of melting temperature and thermal conductivity \cite{huber2016synthesis}, preclude it from any application as a PFM in a fusion tokamak reactor.

The case of aluminum nitride (\ce{AlN}) can be compared to that of gallium nitride (\ce{GaN}). For both of these materials, scientific literature on their potential use as divertor armor is limited (\ce{AlN}) or absent (\ce{GaN}). One of the biggest limitation in terms of applicability as PFM is dictated by their decomposition temperature in vacuum or at low \ce{N2} level. Group III nitrides start to decompose in vacuum at temperatures that are in the order of one third of their melting temperature, and for \ce{AlN} and \ce{GaN} are calculated to be 1,040 and 850 \si{\kelvin} respectively \cite{ambacher1998growth}. This sets a temperature limit for high-temperature applications that is much lower than the estimate presented here; therefore, they are discarded from the list of promising PFMs.
Despite this, \ce{AlN} has also been proposed as a substrate for bolometric systems, which provide information on the spatial distribution of radiated power from the main plasma and the divertor region \cite{gonzalez2005radiation, gusarov2007situ}. It shows good impermeability to hydrogen isotopes \cite{wang2016performances, cherkez2019deuterium} and low susceptibility of its composition under low-energy H and He bombardment \cite{patino2020exposure}. However, both neutron irradiation and high temperature have a detrimental effect on its thermal conductivity \cite{yano1991thermal, snead2005thermal,watari2003microstructure}. In contrast, \ce{GaN} is mainly investigated for its electrical and optical properties for semiconductor device applications \cite{strite1992gan} and its thermo-mechanical behaviour under plasma and neutron bombardment is mostly unknown.

\ce{HfN} shows thermo-mechanical properties that are not dissimilar from those of \ce{HfC} \cite{wuchina2004designing}, but a lower melting temperature which does not consent its application as primary PFM. However, alloy or composites of this material might be investigated in the future.

An ultra-high temperature ceramics that has only recently been identified as a ultra-high thermal conductive material is the hexagonal $\theta$-phase of \ce{TaN} \cite{kundu2021ultrahigh}. Its astonishing thermal conductivity, estimated between 800-1,000 \si{\watt\per\square\meter\per\kelvin} at room temperature (above 200 \si{\watt\per\square\meter\per\kelvin} at 800 \si{\kelvin}), is just below that of c-\ce{BN}, and more than double than the largest for a metal, silver. This might place $\theta$-phase \ce{TaN} at the top of the thickness ranking methods. To the best of our knowledge, its application as PFM has not been discussed yet, and a careful and critical analysis of its mechanical and thermal properties and their evaluation under plasma and neutron irradiation is thus needed.
As stated in the reference article \cite{kundu2021ultrahigh}, phonons are the main heat carriers for this material, and the high thermal conductivity of the $\theta$-phase \ce{TaN} is caused by a particular combination of factors: (i) weak electron-phonon scattering, (ii) low isotopic mass disorder, (iii) large frequency gap between acoustic and optical phonon modes \cite{garg2011high} that, together with (iv) acoustic bunching, impedes three-phonon processes (which is however dominated by four-phonon scattering).
$\theta$-\ce{TaN} is also added to the list of top PFM candidates.

Thanks to its decent thermal properties, \ce{TiN} is selected by the $d^{ELMs}$ ranking as one of the lowest performing material, but still able to sustain divertor conditions in a thin layer. Consistently, \ce{TiN} was used in a stellarator fusion device as coating layer of multiple plasma-facing components \cite{glazunov2001erosion} and has been recently proposed for DEMO as tritium permeation barrier of the first-wall in a sandwich-like structure \cite{zhang2022fabrication}. It shows the capability to withstand hydrogen plasma shots in IR–T1 tokamak with no significant morphology defects \cite{ghoranneviss2016study} and its permeability to hydrogen is also confirmed computationally \cite{bull2022ab}. Although it may not be considered a favorable choice for a primary plasma-facing material, \ce{TiN} has the potential to be used as a coating layer on plasma-facing components. Therefore, it is included to expand the list of the most promising candidates.

Finally, all the nitrides presented here have in common the material activation issues associated with the presence of nitrogen. The transmutation of nitrogen ($^{14}$N) under neutron irradiation produces long-lived carbon isotopes $^{14}$C \cite{wallner2011production, kolbasov2014some} that possess a time-to-low-level-waste (time-to-LLW) in the order of thousands of years \cite{gilbert2019waste}.

\subsection{Revisited thickness ranking}

The collection of materials obtained through the two-step ranking procedure - thickness (ELMs and SS) and Pareto ($3p$, $2p$, $1p$) rankings - has been qualitatively evaluated with the indication of the comparative ranking, which highlighted potential false negatives, and then with the review of the literature for their application as PFMs. The finalized list of PFM candidates is presented in Tab.~\ref{tab:topPFM}. Their capacity to withstand heat loads is re-evaluated by calculating $d^{SS}$ and $d^{ELMs}$ with heat properties extracted directly from the literature references reported in the table. Both high and low temperature properties are reported, in order to calculate the thickness values at the two temperature extrema. 
Considering the two temperature ranges reported in Tab.~\ref{tab:topPFM}, high-temperature conditions are more representative of the thermal demands that the designated PFMs in the divertor region must withstand. Therefore, the candidates are ranked in descending order of maximum $d^{ELMs}$ at high-temperature. The candidates with a larger maximum thicknesses that can avoid melting are prioritized because their surface temperature can be further decreased by reducing their thickness, while keeping a minimum width that can sustain erosive processes.

Carbon-based materials are ranked at the top of the list, diamond and graphite, followed closely by the ceramic materials \ce{C3N4} and \ce{BC2N}, which are novel in the field of PFM, and the well-known \ce{BN}. Within the metal-based materials, \ce{W} shows the highest performance, followed by \ce{Mo} and three transition-metal borides: \ce{TiB2}, \ce{ZrB2}, and \ce{HfB2}. The extensively researched \ce{WC} \cite{traskelin2007radiation,humphry2019shielding} and the recently investigated $\theta$-\ce{TaN} \cite{kundu2021ultrahigh} exhibit adequate thermal properties, with maximum thicknesses of approximately 4 \si{\milli\meter}. All the other materials possess $d^{ELMs}$ values $\leq$ 2.5 \si{\milli\meter}, significantly lower than the current minimum reported thickness design for a \ce{W} divertor of 6 \si{\milli\meter} \cite{hirai2018design}. Consequently, pure \ce{Ta}, \ce{Re}, \ce{VB2}, \ce{TaC}, \ce{Cr}, \ce{TaW}, \ce{TiN} are likely unsuitable for use as primary PFMs. However, they could also be utilized as coating layers or dopants of the primary PFM. Additionally, they might be incorporated as constituents within alloy systems or composite materials.
Only two materials within the selected set showed $d^{ELMs} \leq$ 0 under both temperature conditions: \ce{W2C} and \ce{HfC}. Although these materials do not exhibit exceptional performance, they are still included in the list. While the former may be considered as a potential contaminant of the high-performing \ce{WC}, the latter represents one component of the two prototype crystals that constitute the aforementioned 4\ce{TaC}-1\ce{HfC} alloy system.

\begin{table*}
\caption{List of the top 21 PFM candidates identified in this study. The four properties required for the calculation of the heat balance equation (Eq.~(\ref{eq2})) are retrieved directly from literature. Thermal conductivity $k$ and heat capacity at constant pressure $C_p$ are reported at both low (around RT) and high temperature. The actual values for low and high temperature may vary depending on the conditions documented in the literature. The maximum thickness to avoid melting in steady-state condition ($d^{SS}$) and during ELMs ($d^{ELMs}$) is calculated at low ($T_L$) and high ($T_H$) temperatures. Computational data are reported only when experimental measures are not available.}
\label{tab:topPFM}
\resizebox{\textwidth}{!}{%

\begin{tabular}{l|cc|ccccc|ccccc|cc|c|cc|cc}\hline
\hline

\multicolumn{1}{c|}{\multirow{2}{*}{\textbf{Formula}}} & $\boldsymbol{T_m}$ & \multirow{2}{*}{\textbf{Ref.}} 
& $\boldsymbol{k(T_L)}$ & $\boldsymbol{T_L}$ & $\boldsymbol{k(T_H)}$ & $\boldsymbol{T_H}$ & \multirow{2}{*}{\textbf{Ref.}} 
& $\boldsymbol{C_p(T_L)}$ & $\boldsymbol{T_L}$ & $\boldsymbol{C_p(T_H)}$ & $\boldsymbol{T_H}$ & \multirow{2}{*}{\textbf{Ref.}} 
& $\boldsymbol{\rho}$ & \multirow{2}{*}{\textbf{Ref.}} 
& \textbf{MM} 
& $\boldsymbol{d^{SS}(T_L)}$ & $\boldsymbol{d^{SS}(T_H)}$ 
& $\boldsymbol{d^{ELMs}(T_L)}$ & $\boldsymbol{d^{ELMs}(T_H)}$\Tstrut \\ 
\multicolumn{1}{c|}{} & (\si{\kelvin}) & 
& (\si{\watt\per\meter\per\kelvin}) & (\si{\kelvin}) & (\si{\watt\per\meter\per\kelvin}) & (\si{\kelvin}) & 
& (\si{\joule\per\mole\per\kelvin}) & (\si{\kelvin}) & (\si{\joule\per\mole\per\kelvin}) & (\si{\kelvin}) & 
& (\si{\gram\per\cubic\centi\meter}) & 
& (\si{\gram\per\mole}) 
& (\si{\milli\meter}) & (\si{\milli\meter}) 
&(\si{\milli\meter}) & (\si{\milli\meter}) \\
\hline

\textbf{\ce{C} (diamond)} & 3,950$^{(1)(2)}$ & \cite{fedoseev1986surface,abrahamson1974graphite} & 2,300 & 300 & 1,540 & 400  & \cite{haynes2016crc}                                          & 6.2 & 300 & 24 & 1,200 & \cite{haynes2016crc}                & 3.51  & \cite{haynes2016crc}           & 12.01  & 406.53 & 272.20 & 361.60 & 253.51 \\
\textbf{\ce{C} (graphite)} & 3,950$^{(2)}$ & \cite{abrahamson1974graphite}                     & 1,950 & 300 & 448  & 1,200 & \cite{haynes2016crc}                                          & 8.6 & 300 & 23 & 1,200 & \cite{haynes2016crc}                & 2.20  & \cite{haynes2016crc}           & 12.01  & 344.66 & 79.18  & 300.28 & 66.17  \\
\textbf{h-\ce{C3N4}}  & 3,500$^{(3)}$ & \cite{ruan2014elastic}                            & 520  & 300 & 520  & 300  & \cite{morelli2002thermal}                                     & 30  & 300 & 80 & 1,200 & \cite{ding2015theoretical}          & 3.53  & \cite{ruan2014elastic}         & 92.06  & 80.21  & 80.21  & 53.39  & 63.78  \\
\textbf{h-\ce{BC2N}}  & 3,500$^{(3)}$ & \cite{cheng2008ab}                                & 2,000 & 300 & 400  & 1,000 & \cite{sadeghi2020ultra}                                       & 28  & 300 & 84 & 900  & \cite{chang2010first}               & 2.90  & \cite{luo2007body}             & 48.84  & 308.50 & 61.70  & 264.74 & 50.40  \\
\textbf{c-\ce{BN}}    & 3,240 & \cite{haynes2016crc}                              & 550  & 300 & 350  & 650  & \cite{corrigan1979thermal}                                    & 20  & 300 & 42 & 1,200 & \cite{solozhenko1993thermodynamics} & 3.49  & \cite{corrigan1979thermal}     & 24.82  & 77.69  & 49.44  & 60.04  & 39.73  \\
\textbf{\ce{W}}       & 3,687 & \cite{haynes2016crc}                              & 174  & 300 & 112  & 1,200 & \cite{haynes2016crc}                                          & 24  & 300 & 28 & 1,200 & \cite{white1984heat}                & 19.26 & \cite{haynes2016crc}           & 183.84 & 28.47  & 18.32  & 17.97  & 10.53  \\
\textbf{\ce{TiB2}}    & 3,193 & \cite{haynes2016crc}                              & 65   & 300 & 70   & 1,300 & \cite{samsonov1973thermophysical}                             & 63  & 300 & 83 & 1,200 & \cite{neuman2021thermal}            & 4.95  & \cite{haynes2016crc}           & 69.49  & 9.03   & 9.72   & 4.23   & 5.38   \\
\textbf{\ce{Mo}}      & 2,895 & \cite{haynes2016crc}                              & 138  & 300 & 96   & 1,200 & \cite{baysinger2015crc,rasor1960thermal}                      & 22  & 300 & 30 & 1,200 & \cite{rasor1960thermal}             & 10.23 & \cite{haynes2016crc}           & 95.94  & 17.11  & 11.90  & 7.44   & 4.99   \\
\textbf{\ce{ZrB2}}    & 3,323 & \cite{haynes2016crc}                              & 58   & 300 & 65   & 1,300 & \cite{samsonov1973thermophysical}                             & 45  & 300 & 75 & 1,200 & \cite{xiang2017first}               & 6.17  & \cite{haynes2016crc}           & 112.85 & 8.43   & 9.45   & 2.31   & 4.43   \\
\textbf{\ce{HfB2}}    & 3,373 & \cite{haynes2016crc}                              & 51   & 300 & 60   & 1,300 & \cite{samsonov1973thermophysical}                             & 50  & 300 & 80 & 1,200 & \cite{xiang2017first}               & 10.50 & \cite{haynes2016crc}           & 200.11 & 7.54   & 8.87   & 1.98   & 4.10   \\
\textbf{$\theta$-\ce{TaN}}     & 3,363 & \cite{haynes2016crc}                              & 70   & 300 & 70   & 300  & \cite{liu2023thermal,kundu2021ultrahigh,lee2023high} & 41  & 300 & 41 & 300  & \cite{tabassum2017correlating}      & 13.70 & \cite{haynes2016crc}           & 194.95 & 10.32   & 10.32   & 4.10   & 4.10   \\
\textbf{\ce{WC}}      & 3,022 & \cite{haynes2016crc}                              & 120  & 300 & 70   & 1,000 & \cite{humphry2019shielding}                                   & 36  & 300 & 52 & 1,000 & \cite{andon1975heat}                & 15.60 & \cite{haynes2016crc}           & 195.85 & 15.64  & 9.12   & 7.48   & 3.94   \\
\textbf{\ce{Ta}}      & 3,290 & \cite{haynes2016crc}                              & 58   & 300 & 60   & 1,200 & \cite{savchenko2008thermal}                                   & 26  & 300 & 28 & 1,200 & \cite{milovsevic1999thermal}        & 16.63 & \cite{haynes2016crc}           & 180.95 & 8.34   & 8.63   & 2.12   & 2.53   \\
\textbf{\ce{Re}}      & 3,458 & \cite{haynes2016crc}                              & 48   & 300 & 45   & 520  & \cite{powell1963thermal}                                      & 23  & 300 & 31 & 1,200 & \cite{taylor1961specific}           & 21.02 & \cite{haynes2016crc}           & 186.21 & 7.30   & 6.85   & 1.88   & 2.32   \\
\textbf{\ce{VB2}}     & 2,723 & \cite{haynes2016crc}                              & 42   & 300 & 42   & 1,300 & \cite{samsonov1973thermophysical}                             & 45  & 300 & 70 & 1,200 & \cite{cao2022theoretical}           & 5.07  & \cite{kravchenko2019nanosized} & 72.56  & 4.85   & 4.85   & 0.24   & 1.15   \\
\textbf{\ce{TaC}}     & 4,153 & \cite{haynes2016crc}                              & 22   & 300 & 22   & 300  & \cite{kieffer1953mischkristallbildung}                        & 30  & 300 & 52 & 1,200 & \cite{frisk1996gibbs}               & 14.65 & \cite{haynes2016crc}           & 192.96 & 4.11   & 4.11   & 0.19   & 1.13   \\
\textbf{\ce{Cr}}      & 2,180 & \cite{haynes2016crc}                              & 94   & 300 & 62   & 1,200 & \cite{haynes2016crc}                                          & 22  & 300 & 32 & 1,200 & \cite{bendick1982heat}              & 7.20  & \cite{haynes2016crc}           & 52.00  & 8.30   & 5.47   & 1.29   & 0.75   \\
\textbf{\ce{TaW}}     & 3,450 & \cite{cahn1991binary}                             & 54   & 673 & 60   & 1,400 & \cite{taylor1971thermophysical,denman1970recent}              & 25  & 300 & 28 & 1,200 &  SWA$^{(4)}$      & 17.78 & \cite{turchi2001first}         & 364.79 & 8.19   & 9.11   & -0.20  & 0.74   \\
\textbf{\ce{TiN}}     & 3,220 & \cite{haynes2016crc}                              & 25   & 500 & 25   & 1,200 & \cite{taylor1964thermal}                                      & 45  & 500 & 55 & 1,200 & \cite{lengauer1995solid}            & 5.21  & \cite{haynes2016crc}           & 61.87  & 3.51   & 3.51   & 0.27   & 0.58   \\
\textbf{\ce{W2C}}     & 3,073 & \cite{haynes2016crc}                              & 30   & 300 & 30   & 300  & \cite{zhang2022interfacial}                                   & 59  & 300 & 77 & 1,000 & \cite{gronvold1988heat}             & 14.80 & \cite{haynes2016crc}           & 379.69 & 3.99   & 3.99   & -0.57  & -0.001 \\
\textbf{\ce{HfC}}     & 3,273 & \cite{haynes2016crc}                              & 30   & 300 & 10   & 1,000 & \cite{jiang2023temperature,lengauer1995solid}       & 37  & 300 & 49 & 1,000 & \cite{lengauer1995solid}            & 12.20 & \cite{haynes2016crc}           & 190.50 & 4.29   & 1.43   & -0.20  & -0.82  \\
\hline
\hline
\end{tabular}
}

\footnotesize
\justify
(1) Diamond $T_{m}$ is the same as graphite due to graphitization process happening above 1,900 \si{\kelvin}. (2) Graphite sublimates at atmospheric pressure. (3) $T_{m}$ of \ce{C3N4} and \ce{BC2N} are defined between those of diamond and boron nitride by looking at the predicted Debye temperatures. (4) SWA indicates that the $C_p$ of \ce{TaW} is calculated as a stoichiometric weighted average of the $C_p$ of its metallic single-element constituents.
\end{table*}

\section{SBE and H-IFE workflows results}\label{dft}

The crystal structures of the 21 PFM candidates are retrieved from MPDS if the selected polymorphs are available; otherwise, they are directly obtained from literature crystallographic data. The crystal structures are then employed as the only input required by the Surface Binding Energy (SBE) and the Hydrogen Interstitial Formation Energy (H-IFE) workflows.

The results of the SBE workflow screening are reported in Tab.~\ref{tab5}. $E_b$ is the SBE of a surface atom, while $E_p(D)$ is threshold energy of physical sputtering, in this case the minimum energy that an impinging deuterium atom (D) must possess to remove the target atom from the surface (see Eq.~\ref{eq10}). For each material, a single value for $\overline{E_b}$ and for $\overline{E_p}(D)$ is calculated by averaging the values of individual elements using stoichiometric weighted average (SWA). The weights for this averaging process are determined from the stoichiometric ratios of the elements present in the material. The PFM candidates are then ranked in descending order of $\overline{E_p}(D)$, which is the most crucial value in assessing the material susceptibility to surface particle sputtering, thereby influencing the erosion rate and the contamination of the plasma core. The highest-performing materials are high-Z transition metals, in particular \ce{Re}, \ce{TaW}, \ce{W}, and \ce{Ta}, followed by their carbide, nitride and boride forms. The lowest $\overline{E_p}(D)$ values are obtained for low-Z transition metal and ceramics.
As shown in Eq.~\ref{eq10}, the mass difference between plasma ions and the target surface atom defines the amount of energy that can be transferred during an elastic collision, which is maximum at equal masses. This implies that, despite the chemistry at the surface, hence the SBE of a given atom, the atomic mass governs its susceptibility to sputtering.
Indeed, even if the highest $\overline{E_b}$ are obtained for \ce{HfC}, \ce{TaC}, and graphite (C-194), low $\overline{E_p}(D)$ are also shown, due to the low atomic mass of the carbon atom. In particular, in the case of \ce{HfC}, \ce{TaC}, and \ce{WC}, the transition metals in the carbide surface display impressive $E_b$ values, higher than in their pure metallic forms.
It is important to emphasize that in this cases preferential sputtering towards the carbon atom would occur, leaving a carbon-depleted surface with modified thermal properties.

\begin{table*}
\caption{Summary of the properties of the 21 PFM candidates, calculated through the SBE workflow. The space group number ($SG$) is reported to define the polymorph selected for each crystal. The surface orientation is chosen by finding the lattice plane with minimum surface energy at a given termination, min$\{E^{hkl,\alpha}_{surf}\}$.
The surface binding energy, $E_b$, and threshold energy of physical sputtering, $E_p(D)$, are calculated for each inequivalent surface atoms of the 21 PFM candidates. The stoichiometric weighted average values are calculated for each material, respectively $\overline{E_b}$ and $\overline{E_p}(D)$. The PFM candidates are reported in descending order of $\overline{E_p}(D)$.}
\label{tab5}
\resizebox{\textwidth}{!}{%
\begin{tabular}{lc|cc|lcc|lcc|lcc|cc}
\hline
\hline
\multirow{2}{*}{\textbf{Formula}} & \multirow{2}{*}{\boldsymbol{$SG$}} & \textbf{min}$\boldsymbol{\{E^{hkl,\alpha}_{surf}\}}$           & \multirow{2}{*}{\textbf{Miller index}} & \multirow{2}{*}{\textbf{element}} & \boldsymbol{$E_b$}                            & \boldsymbol{$E_p(D)$}                         & \multirow{2}{*}{\textbf{element}} & \boldsymbol{$E_b$}                            & \boldsymbol{$E_p(D)$}                         & \multirow{2}{*}{\textbf{element}} & \boldsymbol{$E_b$}                            & \boldsymbol{$E_p(D)$}                         & \boldsymbol{$\overline{E_b}$}                 & \boldsymbol{$\overline{E_p}(D)$}\Tstrut              \\
                                  &                                          & \textbf{\si{\electronvolt\per\square\angstrom}} &                                        &                                   & \multicolumn{1}{c}{\si{\electronvolt}} & \multicolumn{1}{c|}{\si{\electronvolt}} &                                   & \multicolumn{1}{c}{\si{\electronvolt}} & \multicolumn{1}{c|}{\si{\electronvolt}} &                                   & \multicolumn{1}{c}{\si{\electronvolt}} & \multicolumn{1}{c|}{\si{\electronvolt}} & \multicolumn{1}{c}{\si{\electronvolt}} & \multicolumn{1}{c}{\si{\electronvolt}} \\
\hline
\ce{Re}                                & 194                                      & 0.19                                            & $\langle$001$\rangle$                  & Re                                & 11.72                                  & 276.81                                 &                                   &                                        &                                        &                                   &                                        &                                        & 11.72                                  & 276.81                                 \\
\ce{TaW}                               & 221                                      & 0.20                                            & $\langle$110$\rangle$                  & Ta                                & 12.34                                  & 283.44                                 & W                                 & 10.90                                  & 254.28                                 &                                   &                                        &                                        & 11.62                                  & 268.86                                 \\
\ce{W}                                 & 229                                      & 0.22                                            & $\langle$110$\rangle$                  & W                                 & 11.16                                  & 260.38                                 &                                   &                                        &                                        &                                   &                                        &                                        & 11.16                                  & 260.38                                 \\
\ce{Ta}                                & 229                                      & 0.17                                            & $\langle$110$\rangle$                  & Ta                                & 10.92                                  & 250.67                                 &                                   &                                        &                                        &                                   &                                        &                                        & 10.92                                  & 250.67                                 \\
\ce{HfC}                               & 225                                      & 0.12                                            & $\langle$100$\rangle$                  & C                                 & 9.83                                   & 19.98                                  & Hf                                & 16.68                                  & 378.03                                 &                                   &                                        &                                        & 13.26                                  & 199.00                                 \\
\ce{TaC}                               & 225                                      & 0.11                                            & $\langle$100$\rangle$                  & Ta                                & 15.12                                  & 347.25                                 & C                                 & 9.39                                   & 19.10                                  &                                   &                                        &                                        & 12.26                                  & 183.17                                 \\
\ce{W2C}                               & 60                                       & 0.19                                            & $\langle$111$\rangle$                  & W                                 & 11.10                                  & 258.83                                 & C                                 & 9.80                                   & 19.93                                  &                                   &                                        &                                        & 10.67                                  & 179.20                                 \\
\ce{WC}                                & 187                                      & 0.25                                            & $\langle$100$\rangle$                  & C                                 & 7.91                                   & 16.09                                  & W                                 & 12.87                                  & 300.11                                 &                                   &                                        &                                        & 10.39                                  & 158.10                                 \\
\ce{TaN}                               & 187                                      & 0.19                                            & $\langle$100$\rangle$                  & Ta                                & 12.63                                  & 289.98                                 & N                                 & 7.48                                   & 17.01                                  &                                   &                                        &                                        & 10.05                                  & 153.50                                 \\
\ce{Mo}                                & 229                                      & 0.19                                            & $\langle$110$\rangle$                  & Mo                                & 8.66                                   & 107.49                                 &                                   &                                        &                                        &                                   &                                        &                                        & 8.66                                   & 107.49                                 \\
\ce{HfB2}                              & 191                                      & 0.21                                            & $\langle$101$\rangle$                  & Hf                                & 11.21                                  & 253.89                                 & B                                 & 8.12                                   & 15.34                                  &                                   &                                        &                                        & 9.15                                   & 94.86                                  \\
\ce{ZrB2}                              & 191                                      & 0.20                                            & $\langle$101$\rangle$                  & B                                 & 8.03                                   & 15.17                                  & Zr                                & 11.13                                  & 131.63                                 &                                   &                                        &                                        & 9.06                                   & 53.99                                  \\
\ce{TiN}                               & 225                                      & 0.09                                            & $\langle$100$\rangle$                  & N                                 & 8.90                                   & 20.25                                  & Ti                                & 11.32                                  & 73.06                                  &                                   &                                        &                                        & 10.11                                  & 46.65                                  \\
\ce{Cr}                                & 229                                      & 0.22                                            & $\langle$110$\rangle$                  & Cr                                & 6.21                                   & 43.26                                  &                                   &                                        &                                        &                                   &                                        &                                        & 6.21                                   & 43.26                                  \\
\ce{C}                                 & 194                                      & 0.001                                           & $\langle$001$\rangle$                  & C                                 & 16.71                                  & 33.96                                  &                                   &                                        &                                        &                                   &                                        &                                        & 16.71                                  & 33.96                                  \\
\ce{TiB2}                              & 191                                      & 0.22                                            & $\langle$101$\rangle$                  & B                                 & 8.16                                   & 15.40                                  & Ti                                & 10.31                                  & 66.51                                  &                                   &                                        &                                        & 8.87                                   & 32.44                                  \\
\ce{VB2}                               & 191                                      & 0.15                                            & $\langle$101$\rangle$                  & B                                 & 8.01                                   & 15.13                                  & V                                 & 9.20                                   & 62.86                                  &                                   &                                        &                                        & 8.41                                   & 31.04                                  \\
\ce{C}                                 & 227                                      & 0.35                                            & $\langle$110$\rangle$                  & C                                 & 11.23                                  & 22.83                                  &                                   &                                        &                                        &                                   &                                        &                                        & 11.23                                  & 22.83                                  \\
\ce{BC2N}                              & 160                                      & 0.23                                            & $\langle$001$\rangle$                  & C                                 & 12.58                                  & 25.57                                  & N                                 & 7.68                                   & 17.46                                  & B                                 & 10.33                                  & 19.52                                  & 10.79                                  & 22.03                                  \\
\ce{BN}                                & 216                                      & 0.04                                            & $\langle$111$\rangle$                  & B                                 & 12.29                                  & 23.20                                  & N                                 & 8.76                                   & 19.93                                  &                                   &                                        &                                        & 10.53                                  & 21.57                                  \\
\ce{C3N4}                              & 176                                      & 0.04                                            & $\langle$100$\rangle$                  & N                                 & 7.76                                   & 17.66                                  & C                                 & 12.30                                  & 25.01                                  &                                   &                                        &                                        & 9.71                                   & 20.81                                    \\     
\hline
\hline
\end{tabular}
}
\end{table*}

The results of the H-IFE workflow are presented in Tab.~\ref{tab6}, where the hydrogen interstitial formation energies $\Delta E^{f}(H_i)$ of the 21 PFM candidates are reported in descending order. Two values are reported, $\Delta E_a^{f}(H_i)$ and $\Delta E_m^{f}(H_i)$; the first is calculated from a reference hydrogen atom ($a$) as a reservoir, while the second from a reference hydrogen molecule ($m$), which is reported for comparison with literature. Their difference represent the $H_2$ binding energy, which in our calculation (excluding ZPE) is equal to -2.312 \si{\electronvolt}. A decrease of $\Delta E^{f}(H_i)$ leads to an exponential increase of the hydrogen solubility in the crystal, as reported in Eq.~\ref{eq12}; this is an undesirable phenomenon for PFMs, which need to minimize tritium inventory for reasons of safety and fuel efficiency. Only 4 materials exhibit a positive $\Delta E_a^{f}(H_i)$: \ce{BN}, \ce{C} graphite, \ce{BC2N}, and \ce{C3N4}, all of which possess high atomic density. Hence, aside from chemical interactions, there appears to be a correlation between smaller interstitial volumes, where atomic hydrogen must fit, and higher $\Delta E^{f}(H_i)$. The trend provided by $\Delta E^{f}(H_i)$ is only an initial indication that must be further refined by selecting more realistic energy references for hydrogen reservoirs, by including kinetic effects to define the hydrogen permeability inside the bulk of the crystal, and by considering damage and diffusion along pores and grains.

\begin{table}
\caption{Formation energy for hydrogen interstitial of the 21 PFM candidates. $\Delta E_a^{f}(H_i)$ and $\Delta E_m^{f}(H_i)$ are calculated using Eq.~\ref{eq13}, hence not including ZPE effects, taking respectively the hydrogen atom and the hydrogen molecule as reference hydrogen reservoirs. The deviations from known literature values range around $\pm$ 0.1 \si{\electronvolt}. This variance can be attributed to both differences in the size of the supercells (larger in this study compared to most of the cited works), the utilization of different functionals, and the choice of a different \ce{H2} chemical potential.}
\label{tab6}
\centering
\begin{tabular}{lc|ccc}
\hline
\hline
\multicolumn{1}{c}{\multirow{3}{*}{\textbf{Formula}}} & \multicolumn{1}{c}{\multirow{3}{*}{\boldsymbol{$SG$}}} & \multicolumn{1}{c}{\boldsymbol{$\Delta E_a^{f}(H_i)$}}& \multicolumn{2}{c}{\boldsymbol{$\Delta E_m^{f}(H_i)$}}\Tstrut \\
\multicolumn{1}{c}{} &
\multicolumn{1}{c}{} &
\multicolumn{1}{c}{\textbf{(\si{\electronvolt})}} &
\multicolumn{2}{c}{\textbf{(\si{\electronvolt})}} \\
\multicolumn{1}{c}{} &
\multicolumn{1}{c}{} &
This work &
This work &
Ref. \\
\hline
\ce{BN}                       & 216                 & 2.81 & 5.12 & -        \\
\ce{C}                       & 227                 & 2.71  & 5.02 & 4.9 \cite{goss2002theory}      \\
\ce{BC2N}                     & 160                 & 1.45   & 3.76 & -     \\
\ce{C3N4}                     & 176                 & 0.27   & 2.58 & -     \\
\ce{TiN}                     & 225                 & -0.13  & 2.18 & -      \\
\ce{VB2}                      & 191                 & -0.42   & 1.89 & -    \\
\ce{TaC}                      & 225                 & -0.54   & 1.77 & -     \\
\ce{HfB2}                     & 191                 & -0.67   & 1.64 & -     \\
\ce{C}                        & 194                 & -0.94   & 1.37 & -    \\
\ce{ZrB2}                     & 191                 & -1.01   & 1.30 & -    \\
\ce{TiB2}                     & 191                 & -1.09   & 1.22 & -     \\
\ce{TaN}                      & 187                 & -1.50   & 0.81 & -     \\
\ce{W}                        & 229                 & -1.53   & 0.78 & 0.87 \cite{johnson2010hydrogen}    \\
\ce{HfC}                      & 225                 & -1.57   & 0.74 & -     \\
\ce{WC}                       & 187                 & -1.77   & 0.54 & 0.7 \cite{zhang2023first}     \\
\ce{Mo}                       & 229                 & -1.80   & 0.51 & 0.62 \cite{duan2010first}      \\
\ce{Cr}                       & 229                 & -1.81   & 0.51 & -    \\
\ce{W2C}                      & 60                  & -1.88   & 0.43 & -     \\
\ce{Re}                       & 194                 & -2.39   & -0.076 & -     \\
\ce{TaW}                      & 221                 & -2.46    & -0.15 & -0.02 \cite{li2023first}   \\
\ce{Ta}                       & 229                 & -2.82    & -0.51 & -   \\     
\hline
\hline
\end{tabular}
\end{table}

\section{Comprehensive evaluation of PFM}\label{comprehensive}

The results obtained through the revisited thickness ranking, $d^{ELMs}(T_H)$, and the SBE and H-IFE workflows, $\overline{E_p}(D)$ and $\Delta E_a^{f}(H_i)$ respectively, are summarized in Tab.~\ref{tab7}. The 21 PFM candidates selected are ranked according to the $WF_i$ equation (see Sec.~\ref{Pareto}, Eq.~\ref{eq5}). 

The best PFM candidate according to this final ranking is carbon diamond (C-227), which outperforms most of the alternatives in terms of thermal properties and hydrogen interstitial formation energy. However, its low resistance to physical sputtering and consequent hydrogen retention remains a bottleneck for its usage as primary PFM. The energy threshold of physical sputtering is even lower than that of graphite, which is notorious for its low resistance to hydrogen sputtering. The application of diamond and diamond-like carbon as coating layer might be more appealing, as it has been already studied in the past \cite{de2011thermal,hughes2021degradation}; nevertheless, the current literature is missing a realistic estimation of diamond erosion rates under the plasma flux of the divertor region, a parameter that constrains the layer lifetime and defines the tritium co-deposition rate and total tritium inventory. Diamond might also be applied in composite materials, but further research on the effects of degradation processes on the diamond thermal properties and intragranular hydrogen solubility at divertor conditions is needed.
\ce{BN} shows similar capabilities than carbon diamond, despite a lower $d^{ELMs}(T_H)$; thus, quantifying how \ce{BN} thermal properties are affected by neutron-induced defects at high temperatures is crucial to assess its applicability as primary PFM. \ce{BN} low sputtering resistance should be investigated to define its tritium co-deposition rate, but less interaction with hydrogen isotopes is expected than for pure carbon. Swelling and cracking of \ce{BN} can be mostly avoided if isotopically separated $^{11}$B is utilized to synthesize the crystal; however, nitrogen transmutation into radioactive $^{14}$C can pose challenges in disposing the material at the end of its lifecycle.
As expected and reassuringly, \ce{W} is ranked at the top of the PFM candidates, showing the best thermal properties among all metal-based materials and significant sputtering resistance, mainly dictated by its high atomic mass. \ce{W} shows a negative $\Delta E_a^{f}(H_i)$; however, the diffusion of hydrogen in the perfect crystal is limited by the energy migration barrier. Indeed, it has been shown that diffusion through vacancy hydrogen traps dominates over interstitial diffusion below 1,500 \si{\kelvin} \cite{fernandez2015hydrogen}. 
Both \ce{BC2N} and \ce{C3N4} show properties that are comparable with the high-performing \ce{BN}. These materials are understudied for their refractory properties due to difficulties with their synthesis, and little is known about their interaction with the extreme environment of a fusion reactor. Their low threshold energy of physical sputtering should be of primary concern, given the presence of carbon atoms in the crystal structure. Therefore, their response to plasma ion bombardment, coupled with tritium retention assessment, is key in evaluating their potential applicability.
\ce{TaC} and \ce{HfC} have both good threshold energies of physical sputtering and reasonable hydrogen solubility compared to the other PFM candidates. However, a bottleneck is presented by their thermal properties: $d^{ELMs}(T_H)$ is slightly above 1 \si{\milli\meter} for \ce{TaC} and even becomes negative for \ce{HfC}. 
Exploring their potential as primary PFMs should then involve their alloy system 4\ce{TaC}-1\ce{HfC}. While availability of thermal properties for this alloy is limited, it has an estimated $d^{ELMs}(T_H)$ of about 5 \si{\milli\meter}, as shown in Sec.~\ref{ceramics}.
\ce{Re} and \ce{Ta} exhibit comparable performance, a high resistance towards physical sputtering due to their high atomic mass, reasonable thermal properties and very high hydrogen solubility. Their use as primary PFMs is limited by a $d^{ELMs}(T_H)$ around 2 \si{\milli\meter}, that might severely restrict their lifetime under erosion conditions. Furthermore, high tritium intake can be excepted given the strongly negative $\Delta E_a^{f}(H_i)$.
\ce{TaW} is the second-best candidate in terms of $\overline{E_p}(D)$, surpassed only by \ce{Re}. Its thermal properties seems to limit its applicability as primary PFM; however, most of the literature sources are based on experimental measures on alloys of various compositions, and reliable data on the ordered crystal are still lacking. \ce{TaW} shows some of the highest potential in terms of hydrogen solubility.
Graphite's $d^{ELMs}(T_H)$ is second only to carbon diamond and its resistance to physical sputtering is the highest among all carbon-based materials, with hydrogen finding its lowest-energy interstitial site when adsorbed on top of a carbon site between two hexagonal planes. The interlayer space can accommodate interstitial hydrogen, in particular when the hexagonal structure is damaged. Despite that, the primary processes governing hydrogen migration into the graphite bulk involve intergranular diffusion on pore surfaces and penetration through grain boundaries.
In the latter section of the ranking, a limited set of materials lacks outstanding performance, but their appealing thermal characteristics render them worthy of attention. \ce{TaN}, \ce{HfB2}, \ce{WC}, \ce{Mo}, \ce{ZrB2}, and \ce{TiB2} have $d^{ELMs}(T_H)$ values around 4 \si{\milli\meter}, and $\overline{E_p}(D)$ equal or higher than graphite. Evaluating their properties under divertor conditions would be of great relevance, since most of have not been tested as primary PFMs, except for \ce{WC} and \ce{Mo}, which are well-studied materials considered as alternatives to pure \ce{W} divertors.
Finally, due to the low $d^{ELMs}(T_H)$, combined with mediocre sputtering and hydrogen solubility resistances, \ce{W2C}, \ce{TiN}, \ce{VB2} and, \ce{Cr} are unlikely to be employed as primary PFMs, protective layers, or composite materials, given that superior alternatives are available.

\begin{table}
\caption{The selected 21 PFM candidates are ranked accordingly to the $WF_i$ equation applied on the three properties calculated in this study: (i) maximum thickness to avoid melting during ELMs, $d^{ELMs}(T_H)$; (ii) threshold energy of physical sputtering, $\overline{E_p}(D)$; (iii) hydrogen interstitial formation energy, $\Delta E_a^{f}(H_i)$. The Pareto layer number $l$ of each material is also reported, showing consistency with the $WF_i$ ranking method.}
\label{tab7}
\centering
\begin{tabular}{cc|lc|ccc}
\hline
\hline
\textbf{$WF_i$}   & \multirow{2}{*}{\textbf{$l$}} & \multirow{2}{*}{\textbf{Formula}} & \multirow{2}{*}{\boldsymbol{$SG$}} & $d^{ELMs}(T_H)$  & \boldsymbol{$\overline{E_p}(D)$}    & \boldsymbol{$\Delta E_a^{f}(H_i)$}  \Tstrut \\
(\%) &                        &                          &                     & (\si{\milli\meter})   & (\si{\electronvolt})   & (\si{\electronvolt})  \\
\hline
86.3 & 1                      & \ce{C}                        & 227                 & 253.51 & 22.83  & 2.71  \\
69.6 & 1                      & \ce{BN}                       & 216                 & 39.73  & 21.57  & 2.81  \\
64.8 & 1                      & \ce{W}                        & 229                 & 10.53  & 260.38 & -1.53 \\
63.1 & 2                      & \ce{BC2N}                     & 160                 & 50.40   & 22.03  & 1.45  \\
60.3 & 1                      & \ce{TaC}                      & 225                 & 1.13   & 183.17 & -0.54 \\
55.9 & 2                      & \ce{C3N4}                     & 176                 & 63.78  & 20.81  & 0.27  \\
54.4 & 1                      & \ce{Re}                       & 194                 & 2.32   & 276.81 & -2.39 \\
51.7 & 2                      & \ce{TaW}                      & 221                 & 0.74   & 267.35 & -2.46 \\
48.5 & 2                      & \ce{HfC}                      & 225                 & -0.82  & 199.00    & -1.57 \\
45.8 & 2                      & \ce{Ta}                       & 229                 & 2.53   & 250.67 & -2.82 \\
45.7 & 1                      & \ce{C}                        & 194                 & 66.17  & 33.96  & -0.94 \\
39.8 & 1                      & \ce{TaN}                      & 187                 & 4.10    & 153.33 & -1.50  \\
39.3 & 2                      & \ce{W2C}                      & 60                  & -0.001 & 177.19 & -1.88 \\
37.6 & 1                      & \ce{HfB2}                     & 191                 & 4.10    & 94.85  & -0.67 \\
36.8 & 2                      & \ce{WC}                       & 187                 & 3.94   & 156.59 & -1.77 \\
33.8 & 1                      & \ce{TiN}                      & 225                 & 0.58   & 46.66  & -0.13 \\
25.6 & 1                      & \ce{VB2}                      & 191                 & 1.15   & 31.04  & -0.42 \\
23.7 & 2                      & \ce{Mo}                       & 229                 & 4.99   & 107.49 & -1.80  \\
21.3 & 1                      & \ce{ZrB2}                     & 191                 & 4.43   & 53.94  & -1.01 \\
15.8 & 2                      & \ce{TiB2}                     & 191                 & 5.38   & 32.44  & -1.09 \\
6.3  & 3                      & \ce{Cr}                       & 229                 & 0.75   & 43.26  & -1.81        \\     
\hline
\hline
\end{tabular}
\end{table}

\section{Conclusions and Outlook} \label{conclusions}

The goal of finding divertor PFMs alternative to \ce{W} has been assessed by defining a target dataset of crystal structures and properties, the inorganic crystal database
MPDS \cite{mpds,paulingfile}, and a set of physical constraints that every candidate needs to satisfy.
Three main loads affect the PFM of the divertor - heat, plasma particles, and neutrons - and following for a few milliseconds these interactions is computationally prohibitive even for a single material. In this work we first focused on the heat balance equation that the PFM is subject to during steady-state conditions and ELM events. Three experimentally measurable thermal properties are used ($T_{melt}$, $k$, and $C_p$), and when all of these are available on MPDS, the heat balance equation is used via the thickness ranking procedure to discriminate between materials that melts or not under divertor heat loads, discarding the latter. When the heat equation could not be solved due to the lack of measure properties, materials were ranked accordingly to the Pareto ranking method on the subset of the available thermal proprieties.
71 PFM candidates have been selected at this stage, and all of them have gone through a selective literature review of their properties under relevant operational conditions. This has led to a further screening of the candidates, which, following the use of the comparative ranking criteria to test for false negatives (i.e.,~materials erroneously discarded) results in a total of 21 potential PFMs. The crystal structures of these materials are used as single input of the two first-principles workflows developed to calculate the candidate surface biding energy (SBE) and the hydrogen-interstitial formation energy (H-IFE). These two properties are used as proxies for two undesired phenomena produced by the plasma bombardment of exposed PFM: (i) the erosion of the surface due to sputtering; (ii) the inventory of tritium inside the material due to hydrogen isotopes diffusion. The materials selected via the screening process together with the properties calculated in this study are summarized in Tab.~\ref{tab7}. Well-known and already proposed PFMs appear in this list, such as tungsten in its pure metallic (\ce{W}) and carbide forms (\ce{WC} and \ce{W2C}), diamond and graphite, boron nitride, and other transition metals, such as \ce{Mo}, \ce{Ta} and \ce{Re}. Other less investigated refractory materials show interesting characteristic that are worth further analysis, such as \ce{TaN} in its $\theta$-phase, \ce{HfB2}, \ce{ZrB2}, and \ce{TiB2}. These do not excel in terms of threshold energy for physical sputtering or hydrogen-interstitial formation energy, but they are still acceptable when compared with those of graphite and tungsten. Especially, their good thermal properties make these appealing as primary PFMs, in particular since the recent possibility to synthesize borides with isotopically separated $^{11}$B, which avoids the boron transmutation reaction. Novel PFMs are also reported, in particular \ce{BC2N}, \ce{C3N4}, \ce{TaW} in its ordered cubic structure, and 4\ce{TaC}-1\ce{HfC}, which, for simplicity, is examined in this study in terms of its carbide components. To the best of the authors' knowledge, there is no literature investigation of the thermal properties of these materials at high temperature under plasma and neutron bombardment. Some limitation come from challenging synthesizability, especially with the goal of obtaining homogeneous phases. In this context, computational studies may offer valuable insights by predicting their thermal properties across different conditions.

This work also highlights the importance of considering the engineering constraints that a PFM in the divertor region needs to satisfy; the combination of the heat, plasma, and neutron loads narrows down the material space available for potential candidates.
The three thermal properties - melting temperature, thermal conductivity, and heat capacity - of a new PFM candidate must always be measured or computed, if the aim is to prevent melting and thermal shock damages. When these constraints are accounted for, many recently proposed W alternatives are excluded from consideration as divertor PFMs. In addition, their evolution during degradation processes induced by plasma and neutron bombardment, as well as changes in composition resulting from elemental transmutation, should be evaluated.
Last, the creation of the two semi-automated workflows is also meant as a proof of concept for the application of the proposed protocol to large high-throughput studies of materials, since the calculated properties have broader utility beyond this specific domain of application.

\section{Acknowledgements}

This work has been carried out within the framework of the EUROfusion Consortium, via the Euratom Research and Training Programme (Grant Agreement No 101052200 - EUROfusion) and funded by the Swiss State Secretariat for Education, Research and Innovation (SERI). Views and opinions expressed are however those of the author(s) only and do not necessarily reflect those of the European Union, the European Commission, or SERI. Neither the European Union nor the European Commission nor SERI can be held responsible for them.
N.M. acknowledges support from NCCR MARVEL, the National Centre of Competence in Research on Computational Design and Discovery of Novel Materials, funded by the Swiss National Science Foundation (Grant No. 205602).

\FloatBarrier

\appendix
\section*{Appendix}

\begin{table*}
\caption{
Ranking of the top 20 PFM candidates based on the minimum thickness ($d^{ELMs}$) required to prevent surface melting under divertor operation conditions during type-I ELMs. The KDE (Kernel Density Estimated) and SD (Standard Deviation) columns provide information on the estimated value and variability of each property in the database. SD values are expressed as a percentage of the KDE value. The calculation of $d^{ELMs}$ is performed using Eq.~(\ref{eq2}) and the input data presented in Tab.~\ref{tab_par}, while its SD is obtained through error propagation.}
\label{tab:d_rank}
\resizebox{\textwidth}{!}{%

\begin{tabular}{l|cc|cc|cc|cc|cc|cc|cc|cc|cc|cc}
\hline
\hline
\textbf{Chemical formula}    & 
\multicolumn{2}{c|}{\textbf{\begin{tabular}[c]{@{}c@{}} $ \boldsymbol{T_m}$\\ (\si{\kelvin})\end{tabular}}} & 
\multicolumn{2}{c|}{\textbf{\begin{tabular}[c]{@{}c@{}}$\boldsymbol{G}$\\ (\si{\giga\pascal})\end{tabular}}} & 
\multicolumn{2}{c|}{\textbf{\begin{tabular}[c]{@{}c@{}}$\boldsymbol{B_S}$\\ (\si{\giga\pascal})\end{tabular}}} & 
\multicolumn{2}{c|}{\textbf{\begin{tabular}[c]{@{}c@{}}$\boldsymbol{B_T}$\\ (\si{\giga\pascal})\end{tabular}}} & 
\multicolumn{2}{c|}{\textbf{\begin{tabular}[c]{@{}c@{}}$\boldsymbol{E}$\\ (\si{\giga\pascal})\end{tabular}}} & 
\multicolumn{2}{c|}{\textbf{\begin{tabular}[c]{@{}c@{}}$\boldsymbol{KH}$\\ (\si{\giga\pascal})\end{tabular}}} & 
\multicolumn{2}{c|}{\textbf{\begin{tabular}[c]{@{}c@{}}$\boldsymbol{MH}$\\ (-)\end{tabular}}} &
\multicolumn{2}{c|}{\textbf{\begin{tabular}[c]{@{}c@{}}$\boldsymbol{k}$\\ (\si{\watt\per\kelvin\per\meter})\end{tabular}}} & 
\multicolumn{2}{c|}{\textbf{\begin{tabular}[c]{@{}c@{}}$\boldsymbol{C_p \cdot \rho}$\\ (\si{\joule\per\kelvin\per\cubic\centi\meter})\end{tabular}}} &
\multicolumn{2}{c}{\textbf{\begin{tabular}[c]{@{}c@{}}$\boldsymbol{d^{ELMs}}$\\ (\si{\milli\meter})\end{tabular}}} \\
\cline{2-21}
& kde                                   & SD (\%)                                  & kde                                   & SD (\%)                                & kde                                  & SD (\%)                                  & kde                                    & SD (\%)                                & kde                                   & SD (\%)                                & kde                                  & SD (\%)                                  & kde                                 & SD (\%)                                 & kde                                     & SD (\%)                                  & kde                                      & SD (\%)                                 & kde                                      & SD (\%)                                                                   \\
\hline
\textbf{C dia} & 4,712.1 & 17.9 & 96.5  & 332.4 &       &          & 443.0   & 7.8  & 1,186.0  & 0.0    & 68.6 & 0.0    & 10.0 & 52.7 & 1,100.2 & 42.8  & 1.8 & 122.5 & 205.0 & 52.2 \\
\textbf{W rt}  & 3,687.1 & 16.9 & 160.2 & 23.7  & 297.0   & 1.9      & 310.7 & 6.1  & 410.1 & 0.0    &      &      &    &      & 174.0    & 15.8  & 2.5 & 25.0    & 18.0  & 37.2 \\
\textbf{Mo rt} & 2,895.5 & 11.9 & 126.5 & 0.0     & 285.0   & 5.7      & 260.0   & 4.1  & 326.9 & 0.0    &      &      &    &      & 138.0    & 0.0     & 2.6 & 26.3  & 7.9 & 34.2 \\
\textbf{Cu}    & 1,357.1 & 10.4 &       &       & 137.0   & 24.3     & 141.9 & 19.5 &       &      &      &      &    &      & 401.0    & 5.6   & 5.2 & 11.0    & 7.9 & 38.0 \\
\textbf{Ir}    & 2,718.9 & 6.3  &       &       &       &          & 385.0   & 13.9 &       &      &      &      &    &      & 138.8  & 3.5   & 2.9 & 0.2   & 7.3 & 17.8 \\
\textbf{Os rt} & 3,307.7 & 9.5  & 320.0   & 0.0     &       &          & 389.8 & 12.4 & 800.0   & 0.0   &      &      &    &      & 86.7   & 54.7  & 2.9 & 0.0     & 5.7 & 91.2 \\
\textbf{AlN}   & 3,286.9 & 13.3 & 3.5   & 0.0     & 202.0   & 0.0        & 207.5 & 8.3  & 353.0   & 0.0    & 12.0   & 0.0    &    &      & 82.4   & 203.9 & 2.4 & 0.0     & 4.4 & 379.5 \\
\textbf{Ru rt} & 2,606.0   & 6.4  & 199.0   & 0.3   &       &          & 314.9 & 8.2  &       &      &      &      &    &      & 103.0    & 11.8  & 2.9 & 6.5   & 3.8 & 34.2 \\
\textbf{Rh}    & 2,236.2 & 6.3  &       &       &       &          & 276.6 & 6.8  &       &      &      &      &    &      & 132.5  & 19.4  & 3.0   & 0.2   & 3.7 & 48.6 \\
\textbf{\ce{ZrB2}}           & 3,323.0   & 0.6  & 220.0   & 4.1   &       &          & 216.0   & 19.5 & 493.5 & 5.8  & 27.6 & 22.7 &    &      & 64.4   & 65.4  & 2.9 & 7.7   & 3.4 & 123.5 \\
\textbf{\ce{TiB2} rt}        & 3,191.4 & 4.1  & 226.7 & 37.8  &       &          & 252.3 & 21.8 & 500.5 & 31.2 & 33.1 & 9.1  &    &      & 64.5   & 49.2  & 3.3 & 0.0     & 3.4 & 91.2 \\
\textbf{\ce{HfB2}}           & 3,373.1 & 3.1  & 224.7 & 4.0     &       &          & 200.0   & 18.2 & 503.1 & 3.5  & 22.2 & 6.0    &    &      & 51.0     & 99.4  & 2.8 & 7.8   & 2.1 & 228.6 \\
\textbf{Re}             & 3,455.1 & 14.9 & 179.0   & 1.3   & 365.2 & 0.0        & 364.0   & 21.5 & 454.0   & 0.0    &      &      &    &      & 47.9   & 3.8   & 2.9 & 0.0     & 2.1 & 57.1 \\
\textbf{Ta}             & 3,279.3 & 26.4 &       &       & 190.0   & 0.0        & 193.6 & 4.2  & 181.0   & 14.1 &      &      &    &      & 54.0     & 4.6   & 2.3 & 13.0    & 1.7 & 141.2 \\
\textbf{TaC}            & 4,153.1 & 1.3  & 200.6 & 10.7  & 265.0   & 5.3      & 344.0   & 4.3  & 424.8 & 28.8 & 22.1 & 10.5 &    &      & 22.2   & 0.0     & 5.0   & 11.1  & 1.5 & 13.3 \\
\textbf{Cr rt}          & 2,179.9 & 4.0    &       &       & 157.0   & 0.0        & 189.4 & 24.5 &       &      &      &      &    &      & 93.4   & 18.2  & 3.2 & 30.1  & 1.5 & 93.3 \\
\textbf{GaN rt}         & 2,782.6 & 26.6 &       &       & 170.0   & 8.3      & 251.7 & 12.3 &       &      &      &      &    &      & 65.7   & 111.7 & 2.5 & 0.0     & 1.3 & 430.8 \\
\textbf{Ag}             & 1,234.6 & 28.1 & 28.2  & 7.0     &     & & 101.0   & 19.8 &       &      &      &      &    &      & 428.6  & 5.8   & 2.5 & 2.4   & 0.9 & 822.2 \\
\textbf{TiN ht}         & 3,220.0   & 1.9  & 196.7 & 57.7  &       &          & 271.4 & 35.5 & 473.6 & 47.7 & 17.4 & 13.2 & 9.0  & 0.0    & 29.0     & 7.3   & 4.5 & 3.7   & 0.9 & 22.2 \\
\textbf{HfC}            & 4,188.1 & 12.6 & 192.0   & 0.0     &       &          & 254.0   & 3.1  & 461.0   & 0.0    & 25.1 & 10.1 &    &      & 20.0     & 0.0     & 2.4 & 0.0     & 0.1 & 500.0 \\
\hline
\hline
\end{tabular}
}
\end{table*}

\begin{table*}
\caption{Ranking of the top 13 PFM candidates based on the minimum thickness ($d^{SS}$) required to prevent surface melting under divertor operation conditions in steady-state. The KDE (Kernel Density Estimated) and SD (Standard Deviation) columns provide information on the estimated value and variability of each property in the database. SD values are expressed as a percentage of the KDE value. The calculation of $d^{SS}$ is performed using Eq.~(\ref{eq1.1})and the input data presented in Tab.~\ref{tab_par}, while its SD is obtained through error propagation.}
\label{tab:d_rank2}
\resizebox{\textwidth}{!}{%

\begin{tabular}{l|cc|cc|cc|cc|cc|cc|cc|cc|cc}
\hline
\hline
\textbf{Chemical formula}    & 
\multicolumn{2}{c|}{\textbf{\begin{tabular}[c]{@{}c@{}} $ \boldsymbol{T_m}$\\ (\si{\kelvin})\end{tabular}}} & 
\multicolumn{2}{c|}{\textbf{\begin{tabular}[c]{@{}c@{}}$\boldsymbol{G}$\\ (\si{\giga\pascal})\end{tabular}}} & 
\multicolumn{2}{c|}{\textbf{\begin{tabular}[c]{@{}c@{}}$\boldsymbol{B_S}$\\ (\si{\giga\pascal})\end{tabular}}} & 
\multicolumn{2}{c|}{\textbf{\begin{tabular}[c]{@{}c@{}}$\boldsymbol{B_T}$\\ (\si{\giga\pascal})\end{tabular}}} & 
\multicolumn{2}{c|}{\textbf{\begin{tabular}[c]{@{}c@{}}$\boldsymbol{E}$\\ (\si{\giga\pascal})\end{tabular}}} & 
\multicolumn{2}{c|}{\textbf{\begin{tabular}[c]{@{}c@{}}$\boldsymbol{KH}$\\ (\si{\giga\pascal})\end{tabular}}} & 
\multicolumn{2}{c|}{\textbf{\begin{tabular}[c]{@{}c@{}}$\boldsymbol{MH}$\\ (-)\end{tabular}}} &
\multicolumn{2}{c|}{\textbf{\begin{tabular}[c]{@{}c@{}}$\boldsymbol{k}$\\ (\si{\watt\per\kelvin\per\meter})\end{tabular}}}& 
\multicolumn{2}{c}{\textbf{\begin{tabular}[c]{@{}c@{}}$\boldsymbol{d^{SS}}$\\ (\si{\milli\meter})\end{tabular}}} \\
\cline{2-19}
& kde                                   & SD (\%)                                  & kde                                   & SD (\%)                                & kde                                  & SD (\%)                                  & kde                                    & SD (\%)                                & kde                                   & SD (\%)                                & kde                                  & SD (\%)                                  & kde                                 & SD (\%)                                 & kde                                     & SD (\%)                                  & kde                                       & SD (\%)                                              \\
\hline
\textbf{\ce{AlP}}     & 2,100.1   & 24.0   &     &   &  &  & 86.5   & 4.2 &       &      &    &   & 5.5     & 0.0  & 92.0   & 0.0    & 7.8 & 29.9    \\
\textbf{\ce{AlAs}}     & 2,013.0   & 0.0   &     &   & 72.0 & 0.0 & 74.0    & 3.2 &       &      &    &   &     &   & 84.0   & 0.0    & 6.7 & 0.0    \\
\textbf{\ce{GaP} rt}   & 1,740.0   & 3.4 &     &   &    &   & 86.9  & 8.3 & 110.9 & 27.8 &    &   &     &   & 75.2 & 0.0    & 5.0   & 4.5  \\
\textbf{\ce{VB2}}      & 2,722.9 & 6.5 & 241.0 & 0.0 &    &   & 281.9 & 7.5 & 340.1 & 46.4 & 24.0 & 0.0 &     &   & 42.1 & 34.8 & 4.9 & 35.6 \\
\textbf{\ce{ZnS} ht}   & 2,100.1 & 2.6 &     &   &    &   & 78.5  & 0.0   &       &      &    &   &     &   & 46.0   & 0.0    & 3.9 & 3.2  \\
\textbf{\ce{W2C} orth} & 3,003.0   & 0.0  &     &   &    &   &       &     &       &      &    &   &     &   & 29.3 & 0.0    & 3.8 & 0.0    \\
\textbf{\ce{CrB2}}     & 2,473.0   & 0.0   &     &   &    &   & 329.0   & 0.0   & 216.0   & 0.0    &    &   &     &   & 34.0   & 37.7 & 3.5 & 37.7 \\
\textbf{\ce{Be13Zr}}   & 2,200.0   & 0.0   &     &   &    &   &       &     &       &      &    &   &     &   & 36.3 & 8.7  & 3.2 & 8.7  \\
\textbf{\ce{BN} rhom}  & 3,240.0   & 0.0   &     &   &    &   & 33.4  & 0.0   &       &      &    &   &     &   & 22.4 & 34.8 & 3.2 & 34.8 \\
\textbf{\ce{Be17Ta2}}  & 2,261.0   & 0.0   &     &   &    &   &       &     &       &      &    &   &     &   & 33.7 & 13.9 & 3.1 & 13.9 \\
\textbf{\ce{WSi2}}     & 2,438.0   & 0.5 &     &   &    &   & 240.0   & 0.0   & 110.1 & 122.0  &    &   &     &   & 30.3 & 3.6  & 3.1 & 3.6  \\
\textbf{\ce{MoSi2} rt} & 2,273.0   & 1.4 &     &   &    &   & 230.0   & 0.0   & 402.6 & 34.1 &    &   &     &   & 32.7 & 39.6 & 3.0   & 39.6 \\
\textbf{\ce{DyIn3}}    & 1,423.0   & 0.0   &     &   &    &   &       &     &       &      &    &   &     &   & 60.0   & 0.0    & 3.0   & 0.0   \\     
\hline
\hline
\end{tabular}
}
\end{table*}

\begin{table*}
\caption{
Top 6 PFM candidates based on the Pareto ranking procedure applied to the $3p$ subset of materials, where only materials with at least one property per category are classified (see Tab.~\ref{tab_category}). The procedure criteria are summarized in Tab.~\ref{tab_pareto}. The KDE (Kernel Density Estimated) and SD (Standard Deviation) columns provide information on the estimated value and variability of each property in the database. SD values are expressed as a percentage of the KDE value. Materials undergoing the thickness rankings are excluded from this procedure.}
\label{tab2}
\resizebox{\textwidth}{!}{%

\begin{tabular}{l|cc|cc|cc|cc|cc|cc|cc|cc|cc|cc}
\hline
\hline
\textbf{Chemical formula}    & 
\multicolumn{2}{c|}{\textbf{\begin{tabular}[c]{@{}c@{}} $ \boldsymbol{T_m}$\\ (\si{\kelvin})\end{tabular}}} & 
\multicolumn{2}{c|}{\textbf{\begin{tabular}[c]{@{}c@{}}$\boldsymbol{G}$\\ (\si{\giga\pascal})\end{tabular}}} & 
\multicolumn{2}{c|}{\textbf{\begin{tabular}[c]{@{}c@{}}$\boldsymbol{B_S}$\\ (\si{\giga\pascal})\end{tabular}}} & 
\multicolumn{2}{c|}{\textbf{\begin{tabular}[c]{@{}c@{}}$\boldsymbol{B_T}$\\ (\si{\giga\pascal})\end{tabular}}} & 
\multicolumn{2}{c|}{\textbf{\begin{tabular}[c]{@{}c@{}}$\boldsymbol{E}$\\ (\si{\giga\pascal})\end{tabular}}} & 
\multicolumn{2}{c|}{\textbf{\begin{tabular}[c]{@{}c@{}}$\boldsymbol{KH}$\\ (\si{\giga\pascal})\end{tabular}}} & 
\multicolumn{2}{c|}{\textbf{\begin{tabular}[c]{@{}c@{}}$\boldsymbol{MH}$\\ (-)\end{tabular}}} &
\multicolumn{2}{c|}{\textbf{\begin{tabular}[c]{@{}c@{}}$\boldsymbol{k}$\\ (\si{\watt\per\kelvin\per\meter})\end{tabular}}} & 
\multicolumn{2}{c|}{\textbf{\begin{tabular}[c]{@{}c@{}}$\boldsymbol{C_p \cdot \rho}$\\ (\si{\joule\per\kelvin\per\cubic\centi\meter})\end{tabular}}} & 
\multicolumn{2}{c}{\textbf{\begin{tabular}[c]{@{}c@{}}$\boldsymbol{C_v \cdot \rho}$\\ (\si{\joule\per\kelvin\per\cubic\centi\meter})\end{tabular}}} \\
\cline{2-21}
& kde                                   & SD (\%)                                  & kde                                   & SD (\%)                                & kde                                  & SD (\%)                                  & kde                                    & SD (\%)                                & kde                                   & SD (\%)                                & kde                                  & SD (\%)                                  & kde                                 & SD (\%)                                 & kde                                     & SD (\%)                                  & kde                                       & SD (\%)                                      & kde                                       & SD (\%)                                      \\
\hline
\textbf{\ce{Ba_{0.3}La_{0.7}MnO3 rt}} &                                       &                                          & 46.4                                 & 0.0                                     &                                      &                                          &                                      &                                          & 85.3                                 & 11.5                                    &                                      &                                          &                                     &                                         & 5.4                                    & 0.0                                       & 4.3                                       & 8.2                                          &                                           &                                              \\   
\textbf{\ce{SrVO3 cub}}         &                                       &                                          & 90.0                                 & 0.0                                     &                                      &                                          &                                      &                                          & 224.0                                & 0.0                                     &                                      &                                          &                                     &                                         & 12.8                                   & 0.0                                       & 4.3                                       & 0.0                                          &                                           &                                              \\             
\textbf{\ce{Ta_{0.5}W_{0.5} ht}}      &                                       &                                          &                                      &                                         & 151.0                                & 0.0                                      & 255.9                                & 4.0                                      &                                      &                                         &                                      &                                          &                                     &                                         & 65.0                                   & 7.4                                       & 3.0                                       & 19.2                                         &                                           &                                              \\            
\textbf{\ce{TiCoSb}}            &                                       &                                          & 91.0                                 & 0.0                                     &                                      &                                          & 142.0                                & 0.0                                      & 224.0                                & 0.0                                     &                                      &                                          &                                     &                                         & 16.6                                   & 43.2                                      & 3.5                                       & 23.6                                         &                                           &                                              \\  
\textbf{\ce{Y_{0.2}Zr_{0.8}O_{1.9} ht2
}}          &                                       &                                          & 105.0                                & 0.0                                     &                                      &                                          &                                 &                                    & 257.0                                & 16.6                                     &                                      &                                          &                                     &                                         & 2.0                                   & 5.3                                      & 3.2                                       & 5.6                                          &                                           &                                              \\ 
\textbf{\ce{Zr2Al3C4}}          &                                       &                                          & 161.0                                & 0.0                                     &                                      &                                          & 196.0                                & 0.0                                      & 379.0                                & 0.0                                     &                                      &                                          &                                     &                                         & 8.8                                   & 52.2                                       & 3.6                                       & 0.0                                          &                                           &                                              \\      
\hline
\hline
\end{tabular}
}
\end{table*}

\begin{table*}
\caption{
Top 6 PFM candidates based on the Pareto ranking procedure applied to the $2p$ subset of materials, where only materials with at least one property in two different categories are classified (see Tab.~\ref{tab_category}). The procedure criteria are summarized in Tab.~\ref{tab_pareto}. The KDE (Kernel Density Estimated) and SD (Standard Deviation) columns provide information on the estimated value and variability of each property in the database. SD values are expressed as a percentage of the KDE value. Materials undergoing the thickness and $3p$ Pareto rankings are excluded from this procedure.}
\label{tab3}
\resizebox{\textwidth}{!}{%

\begin{tabular}{l|cc|cc|cc|cc|cc|cc|cc|cc|cc|cc}
\hline
\hline
\textbf{Chemical formula}    & 
\multicolumn{2}{c|}{\textbf{\begin{tabular}[c]{@{}c@{}} $ \boldsymbol{T_m}$\\ (\si{\kelvin})\end{tabular}}} & 
\multicolumn{2}{c|}{\textbf{\begin{tabular}[c]{@{}c@{}}$\boldsymbol{G}$\\ (\si{\giga\pascal})\end{tabular}}} & 
\multicolumn{2}{c|}{\textbf{\begin{tabular}[c]{@{}c@{}}$\boldsymbol{B_S}$\\ (\si{\giga\pascal})\end{tabular}}} & 
\multicolumn{2}{c|}{\textbf{\begin{tabular}[c]{@{}c@{}}$\boldsymbol{B_T}$\\ (\si{\giga\pascal})\end{tabular}}} & 
\multicolumn{2}{c|}{\textbf{\begin{tabular}[c]{@{}c@{}}$\boldsymbol{E}$\\ (\si{\giga\pascal})\end{tabular}}} & 
\multicolumn{2}{c|}{\textbf{\begin{tabular}[c]{@{}c@{}}$\boldsymbol{KH}$\\ (\si{\giga\pascal})\end{tabular}}} & 
\multicolumn{2}{c|}{\textbf{\begin{tabular}[c]{@{}c@{}}$\boldsymbol{MH}$\\ (-)\end{tabular}}} &
\multicolumn{2}{c|}{\textbf{\begin{tabular}[c]{@{}c@{}}$\boldsymbol{k}$\\ (\si{\watt\per\kelvin\per\meter})\end{tabular}}} & 
\multicolumn{2}{c|}{\textbf{\begin{tabular}[c]{@{}c@{}}$\boldsymbol{C_p \cdot \rho}$\\ (\si{\joule\per\kelvin\per\cubic\centi\meter})\end{tabular}}} & 
\multicolumn{2}{c}{\textbf{\begin{tabular}[c]{@{}c@{}}$\boldsymbol{C_v \cdot \rho}$\\ (\si{\joule\per\kelvin\per\cubic\centi\meter})\end{tabular}}} \\
\cline{2-21}
& kde                                   & SD (\%)                                  & kde                                   & SD (\%)                                & kde                                  & SD (\%)                                  & kde                                    & SD (\%)                                & kde                                   & SD (\%)                                & kde                                  & SD (\%)                                  & kde                                 & SD (\%)                                 & kde                                     & SD (\%)                                  & kde                                       & SD (\%)                                      & kde                                       & SD (\%)                                      \\
\hline
\textbf{\ce{CaSiO3}   ht2}     & 1,813.3                                  & 23.8                                   & 50.5                                   & 0.0                                   &                                        &                                        & 60.4                                   & 178.3                                    & 118.5                                  & 0.0                                   &                                        &                                        &                                       &                                       &                                          &                                         &                                             &                                            & 4.6                                         & 0.0                                        \\                      
\textbf{\ce{GeO2} rt}          &                                         &                                        &                                        &                                       &                                        &                                        & 247.0                                  & 0.0                                    &                                        &                                       &                                        &                                        &                                       &                                       &                                          &                                         & 4.8                                         & 13.4                                       &                                             &                                            \\                      
\textbf{\ce{NbN} ht2}          & 2,323.3                                  & 11.3                                   & 333.0                                  & 0.0                                   &                                        &                                        & 148.1                                  & 99.8                                    & 385.0                                  & 0.0                                   &                                        &                                        &                                       &                                       &                                          &                                         &                                             &                                            & 3.4                                         & 0.0                                        \\                      
\textbf{\ce{VC}}               & 3,083.0                                  & 0.4                                    & 223.0                                  & 1.6                                   &                                        &                                        & 340.1                                  & 7.2                                    & 545.9                                  & 11.6                                  & 21.8                                   & 11.6                                   &                                       &                                       &                                          &                                         & 3.0                                         & 0.0                                        & 3.5                                         & 0.0                                        \\       
\textbf{\ce{VN rt}}            & 2,323.0                                  & 0.0                                    & 157.0                                  & 0.0                                   &                                        &                                        & 346.0                                  & 15.5                                   & 409.0                                  & 0.0                                   & 15.2                                   & 0.0                                    &                                       &                                       &                                          &                                         &                                             &                                            & 4.1                                         & 0.0                                       \\           
\textbf{\ce{Zn_{0.5}Mn_{0.5}Fe2O4}}           &                                         &                                        & 66.1                                   & 0.0                                   &                                        &                                        &                                        &                                        & 153.0                                  & 0.0                                   &                                        &                                        &                                       &                                       &                                          &                                         & 5.3                                         & 0.0                                        &                                             &                                            \\                      
\hline
\hline
\end{tabular}
}
\end{table*}

\begin{table*}
\caption{Top 21 PFM candidates based on the Pareto ranking procedure applied to the $1p$ subset of materials, where only materials with at least one property are classified (see Tab.~\ref{tab_category}). The procedure criteria are summarized in Tab.~\ref{tab_pareto}. The KDE (Kernel Density Estimated) and SD (Standard Deviation) columns provide information on the estimated value and variability of each property in the database. SD values are expressed as a percentage of the KDE value. Materials undergoing the thickness, $3p$ and $2p$ Pareto rankings are excluded from this procedure.
The full list consists of 26 materials, but only 21 are shown in the table. The remaining 5 materials, which belong to the same polymorphs (\ce{BC2N} and \ce{C3N4}), have been omitted as they exhibit highly similar properties. \ce{CoC36Cl2} sta4, \ce{FeC24Cl3}, and \ce{SbC_{16.09}C} are the MPDS chemical formulas for graphite intercalated with \ce{CoCl2}, \ce{FeCl3}, and \ce{SbCl5}.}
\label{tab4}
\resizebox{\textwidth}{!}{%

\begin{tabular}{l|cc|cc|cc|cc|cc|cc|cc|cc|cc|cc}
\hline
\hline
\textbf{Chemical formula}    & 
\multicolumn{2}{c|}{\textbf{\begin{tabular}[c]{@{}c@{}} $ \boldsymbol{T_m}$\\ (\si{\kelvin})\end{tabular}}} & 
\multicolumn{2}{c|}{\textbf{\begin{tabular}[c]{@{}c@{}}$\boldsymbol{G}$\\ (\si{\giga\pascal})\end{tabular}}} & 
\multicolumn{2}{c|}{\textbf{\begin{tabular}[c]{@{}c@{}}$\boldsymbol{B_S}$\\ (\si{\giga\pascal})\end{tabular}}} & 
\multicolumn{2}{c|}{\textbf{\begin{tabular}[c]{@{}c@{}}$\boldsymbol{B_T}$\\ (\si{\giga\pascal})\end{tabular}}} & 
\multicolumn{2}{c|}{\textbf{\begin{tabular}[c]{@{}c@{}}$\boldsymbol{E}$\\ (\si{\giga\pascal})\end{tabular}}} & 
\multicolumn{2}{c|}{\textbf{\begin{tabular}[c]{@{}c@{}}$\boldsymbol{KH}$\\ (\si{\giga\pascal})\end{tabular}}} & 
\multicolumn{2}{c|}{\textbf{\begin{tabular}[c]{@{}c@{}}$\boldsymbol{MH}$\\ (-)\end{tabular}}} &
\multicolumn{2}{c|}{\textbf{\begin{tabular}[c]{@{}c@{}}$\boldsymbol{k}$\\ (\si{\watt\per\kelvin\per\meter})\end{tabular}}} & 
\multicolumn{2}{c|}{\textbf{\begin{tabular}[c]{@{}c@{}}$\boldsymbol{C_p \cdot \rho}$\\ (\si{\joule\per\kelvin\per\cubic\centi\meter})\end{tabular}}} & 
\multicolumn{2}{c}{\textbf{\begin{tabular}[c]{@{}c@{}}$\boldsymbol{C_v \cdot \rho}$\\ (\si{\joule\per\kelvin\per\cubic\centi\meter})\end{tabular}}} \\
\cline{2-21}
& kde                                   & SD (\%)                                  & kde                                   & SD (\%)                                & kde                                  & SD (\%)                                  & kde                                    & SD (\%)                                & kde                                   & SD (\%)                                & kde                                  & SD (\%)                                  & kde                                 & SD (\%)                                 & kde                                     & SD (\%)                                  & kde                                       & SD (\%)                                      & kde                                       & SD (\%)                                      \\
\hline
\textbf{\ce{BC2N} hyp}           &                 &     & 444.4 & 0.0   &  &  & 391.4 & 0.0    &       &      &      &   &     &   &       &      &      &      &     &       \\
\textbf{\ce{BeAl2O4}}            &                 &     &       &     &  &  &       &      &       &      &      &   & 8.5 & 0.0 &       &      &      &      &     &       \\
\textbf{\ce{C3N4} tf}            &                 &     & 60.5  & 0.0   &  &  & 432.0   & 0.0    &       &      &      &   &     &   &       &      &      &      &     &       \\
\textbf{\ce{CoC36Cl2} sta4}  &     &       &     &  &  &     &  &      &       &      &      &   &     &   & 500.0   & 0.0    &      &      &     &       \\
\textbf{\ce{FeC24Cl3}}           &                 &     &       &     &  &  &       &      &       &      &      &   &     &   & 484.9 & 52.8 &      &      &     &      \\
\textbf{HfN}                & 3,583.0            & 0.0   &       &     &  &  &       &      &       &      & 16.7 & 0.0 &     &   &       &      &      &      &     &       \\
\textbf{\ce{Mg2B25C4}}        &     &       &     &  &  &       &      &       &    &  & 31.5 & 0.0 &     &   &       &      &      &      &     &       \\
\textbf{\ce{ReB2}}               &                 &     & 283.0   & 4.4 &  &  & 359.9 & 2.8  & 599.2 & 24.2 &      &   &     &   &       &      &      &      &     &       \\
\textbf{\ce{Ru_{0.25}Co_{0.75}}}     &     &       &     &  &  & & 427.0   & 0.0    &       &      &      &   &     &   &       &      &      &      &     &       \\
\textbf {\ce{SbC_{16.09}C}} &     &       &     &  &  &       &      &       &      &      & &  &     &   & 295.0   & 63.2 &      &      &     &       \\
\textbf{SiC 2H}            & 3,103.0            & 0.0   &       &     &  &  & 260.0   & 9.5  & 484.0   & 0.0    & 27.1 & 0.0 &     &   &       &      &      &      &     &       \\
\textbf{\ce{Ta2C rt}}            & 3,600.0            & 0.0   &       &     &  &  &       &      &       &      &      &   &     &   &       &      &      &      &     &      \\
\textbf{TaN ht}             & 3,363.0            & 0.0   & 179.0   & 4.1 &  &  & 367.5 & 13.2 & 772.7 & 32.1 &      &   &     &   &       &      &      &      &     &      \\
\textbf{Te tf}              &                 &     &       &     &  &  & 425.0   & 0.0    &       &      &      &   &     &   &       &      &      &      &     &      \\
\textbf{\ce{TiFe3N}}             &                 &     &       &     &  &  & 596.0   & 0.0    &       &      &      &   &     &   &       &      &      &      &     &       \\
\textbf{\ce{TiO2 ram}}           &                 &     &       &     &  &  & 431.0   & 0.0   &       &      &      &   &     &   &       &      &      &      &     &       \\
\textbf{\ce{TiVO3}}              &                 &     &       &     &  &  &       &      &       &      &      &   &     &   &       &      & 3.8  & 14.6 & 3.9 & 6.7   \\
\textbf{\ce{VFe3N}}              &                 &     &       &     &  &  & 489.9 & 36.1 &       &      &      &   &     &   &       &      &      &      &     &       \\
\textbf{\ce{W2C ht1}}            &                 &     &       &     &  &  & 490.0   & 10.1 &       &      &      &   &     &   &       &      &      &      &     &       \\
\textbf{WC}                & 3,140.6          & 1.3 & 405.1 & 0.0   &  &  & 576.9 & 34.9 & 848.0   & 2.9  & 18.4 & 0.0 &     &   &       &      &      &      &     &       \\
\textbf{\ce{YB66}}               &                 &     &       &     &  &  &       &      &       &      & 25.7 & 0.0 &     &   &       &      &      &      &     &       \\
\hline
\hline
\end{tabular}
}
\end{table*}

\begin{figure*}
  \includegraphics[width=\textwidth,height=0.9\textheight,keepaspectratio]{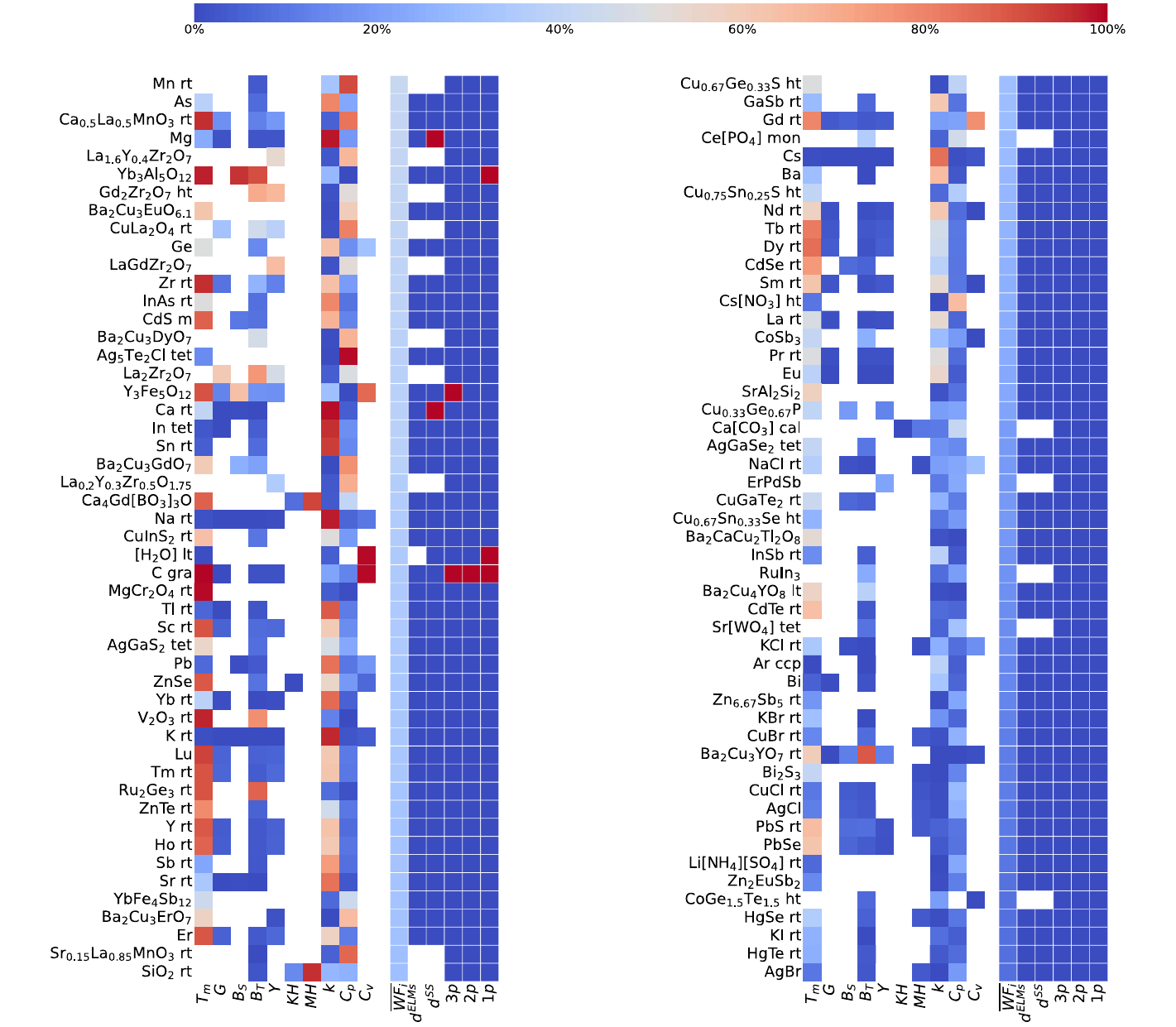}
  \caption{Heatmap representation of the materials ranked from 101 (top left) to 200 (bottom right) according to the comparative ranking procedure. The materials are classified into the $3p$ subset and arranged in order based on their weighted average win-fraction ($\overline{WF_{i}}$) scores, with a maximum potential score of 100\%. The figure also includes the individual win-fraction ($WF_{i,k}$) values for each property of the materials.}
  \label{fig:wf_3.2}
\end{figure*}

\begin{figure*}
  \includegraphics[width=\textwidth,height=0.9\textheight,keepaspectratio]{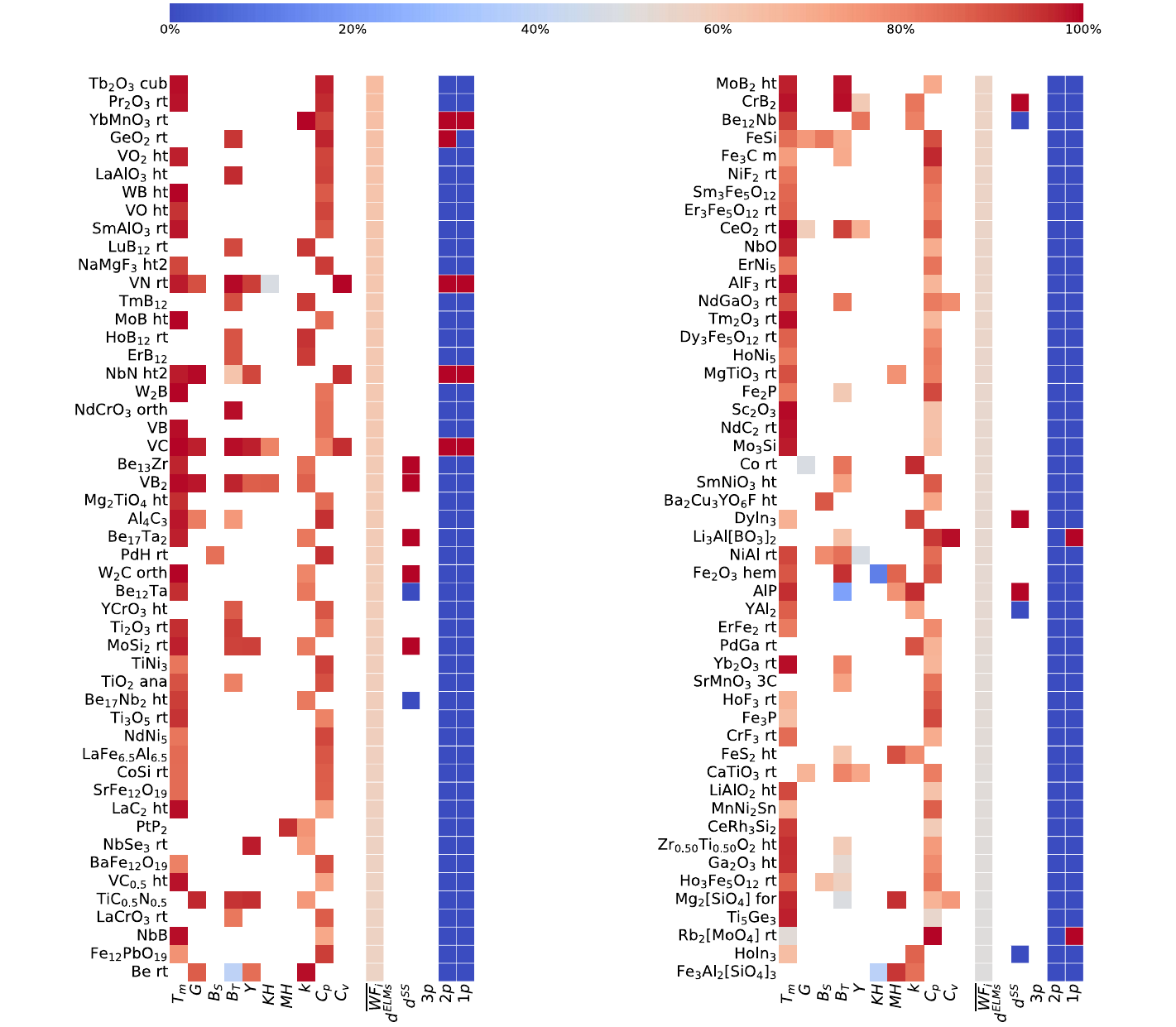}
  \caption{Heatmap representation of the materials ranked from 1 (top left) to 100 (bottom right) according to the comparative ranking procedure. The materials are classified into the $2p$ subset and arranged in order based on their weighted average win-fraction ($\overline{WF_{i}}$) scores, with a maximum potential score of 66.7\%. The figure also includes the individual win-fraction ($WF_{i,k}$) values for each property of the materials.}
  \label{fig:wf_2.1}
\end{figure*}

\begin{figure*}
  \includegraphics[width=\textwidth,height=0.9\textheight,keepaspectratio]{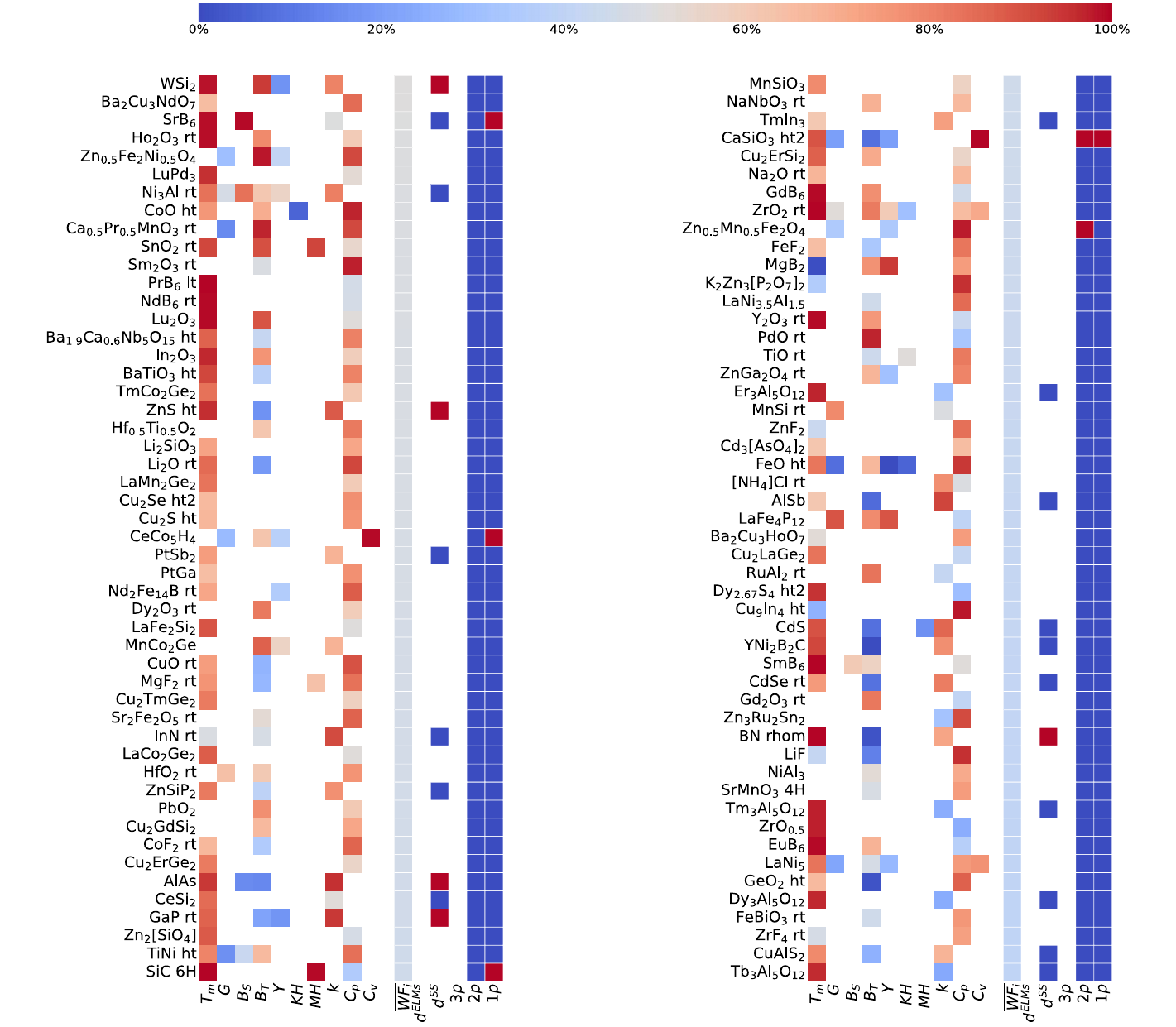}
  \caption{Heatmap representation of the materials ranked from 101 (top left) to 200 (bottom right) according to the comparative ranking procedure. The materials are classified into the $2p$ subset and arranged in order based on their weighted average win-fraction ($\overline{WF_{i}}$) scores, with a maximum potential score of 66.7\%. The figure also includes the individual win-fraction ($WF_{i,k}$) values for each property of the materials.}
  \label{fig:wf_2.2}
\end{figure*}

\begin{figure*}
  \includegraphics[width=\textwidth,height=0.9\textheight,keepaspectratio]{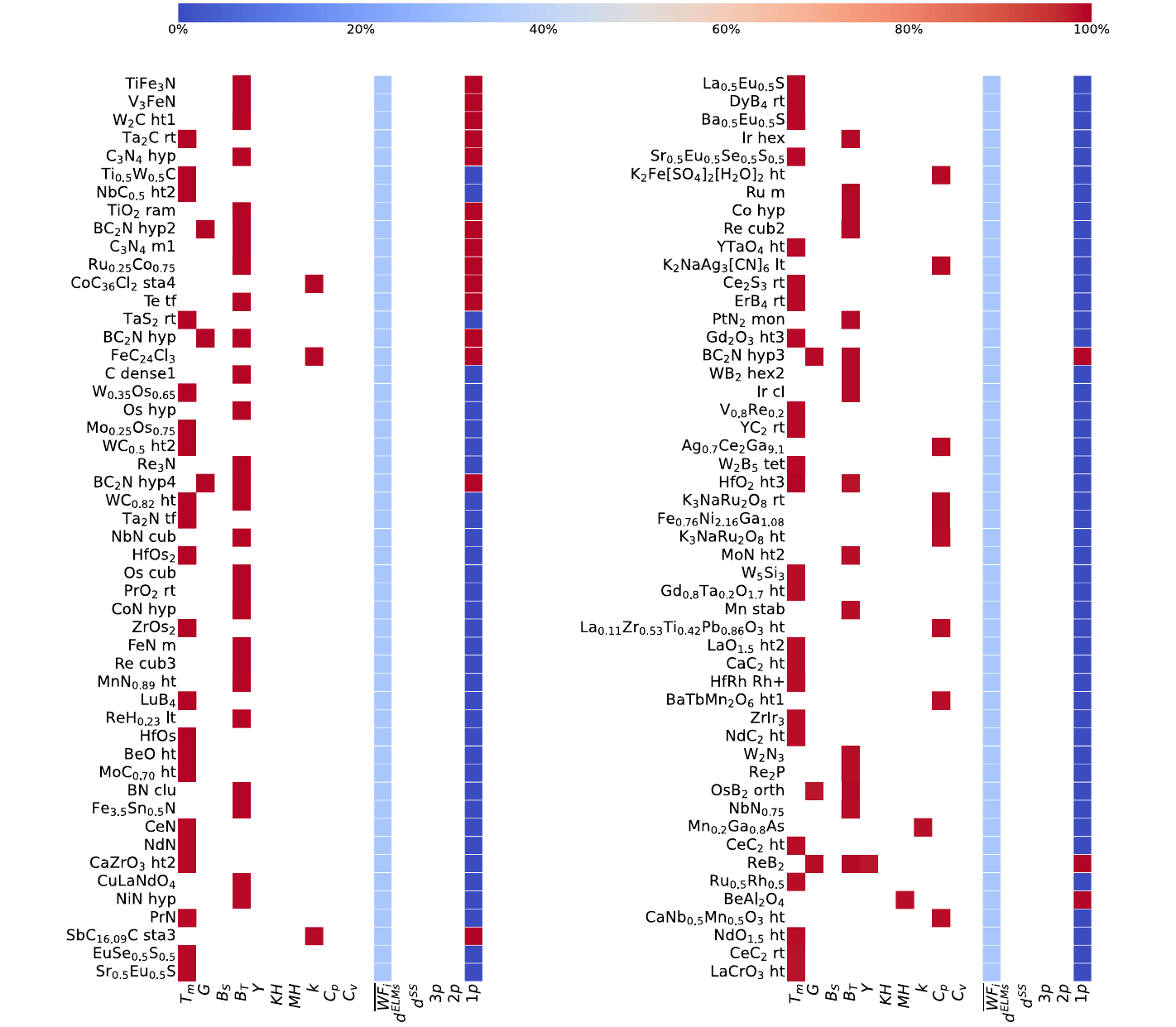}
  \caption{Heatmap representation of the materials ranked from 1 (top left) to 100 (bottom right) according to the comparative ranking procedure. The materials are classified into the $1p$ subset and arranged in order based on their weighted average win-fraction ($\overline{WF_{i}}$) scores, with a maximum potential score of 33.3\%. The figure also includes the individual win-fraction ($WF_{i,k}$) values for each property of the materials.}
  \label{fig:wf_1.1}
\end{figure*}

\begin{figure*}
  \includegraphics[width=\textwidth,height=0.9\textheight,keepaspectratio]{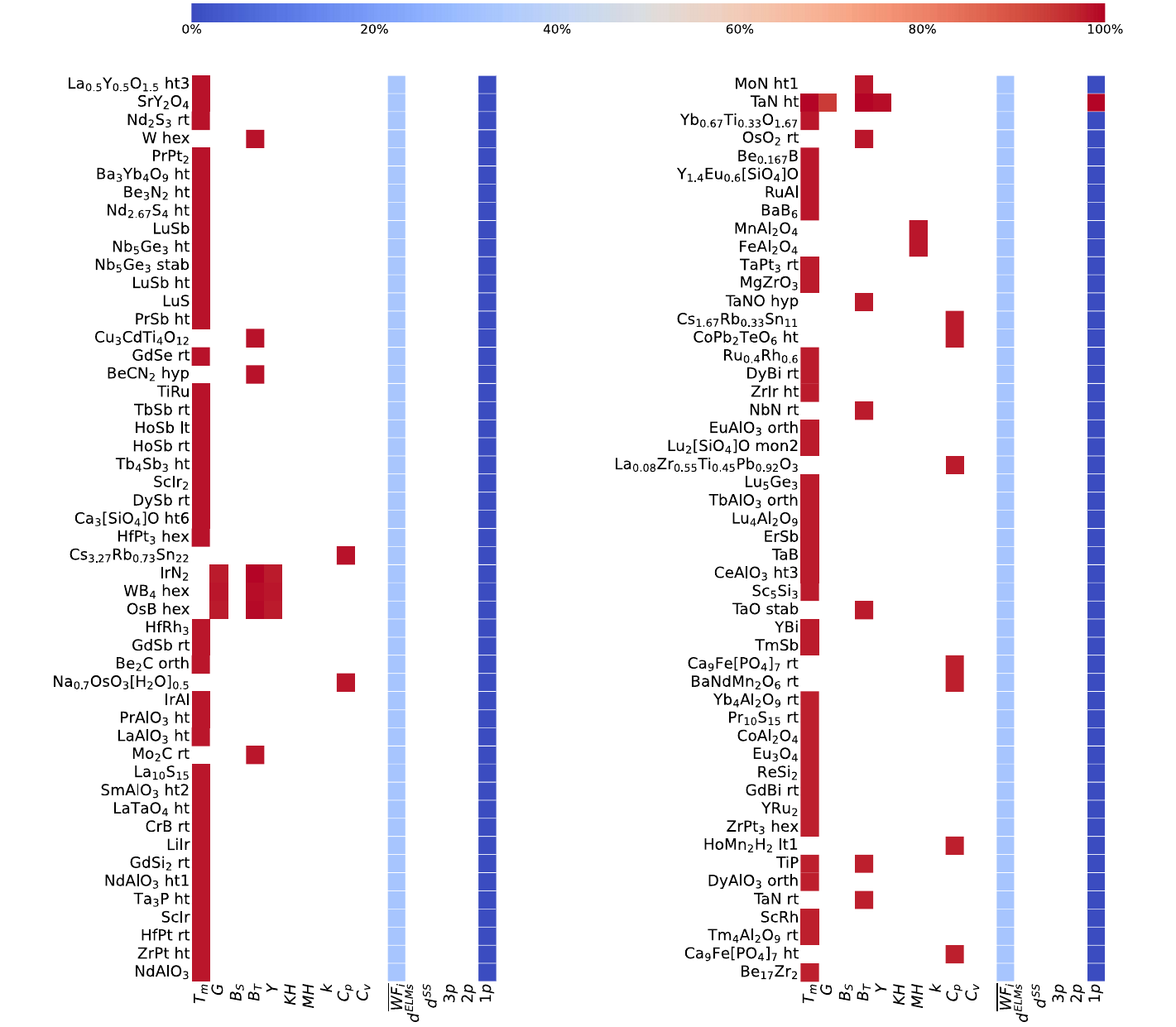}
  \caption{Heatmap representation of the materials ranked from 101 (top left) to 200 (bottom right) according to the comparative ranking procedure. The materials are classified into the $1p$ subset and arranged in order based on their weighted average win-fraction ($\overline{WF_{i}}$) scores, with a maximum potential score of 33.3\%. The figure also includes the individual win-fraction ($WF_{i,k}$) values for each property of the materials.}
  \label{fig:wf_1.2}
\end{figure*}

\FloatBarrier

\bibliography{apssamp}

\end{document}